\documentclass{amsart}
\usepackage{amssymb}
\usepackage{amssymb,amsmath,amscd,mathrsfs,amsbsy,young,amsfonts}
\usepackage{pstricks,pst-node}
\usepackage[metapost]{mfpic}
\usepackage{mflogo}

\psset{unit=.5cm}

\usepackage{graphicx}
\usepackage{amscd}
\usepackage{graphics}
\newtheorem{Theorem}{Theorem}[section]
\newtheorem{Proposition}{Proposition}[section]
\newtheorem{Corollary}{Corollary}[section]
\newtheorem{Lemma}{Lemma}[section]

\def\proof{\par{\it Proof}. \ignorespaces}
\def\endproof{{\ \vbox{\hrule\hbox{%
     \vrule height1.3ex\hskip0.8ex\vrule}\hrule }}\par}
\newenvironment{Proof}{\proof}{\endproof}
\def\sf#1#2{{\textstyle\frac{#1}{#2}}}

\theoremstyle{definition}
\newtheorem{Definition}[Theorem]{Definition}
\newtheorem{Example}[Theorem]{Example}

\theoremstyle{remark}
\newtheorem{Remark}[Theorem]{Remark}

\numberwithin{equation}{section}

\numberwithin{figure}{section}

\let\trueint=\int
\let\truesum=\sum
\def\int{\mathop{\textstyle\trueint}\limits}
\def\sum{\mathop{\textstyle\truesum}\limits}

\let\<=\langle
\let\>=\rangle

\def\sech{\mathop{\rm sech}\nolimits}

\renewcommand\labelitemi{\ifmmode\circ\else$\circ$\fi}

\begin{document}

\title[Soliton solutions of the KP equation]
{Soliton solutions of the KP equation and application to shallow water waves}

\author{Sarbarish Chakravarty$^\dagger$}
%  Address of record for the research reported here
\address{Department of Mathematics, University of Colorado, Colorado Springs, CO 80933}
\email{chuck@math.uccs.edu}

\author{Yuji Kodama$^*$}
\address{Department of Mathematics, Ohio State University, Columbus,
OH 43210}
\email{kodama@math.ohio-state.edu}

\thanks{$^\dagger$ Partially supported by NSF grant DMS-0807404,  $^*$ Partially
supported by NSF grant DMS0806219}

\maketitle
\begin{abstract}
The main purpose of this paper is to give a survey of recent developments on a classification
of soliton solutions of the KP equation. The paper is self-contained, and
we give complete proofs of theorems needed for
the classification. The classification is based on the totally non-negative cells in the Schubert
decomposition of the real Grassmann manifold, Gr$(N,M)$,
the set of $N$-dimensional subspaces in $\mathbb{R}^M$.
Each soliton solution defined on Gr$(N,M)$ asymptotically consists of the $N$ number of 
 line-solitons for $y\gg 0$ and
the $M-N$ number of line-solitons for $y\ll 0$. In particular, we give
detailed description of the soliton solutions associated with Gr$(2,4)$, which
play a fundamental role in the study of multi-soliton solutions.
 We then consider a physical application of some  of  those
solutions related to the Mach reflection discussed by
J. Miles in 1977.
\end{abstract}

\tableofcontents
\thispagestyle{empty}

%%%%%%%%%%%%%%%%%%%%%%%%%%%%%%%%%%%%%%%%%%%%%%%%%%%%%%%%%%%%%%%%%%%%%%%%%%%%%%%
\section{Introduction}
The purpose of this paper is to give a survey of our recent works on a classification
theory of soliton solutions of the Kadomtsev-Petviashvili (KP) equation \cite{BK:03, K:04, BC:06, CK:08, CK:08b}.
We also explain the detailed structure of some of the soliton solutions obtained
in this classification theory, and
discuss a physical application of a subclass of those solutions, which is related to the resonant interactions of the solitary
waves in shallow water.
Most of the materials in  this paper are based on a series of lectures given by one of the authors (YK)
at Chinese Academy of Science in Beijing after the international conference ``Nonlinear Waves - Theory and Applications'', on June 9-12, 2008.
It is a pleasure to thank Professors Qing-Ping Liu,  Xing-Biao Hu and Ke Wu 
for the arrangement of the lectures,
and Mr. Kai Tian for making lecture notes.

In 1970, Kadomtsev and Petviashvili \cite{KP:70} proposed a two-dimensional dispersive wave equation
to study the stability of one soliton solution of the KdV equation under the influence of
weak transverse perturbations. This equation is now referred to as the KP equation, and
considered to be a prototype of the integrable nonlinear dispersive wave equations in two-dimension.
The KP equation is a completely integrable system with remarkably rich mathematical
structure which is well documented in several monographs (see for examples,
\cite{AS:81, AC:91, D:91, H:04, MS:91, MJD:00, N:85, NMPZ:84}). These mathematical structures include the existence of $N$-soliton solutions, the Lax formulation of
the inverse scattering transform, and the existence of an infinite dimensional symmetries.
However the real breakthrough in the KP theory occurred in 1981 by Sato \cite{S:81}  who realized that the solution
of the KP equation is given by a GL$(\infty)$-orbit on an infinite dimensional
Grassmann manifold (Sato universal Grassmannian), and the KP equation is just a Pl\"ucker relation
on the Grassmannian.
The present paper deals with a finite dimensional version of the Sato theory, and
in particular, we are interested in the solutions which are real and non-singular
in the entire $xy$-plane where they are localized along certain rays.

It is quite important to recognize that the resonant interaction plays a fundamental role in
multi-dimensional waves. The original description of the soliton interaction of the KP equation
 was based on a two-soliton solution found in the Hirota form, which has the shape of ``X" describing an
intersection of two lines with some angle and a shift of lines at the intersection
(the phase shift). In 1977, Miles \cite{M:77} pointed out that this two-soliton solution
(referred to as the ``O-type" solution, where ``O" stands for {\it original}) becomes singular
if the angle of the intersection is smaller than certain critical value.
Since the KP equation was proposed to describe {\it quasi}-two dimensional waves,
it should admit solutions that provide good approximations for two soliton interaction with smaller angle.
Thus, it seems strange that one does not have a reasonable solution in the parameter regimes where the KP equation
is supposed to give a better approximation.
Miles also found that at the critical angle the two line-solitons of the O-type solution interact resonantly, and
a third wave (soliton) is created to make a ``Y-shape" solution. Indeed, it turns out that such Y-shape resonant solutions are exact solutions of the KP equation
(see also \cite{NR:77}).

After the discovery of the resonant phenomena in the KP equation, several
numerical and experimental studies were performed to investigate
 resonant interactions in other {\it real}
two dimensional equations such as the ion-acoustic and shallow water
wave equations under the Boussinesq approximation (see for examples
\cite{KY:80, KY:82, NN:83, FID:80, MK:96}). However, after these activities, no significant progress 
has been made in the study of the solution space or real applications of the KP equation. 
It would appear that the general perception was that there were not many new and significant results left 
to be uncovered in
the soliton solutions of the KP theory. For the last 5 years, we have been working on the classification
problem of the soliton solutions of the KP equation \cite{BK:03, K:04, BC:06, CK:08, CK:08b}, and 
our studies have revealed a large variety of soliton
solutions which were totally overlooked in the past.
In this paper, we give a brief survey of our research and explain an application of our new soliton
solutions to describe the resonant interaction of solitons in shallow water.

The paper is organized as follows:

In Section \ref{C:KP}, we describe an interesting connection between the KP equation and the Burgers equation (Theorem \ref{KP-B}), and then construct some exact solutions based on this connection.
We show that the resonant interaction in the KP equation is a natural consequence of the
mathematical structure of the KP equation, that is, the Burgers equation gives a symmetry of
the KP equation in the $y$-direction. The Wronskian determinant structure of the solutions
is derived from an extension of the Burgers hierarchy (Theorem \ref{PHrelation}, see also \cite{FN:83, Ha:87}).

In Section \ref{C:Sato}, we present a brief introduction of the Sato theory, and give  a direct
construction of the Wronskian structure of the $\tau$-function (see also
\cite{S:81, MJD:00}). The KP hierarchy is
 the set of the symmetries of the KP equation, and the geometric structure
 of the solutions of the KP hierarchy is given by the $\tau$-function. We explain
 the important role that the $\tau$-function plays in the integrability of the KP equation.
 
 In Section \ref{Grassmann}, we give an elementary introduction of the real Grassmann manifold
 Gr$(N,M)$, the set of $N$-dimensional subspaces of $\mathbb{R}^M$. This is necessary
  background information for the classification problem of the soliton solutions of the KP equation.
 In particular, we explain the basic structure of the $\tau$-function in terms of the Schubert
 decomposition of the Grassmannian Gr$(N,M)$.
 
 In Section \ref{C:Classification}, we present the main theorem of the classification problem
 of the soliton solution (Theorem \ref{derangement}). This theorem states that the $\tau$-function identified as a point on  Gr$(N,M)$ generates a solution that has asymptotically $M-N$ line-solitons
 for $y\ll 0$ and $N$ line-solitons for $y\gg 0$. Moreover, these soliton solutions can be
 parametrized by the chord diagrams associated with certain permutation called {\it derangements}
  of the symmetry group
 $S_M$. This type of solutions is called $(M-N,N)$-soliton solution.
 We give a new proof of the Theorem, some of which did not appear in
 the previous papers. Specifically in Corollary \ref{pairing}, we show how the conserved densities of the KP equation
 lead directly to establish the correspondence between the line-soliton solutions and the derangements of $S_M$.
 
 In Section \ref{C:2-2soliton}, we study the detailed structure of $(2,2)$-solitons defined on Gr$(2,4)$ as a fundamental
 class of soliton solutions of the KP equation. We present the detailed description of the solutions
 based on the $2\times 4$ matrix which marks the point of Gr$(2,4)$.  The solution is completely determined if
 one prescribes a specific data for this matrix.
 We then show that for some classes of $(2,2)$-soliton solutions, the matrix data is equivalent to {\it asymptotic} data
 such as the soliton locations and phase shifts, while for some other solution classes,  one also need to
 specify  {\it internal} data given by a local structure.
 
 In Section \ref{IVP},  we discuss
 an application of our new solutions to the Mach reflection in shallow water with a rigid wall,
 based on the results of Section \ref{C:2-2soliton}.
 The resonant interaction among the incident wave, the reflection wave and the Mach stem  
 is well described by our soliton solution, if we ignore the effect of the boundary layer at the wall.
 We also present a direct numerical simulation of the KP equation with a V-shape initial wave,
 which represents the reflection problem of the incident wave on an inclined wall (see also
 \cite{F:80, PTLO:05, TO:07}).

\section{The KP equation}\label{C:KP}
The KP equation is the following partial differential 
equation in $2+1$ dimensions,
\begin{equation}\label{kp}
(-4u_t+6uu_x+u_{xxx})_x+3\beta u_{yy}=0.
\end{equation}
When $\beta=1$, this equation is referred to as
the KPII equation, while the equation with $\beta=-1$ is called the KP I equation.
Since we consider only the KPII equation in this paper, we simply refer to
the KPII as the KP equation throughout the text.

Note when $u$ is independent of $y$, the well-known Kortweg-de Vries (KdV)
equation is obtained from (\ref{kp})
\begin{equation*}
-4u_t+6uu_x+u_{xxx}=0 \,,
\end{equation*}
which is why the KP equation is regarded as a $(2+1)$-dimensional extension
of the KdV equation. Physically, the KP equation was introduced to study the stability
of the KdV soliton under the influence of weak transverse perturbations \cite{KP:70}.
So the KP equation should give a better approximation of the original physical system
if the system depends weakly on the transverse direction, i.e. the waves are almost
parallel to the $y$-axis.

Let $w = w(x,y,t)$ be a function defined by
$$u=2w_x \,,$$
then the KP equation (\ref{kp}) takes the form
\begin{equation*}
(-4w_{xt}+12w_xw_{xx}+w_{xxxx})_x+3w_{xyy}=0 \,.
\end{equation*}
Integrating the above equation once with respect to $x$ and setting the
integration constant to zero
(i.e. $w$ is assumed to be constant as $|x|\to\infty$), give the {\it potential} form 
of the KP equation, namely,
\begin{equation}\label{pkp}
(-4w_t+6w_x^2+w_{xxx})_x+3w_{yy}=0.
\end{equation}
We will henceforth refer (\ref{pkp}) as the pKP equation for the function $w$
which plays an important role in our description of the soliton solutions.

We next review an interesting connection between the KP equation
and the Burgers hierarchy, first established systematically in \cite{Ha:87}.
This connection is the basis of our description of the fundamental structure of the      
soliton solutions discussed in this paper.

\subsection{The Burgers hierarchy and the KP equation}\label{B-KPrelation}
Let us first briefly review the Cole-Hopf transformation linking
the heat (linear) and Burgers hierarchies:
\begin{Proposition} \label{Heat}
If $f$ is a solution of the set of linear equations, i.e. the heat hierarchy,
\begin{equation}\label{heat}
\partial_{t_n}f=\partial_x^n f 
\qquad {\rm for}\quad n=1,2,\ldots\,,
\end{equation}
then $w :=\partial_x(\ln f)$ satisfies the following set of nonlinear equations,
i.e. the Burgers hierarchy, 
\begin{equation}\label{bh}
\partial_{t_n} w=\partial_x(\partial_x+w)^{n-1}w\qquad{\rm for}\quad n=1,2,\ldots\,.
\end{equation}
\end{Proposition}
\begin{Proof} First, by induction, we prove that
\[
\partial_x^nf=[(\partial_x+w)^{n-1}w]f := P_nf.
\]
When $n=1$, the above formula is just the definition of the function $w$.
Now, suppose
$\partial_x^{n-1}f=[({\partial_x}+w)^{n-2}w]f := P_{n-1}f$.
Then,
\[ \partial_x^nf = \partial_x(P_{n-1}f) = \partial_x(P_{n-1})f+P_{n-1}wf
= [(\partial_x + w)P_{n-1}]f = P_nf\,.
\]
Therefore, 
$\partial_{t_n} w=\partial_{t_n}(\partial_x \ln f) =
\partial_x(\partial_{t_n}ff^{-1})= \partial_x(\partial_x^n ff^{-1}) 
= \partial_x(\partial_x+w)^{n-1}w $.
\end{Proof}
Note here that the Burgers hierarchy is well-defined in the sense that all the members
commute, because the equations for $f$ in (\ref{heat}) are commute trivially.
The Burgers equation corresponds to $n=2$ in (\ref{bh}), i.e.
the first nontrivial member of the hierarchy (\ref{bh}), and the members
for $n=2,3$ are given by
\begin{equation}\label{bh23}\left\{
\begin{array}{lll}
\partial_{t_2}w&=\partial_x(w_x+w^2),\\[1.5ex]
\partial_{t_3}w&=\partial_x(w_{xx}+3ww_x+w^3).
\end{array}\right.
\end{equation}
Note that the even members of the hierarchy are {\it dissipative}, while the odd ones
are {\it dispersive}. This feature is a key for the resonance phenomena in the soliton
solutions of the KP equation (see Example~\ref{ex:12soliton}).

Now we show the connection between the KP equation and the Burgers hierarchy (\ref{bh}):
\begin{Proposition}\label{Burgers-KP}
Suppose $w(x,t_2,t_3)$ is a common solution of (\ref{bh23}) with $t_2=y$ and $_3=t$, 
then $w=w(x,y,t)$ solves the   
pKP equation (\ref{pkp}).
\end{Proposition}
\begin{Proof} Expressing $w_{yy}, \, w_{xt}$ in terms of only
the $x$-partial derivatives of $w$, we have
\begin{align*}
w_{yy}&=\partial_x(w_{xxx}+2w_x^2+4ww_{xx}+4w^2w_x),\\
 w_{xt}&=\partial_x(w_{xxx}+3w_x^2+3ww_{xx}+3w^2w_x).
 \end{align*}
Then eliminating the common terms including $(ww_{xx}+w^2w_x)$, i.e.
calculate $3w_{yy}-4w_{xt}$,
we obtain the pKP equation (\ref{pkp}).
\end{Proof}
Thus the KP equation contains the Burgers hierarchy, and
the next theorem combines the results of Propositions \ref{Heat} and \ref{Burgers-KP},
which gives a {\it linearization} of the KP equation:
\begin{Theorem}\label{KP-B}
If $f(x, t_2, t_3, \ldots) $ satisfies the set of linear equations,
$\partial_{t_n}f=\partial_x^n f \,, \,\, n=1,2,\ldots $ 
with $t_2 := y, t_3 := t$, then $w=\partial_x(\ln f)$ satisfies the pKP equation (\ref{pkp}).
\end{Theorem}
This implies that a solution of the KP equation obtained from the Burgers hierarchy
(equivalently the set of linear equations) has a dissipative behavior in the $y$-direction.
In particular, one should note that a confluence of shocks~\cite{Wh:74} 
in the Burgers equation corresponds to a fusion of solitons as shown 
by Example~\ref{ex:12soliton} of the next section.

\begin{Remark}
One should also note that the infinitely many commuting symmetries of the Burgers hierarchy
(\ref{bh}) induces higher flows of the pKP hierarchy. In other words, proceeding in the 
same way as in Proposition \ref{Burgers-KP} one could also obtain an equation
for $w(x,t_2,\ldots)$ in the first $n$ variables $x, t_2, t_3, \ldots, t_n$, which
form the $n$th member of the pKP hierarchy. The substitution $u=2w_x$ then gives
the corresponding member of the KP hierarchy. Thus, the KP equation (\ref{kp})
corresponds to the $n=3$ flow of the KP hierarchy, and the higher flows generate
the infinitely many symmetries of the KP equations.
The KP hierarchy plays an important role when we discuss the multi-soliton solutions
as we show in this paper.
\end{Remark}

\subsection{Some exact solutions}
We now consider some solutions of the KP equation obtained from the linear system
(\ref{heat}). The general solution for this linear system 
with $t_2=y$ and $t_3=t$ admits the integral representation (Ehrenpreis principle)
\[
f(x,y,t)=\int_C e^{kx+k^2y+k^3t}\rho(k)\,dk, 
\]
with an appropriate measure $\rho(k)\,dk$ and a proper contour $C$ in the complex plane.
A particularly simple finite dimensional solution is recovered by choosing 
\[
\rho(k)\,dk=\sum^M_{j=1}a_j\delta(k-k_j)\,dk
\]
with arbitrary real constants $k_j$ and $a_j$ for $j=1,2,\ldots,M$, and the contour
$C=\mathbb{R}$. Then
\begin{equation}\label{E}
f(x,y,t) = \sum_{j=1}^M a_j \,E_j(x,y,t)\qquad  E_j(x,y,t):=e^{\theta_j(x,y,t)}\,, 
\end{equation}
with $\theta_j :=k_jx+k_j^2y+k_j^3t$.
Here we assume $a_j$ to be positive, so that $f$ is positive definite, and also assume the
ordering in the $k$-parameters,
\begin{equation}\label{ordering}
k_1<k_2<\cdots<k_M.
\end{equation}
Now let us give some explicit examples:
\begin{Example}\label{onesolitonexample}
Consider the case with $M=2$, i.e. $f=a_1E_1+a_2E_2$.  Since $w=\partial_x(\ln f)$, we can choose
$a_1=1$ without altering the solution. So we take, with $a_1=a$,

\[
f= E_1+aE_2=2\sqrt{a}\,e^{\frac{1}{2}(\theta_1+\theta_2)}\,\cosh\sf{1}{2}(\theta_1-\theta_2+\theta_{12}),
\]
where $\theta_{12}=-\ln a$.
Then we have
\[
u=2\partial_x^2\,\ln f= \sf{1}{2}(k_1-k_2)^2 \mathrm{sech}^2\sf{1}{2}(\theta_1-\theta_2+\theta_{12})\,,
\]
which is called the one-soliton solution. The solution
$u(x,y,t)$ is localized in the $xy$-plane along the line $\theta_1-\theta_2+\theta_{12}=0$,
whose location is determined by the constant $a$.
In particular, at $t=0$ the peak of the soliton determines the line,
\[
 x+(k_1+k_2)y=\frac{1}{k_2-k_1}\,\ln a,
\]
so that choosing $a=1$ the line of the peak crosses the origin.
Because of this, we often refer the one-soliton as a line-soliton 
(in a local sense in general). 

We also note that in terms of the function $w(x,y,t)$, we have the following 
asymptotic values for $x\to\pm\infty$ with
the ordering $k_1<k_2$, 
$$ 
w=\frac{f_x}{f}=\frac{k_1 E_1+k_2 E_2}{E_1+E_2}~
\rightarrow~\left\{\begin{array}{lll}k_2\quad&{\rm as}~ &x\to +\infty,\\
k_1\quad&{\rm as}~ &x\to -\infty. \end{array}\right. 
$$
and the solution $u =2w_x \to 0$, exponentially as $|x|\to\infty$. The change in the asymptotic 
value of the potential function $w$ for the one-soliton 
solution is due to fact that the exponential $E_1$ dominates over $E_2$
as $x \to -\infty$, whereas the exponential $E_2$ dominates over $E_1$
as $x \to \infty$. Consequently, the solution $u$ is localized along
the phase transition line: $\theta_1-\theta_2=$constant, where both exponential terms are in balance.
This motivates labeling the one-soliton solution as the $[1,2]$-soliton which
 also represents a permutation $\pi=\binom{1~2}{2~1}$ of
the index set $\{1, 2\}$, i.e. exchanging the values $k_1$ and $k_2$ by crossing the line-soliton.
\end{Example}
%%%%%%%%%%%%%%%%%%%%%%%%%%%
\begin{figure}[t]
\begin{center}
\includegraphics[height=2in]{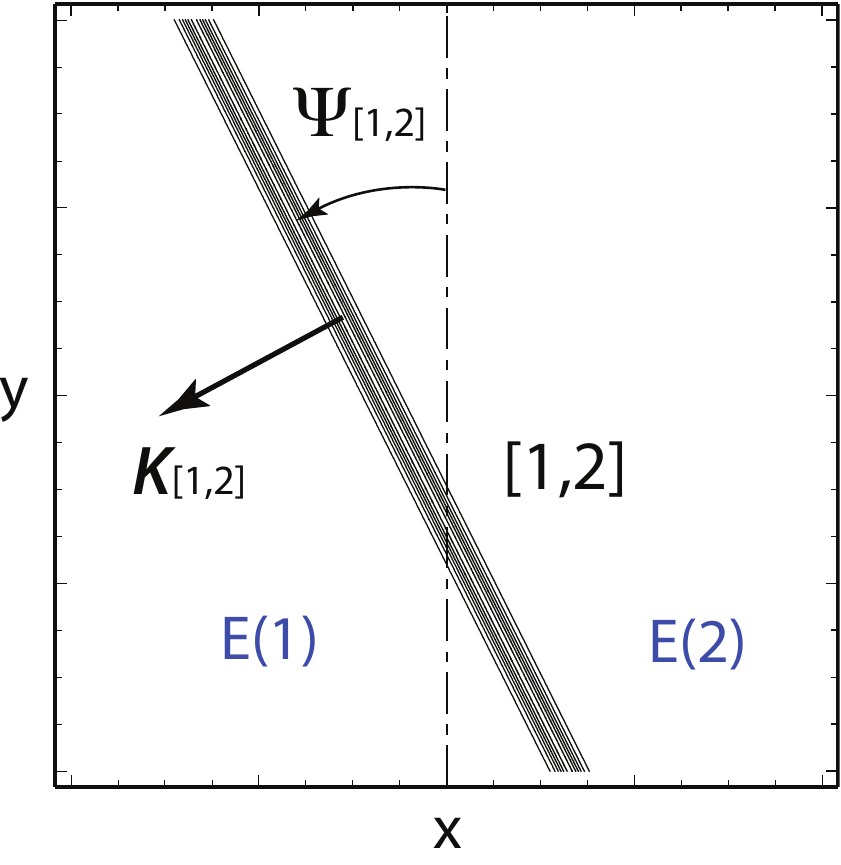}
\end{center}
\caption{A One-soliton solution. The $[1,2]$ is the label of this
line-soliton, and each $E(j)$ for $j=1,2$ indicates the dominant exponential
term $E_j$ in this region. The amplitude $A_{[1,2]}$ and the angle $\Psi_{[1,2]}$
are given by $A_{[1,2]}=\sf{1}{2}(k_2-k_1)^2$ and $\tan\Psi_{[1,2]}=k_1+k_2$.
${\bf K}_{[1,2]}=(K^x_{[1,2]},K^y_{[1,2]})$ shows the wave-vector,
and the slope is given by $\tan\Psi_{[1,2]}=K^y_{[1,2]}/K^x_{[1,2]}$.
Throughout the paper, the graphs of soliton solutions illustrate contour lines of
the function $u(x,y,t)$.
\label{fig:1soliton}}
\end{figure}
%%%%%%%%%%%%%%%%%%%%%%%%%

It is also convenient to introduce the following notations to
describe a {\it line}-soliton physically as a traveling wave:
We say that a line-soliton is of $[i,j]$-type (or simply $[i,j]$-soliton), 
if the soliton solution $u$ has the form (locally),
\begin{equation}\label{Onesoliton}
u=A_{[i,j]}\sech^2\left({\bf K}_{[i,j]}\cdot {\bf x}+\Omega_{[i,j]}t+\Theta^0_{[i,j]}\right)
\end{equation}
The amplitude $A_{[i,j]}$, the wave-vector ${\bf K}_{[i,j]}$ and the frequency 
$\Omega_{[i,j]}$ are defined by
\begin{align*}
A_{[i,j]}&=\frac{1}{2}(k_j-k_i)^2\\
{\bf K}_{[i,j]}&=\left(\frac{1}{2}(k_j-k_i),\frac{1}{2}(k_j^2-k_i^2)\right)=
\frac{1}{2}(k_j-k_i)\left(1,k_i+k_j\right),\\
\Omega_{[i,j]}&=\frac{1}{2}(k_j^3-k_i^3)=\frac{1}{2}(k_j-k_i)(k_i^2+k_ik_j+k_j^2).
\end{align*}
The direction of the wave-vector ${\bf K}_{[i,j]}=(K_{[i,j]}^x,K_{[i,j]}^y)$ is measured 
in the counterclockwise from the $x$-axis (see Figure \ref{fig:1soliton}), and
it is given by
\[
\frac{K^y_{[i,j]}}{K^x_{[i,j]}}=\tan\Psi_{[i,j]}=k_i+k_j.
\]
Note here that the $\Psi_{[i,j]}$ also gives the angle of the line $[i,j]$ from the $y$-axis in the counterclockwise,
and $-\sf{\pi}{2}<\Psi_{[i,j]}<\sf{\pi}{2}$.
For each soliton solution of (\ref{Onesoliton}),  the wave vector ${\bf K}_{[i,j]}$ and 
the frequency $\Omega_{[i,j]}$ satisfy the soliton-dispersion relation,
\begin{equation}\label{Sdispersion}
4\Omega_{[i,j]} K_{[i,j]}^x=4(K_{[i,j]}^x)^4+3(K_{[i,j]}^y)^2.
\end{equation}
The soliton velocity ${\bf V}_{[i,j]}$ defined by  
${\bf K}_{[i,j]}\cdot{\bf V}_{[i,j]}=-\Omega_{[i,j]}$ is given by
\[
{\bf V}_{[i,j]}=-\frac{\Omega_{[i,j]}}{|{\bf K}_{[i,j]}|^2}{\bf K}_{[i,j]}=
\frac{k_i^2+k_ik_j+k_j^2}{1+(k_i+k_j)^2}\,(-1,\,-(k_i+k_j)).
\]
Note in particular that $(k_i^2+k_ik_j+k_j^2)>0$, and this implies that 
the $x$-component of the velocity is {\it always} negative. That is, a line-soliton
propagates in the negative $x$-direction. On the other hand, any small perturbation propagates
in the positive $x$-direction. This can be seen from the dispersion relation of the KP equation
for a linear wave $\phi=\exp(i{\bf k}\cdot{\bf x}-i\omega t)$ with wave-vector ${\bf k}=(k_x,k_y)$ and
frequency $\omega$, 
\[
\omega=\frac{1}{4}k_x^3-\frac{3}{4}\,\frac{k_y^2}{k_x},
\]   from which the group velocity of the wave is given by
\[
{\bf v}=\nabla \omega=\left(\frac{\partial\omega}{\partial k_x},\frac{\partial \omega}{\partial k_y}\right)=
\left(\frac{3}{4}\left(k_x^2+\frac{k_y^2}{k_x^2}\right),- \frac{2}{3}\,\frac{k_y}{k_x}\right).
\]
Thus,  it is reasonable to expect from a first-order perturbation theory that the 
soliton separates from small radiations asymptotically, similar to the case of the KdV equation
(see for example \cite{AS:81, N:85}).

We also remark that the formula (\ref{Onesoliton}) for one-soliton solution of the KP 
equation can be extended to the one-soliton solution of the KP hierarchy by including 
the higher times $t_n$ (see Remark 2.2),
\begin{equation}\label{OneHsoliton}
u(t_1,t_2,t_3,\ldots)=\mathcal{A}\sech^2\left(
\sum_{n=1}^{\infty}\mathcal{K}_nt_n+\Theta^0\right)\,,
\end{equation}
with some constant $\Theta^0$. Here the amplitude and the {\it infinite} dimensional 
wave-vector are given by
\[\left\{
\begin{array}{llll}
\mathcal{A}&=&\displaystyle{\frac{1}{2}(k_j-k_i)^2,}\\[2.0ex]
\mathcal{K}_n&=&\displaystyle{\frac{1}{2}\left(k_j^n-k_i^n\right),\qquad {\rm for}\quad n=1,2,\ldots.}
\end{array}\right.
\]
This can be easily seen from the structure of $f$-function, i.e.
\[
f=E_i+aE_j,\qquad {\rm with}\quad E_i=\exp\left(\sum_{n=1}^{\infty}k_i^nt_n\right).
\]
Although the higher times $t_n$ for $n>3$ do not have a direct physical meaning,
those parameters are related to the existence of the multi-soliton solutions
through the symmetries of the KP equation as mentioned in Remark 2.2.

\begin{Example}\label{ex:12soliton}
Now we consider the case with $M=3$. We again take $a_1=1$, so that we have
\[
f= E_1+aE_2+bE_3
\]
with some positive constants $a$ and $b$. 
As in the previous example, it is also possible here to determine the dominant
exponentials and analyze the structure of the solution in the $xy$-plane. Let us
consider the function $f$ along the line $x=-cy$ with $c=\tan\Psi$ where
$\Psi$ is the angle measured counterclockwise from the $y$-axis (see Figure \ref{fig:1soliton}). 
Then along $x=-cy$, we have the exponential function $E_j =\exp[\eta_j(c)y+k_j^3t]$ with
\begin{equation}\label{eta}
\eta_j(c)=k_j(k_j-c).
\end{equation}
%%%%%%%%%%%%%%%%%%%%%%%%%%%
\begin{figure}[t]
\begin{center}
\includegraphics[height=1.5in]{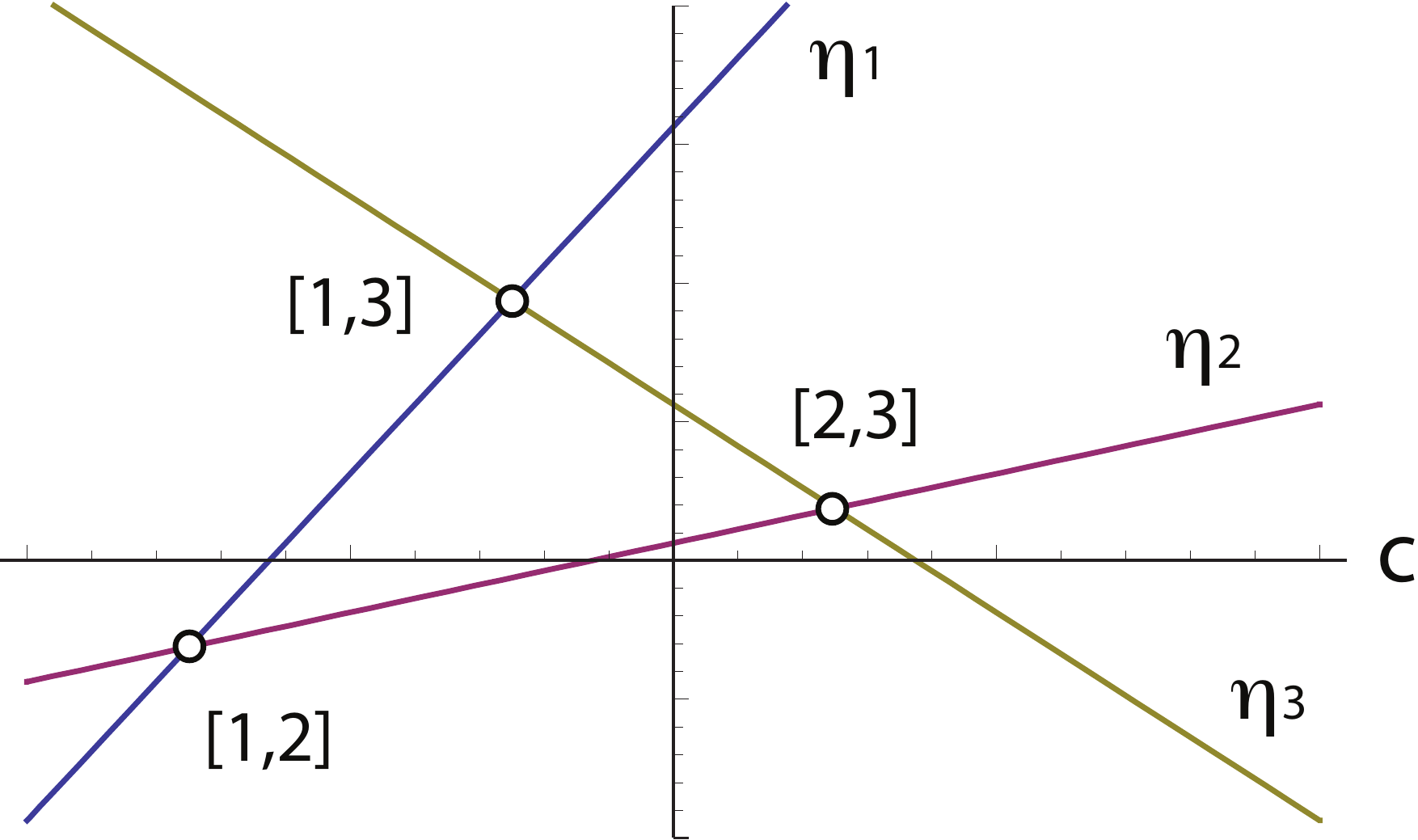}
\end{center}
\caption{The graphs of $\eta_j(c)=k_j(k_j-c)$. Each $[i,j]$ represents the exchange of the order
between $\eta_i$ and $\eta_j$.
The parameters are given by $(k_1,k_2,k_3)=
(-\frac{5}{4},-\frac{1}{4},\frac{3}{4})$.\label{fig:1}}
\end{figure}
%%%%%%%%%%%%%%%%%%%%%%%%%
It is then seen from Figure \ref{fig:1}  that for $y\gg 0$ and a fixed $t$, the exponential term
$E_1$ dominates when $c$ is large positive ($\Psi\approx \frac{\pi}{2}$, i.e. $x\to-\infty$).
Decreasing the value of $c$ (rotating the line clockwise), the dominant term changes to $E_3$.
Thus we have
\begin{equation*}\label{ypositive}
w=\partial_x\,\ln f\longrightarrow\left\{\begin{array}{lll}
k_1\quad& {\rm as}\quad& x\to -\infty,\\
k_3\quad &{\rm as}\quad &x\to\infty.
\end{array}\right. 
\end{equation*}
The transition of the dominant exponentials $E_1\to E_3$ 
is characterized by the condition $\eta_1 = \eta_3$, which corresponds the 
direction parameter value $c=\tan\Psi_{[1,3]}=k_1+k_3$. In the neighborhood of this line, 
the function $f$ can be approximated as
\[
 f\approx E_1+bE_3,
 \]
which implies that there exists a $[1,3]$-soliton for $y\gg 0$. The constant $b$ can be used to choose
a specific location of this soliton.

Next consider the case of $y\ll 0$. The dominant exponential corresponds to the
{\it least} value of $\eta_j$ for any given value of $c$.
For large positive $c$ ($\Psi\approx\sf{\pi}{2}$, i.e. $x\to \infty$),
$E_3$ is the dominant term. Decreasing the value of $c$ (rotating the line $x=-cy$ clockwise),
the dominant term changes to $E_2$ when $k_2+k_3>c>k_1+k_2$,
and $E_1$ becomes dominant for $c< k_1+k_2$. Hence, we have for $y\ll 0$
\begin{equation*}\label{ynegative}
w\longrightarrow\left\{\begin{array}{llll}
k_1\quad &{\rm as}\quad &x\to-\infty,\\
k_2\quad &{\rm for}\quad  & -(k_1+k_2)y<x<-(k_2+k_3)y, \\
k_3\quad &{\rm as}\quad &x\to\infty.
\end{array}\right.
\end{equation*}
In the neighborhood of the line $x+(k_1+k_2)y=$constant, 
\[
f\approx E_1+aE_2,
\]
which corresponds to a $[1,2]$-soliton and its location is fixed by the constant $a$.
The solutions also consists of a $[2,3]$-soliton in the neighborhood
of the line $x+(k_2+k_3)y=$constant, and whose location is determined by the 
locations of other line-solitons. Therefore, we need only two parameters $a,\,b$
(besides the $k$-parameters) to specify the solution uniquely. These free parameters 
may be considered as the asymptotic data for the solution $u$.
The shape of solution generated by $f= E_1+aE_2+bE_3$ with $a=b=1$ (i.e. at $t=0$ three
line-solitons meet at the origin)
is illustrated via the contour plot in Figure \ref{fig:2}.
%%%%%%%%%%%%%%%%%%%%%%%%%%%
\begin{figure}[t]
\begin{center}
\includegraphics[height=4.5cm]{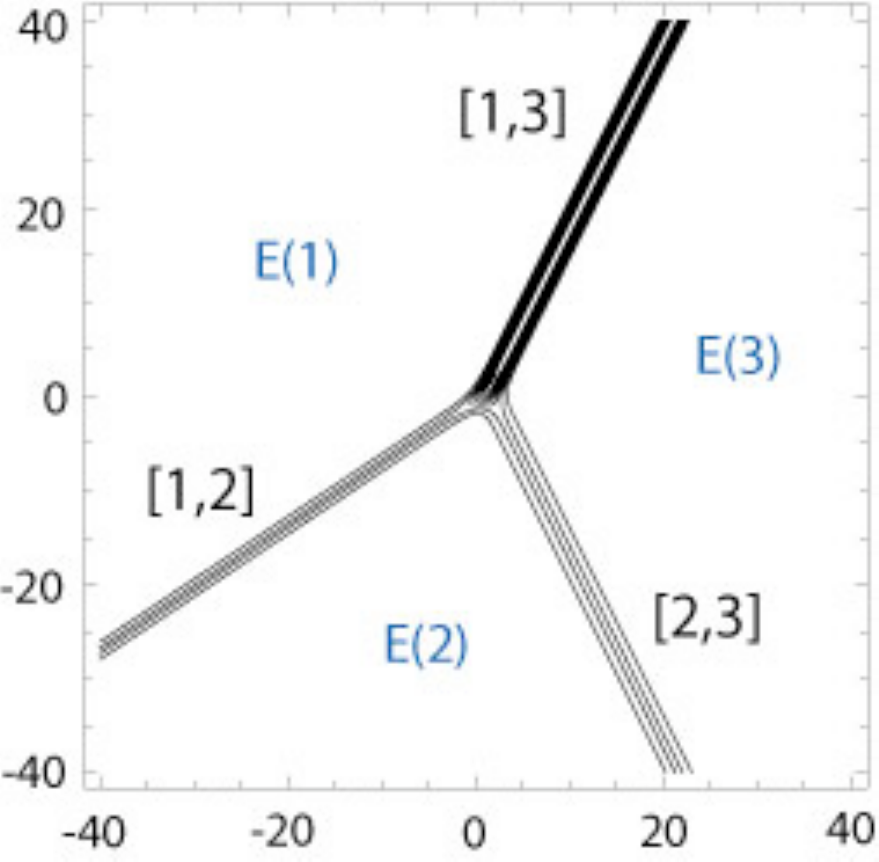}
\caption{A $(2,1)$-soliton solution.
Each $E(j)$ with $j=1,2$ or $3$ indicates the dominant exponential
term $E_j$ in that region. The boundaries of any two adjacent regions give 
the line-solitons indicating the transition
of the dominant terms $E_j$. The $k$-parameters are the same as those
in Figure \ref{fig:1}, and the line-solitons are determined from
the intersection points of the $\eta_j(c)$'s in Figure \ref{fig:1}. 
Here $a=b=1$ (i.e. $\tau=E_1+E_2+E_3$) so that the three solitons 
meet at the origin at $t=0$.\label{fig:2}}
\end{center}
\end{figure}
%%%%%%%%%%%%%%%%%%%%%%%%%%%%%%%%
%\end{Example}
In this Figure, one can see that the line-soliton in $y\gg 0$ labeled by $[1,3]$,
is localized along the phase transition line $\theta_1 = \theta_3$ (equivalently, $\eta_1=\eta_3$)
with direction parameter $c = k_1+k_3$; two other line-solitons in $y\ll 0$ labeled by 
$[1,2]$ and $[2,3]$ are localized respectively, along the phase transition lines 
with $c=k_1+k_2$ and $c=k_2+k_3$.
This solution represents a resonant solution of three line-solitons. In terms of the function $w$,
which is a solution of the Burgers equation in the $y$-direction, this corresponds to a confluence
of two shocks (see p.110 in \cite{Wh:74}). The resonant condition among those three line-solitons is
given by
\begin{align*}
{\bf K}_{[1,3]}={\bf K}_{[1,2]}+{\bf K}_{[2,3]},\qquad \Omega_{[1,3]}=\Omega_{[1,2]}+\Omega_{[2,3]},
\end{align*}
which are identically satisfied with 
${\bf K}_{[i,j]}=\sf{1}{2}(k_j-k_i,k_j^2-k_i^2)$ and $\Omega_{[i,j]}=\sf{1}{2}
(k_j^3-k_i^3)$.
The resonant condition may be symbolically written as
\[
[1,3]=[1,2]+[2,3].
\]
One can also represent this line-soliton solution by a permutation of
three indices: $\{1, 2, 3\}$ which is illustrated by a (linear) {\it chord diagram}
shown below.
%%%%%%%%%%%%%%%%%%%%%%%%%%%
\begin{figure}[h]
\begin{center}
\includegraphics[height=1.8cm]{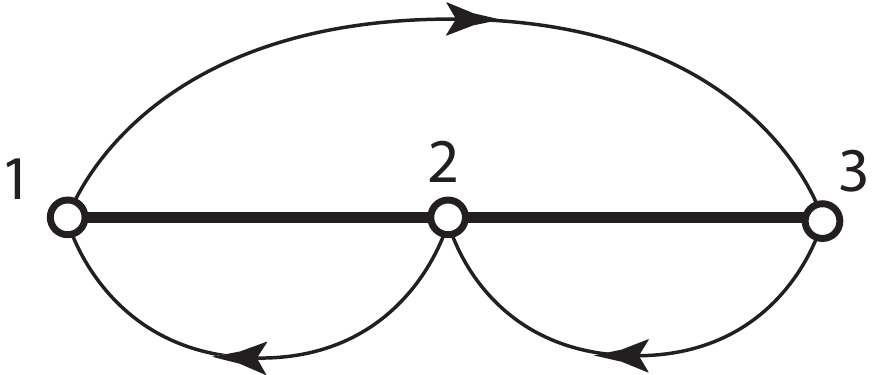}
\end{center}
\end{figure}
%%%%%%%%%%%%%%%%%%%%%%%%%%%%%%%%
Here, the upper chord represents the $[1,3]$-soliton in $y\gg 0$ and 
the lower two chords represent $[1,2]$ and $[2,3]$-solitons in $y\ll 0$. 
Following the arrows in the chord diagram, one recovers the permutation,
\[
\pi=\begin{pmatrix}1 & 2& 3\\3 &1&2\end{pmatrix} \qquad {\rm or~simply}\quad \pi=(312).
\]
In general, each line-soliton solution of the KP equation can be parametrized 
by a {\it unique} permutation corresponding to a chord diagram. We will
discuss this issue in Section 5.
\end{Example}

The results described in the previous examples can be easily extended to the
general case where $f$ has arbitrary number of exponential terms (see also \cite{Me:02, BK:03}). 
\begin{Proposition} \label{Mm11sol}
If $f=a_1E_1+a_2E_2+\cdots+a_ME_M$ with $a_j > 0$ for $j=1,2,\ldots,M$, then the 
solution $u$ consists of $M-1$ line-solitons for $y\ll 0$ and one 
line-soliton for $y\gg 0$.
\end{Proposition} 
Such solutions are referred to as the $(M-1,1)$-soliton solutions. Note that the 
line-soliton for $y\gg 0$ is labeled by $[1,M]$, whereas the other 
line-solitons in $y\ll 0$ are labeled by $[k,k+1]$ for $k=1,2,\ldots,M-1$,
counterclockwise from the negative to the positive $x$-axis, i.e.
increasing $\Psi$ from $-\sf{\pi}{2}$ to $\sf{\pi}{2}$. As in the previous
examples one can set $a_1 =1$ without any loss of generality, then the remaining
$M-1$ parameters $a_2, \ldots, a_M$ determine the locations of the $M$ line-solitons.
Also note that the $xy$-plane is divided into $M$ sectors for the asymptotic region with $x^2+y^2\gg0$,
and the boundaries of those sectors are given by the asymptotic line-solitons.
This feature is common even for the general case.
%\end{Proposition}
%%%%%%%%%%%%%%%%%%%%%%%%%%%%%%
\begin{figure}[t]
\begin{center}
\includegraphics[height=7cm]{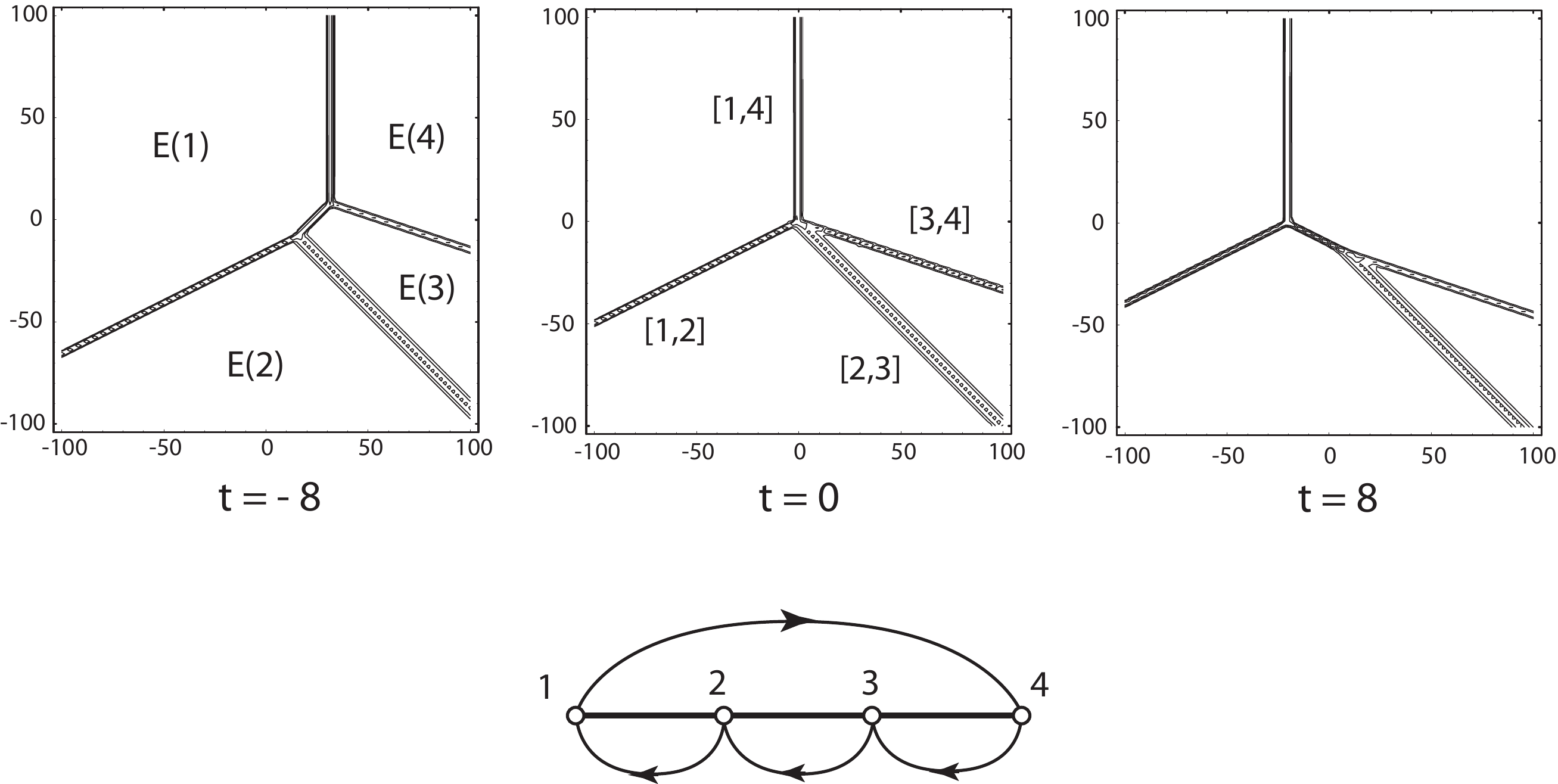}
\caption{The time evolution of a $(3,1)$-soliton solution and the corresponding chord diagram.
The upper chord represents the $[1,4]$-soliton, and the lower ones 
represent $[1,2]$-, $[2,3]$- and $[3,4]$-solitons.
The chord diagram shows $\pi=(4123)$.\label{fig:31soliton}}
\end{center}
\end{figure}
%%%%%%%%%%%%%%%%%%%%%%%%%%%%%%%%%%
Figure \ref{fig:31soliton} illustrates the case for a $(3,1)$-soliton solution
with $f= E_1+E_2+E_3+E_4$. The chord diagram for this solution
represents the permutation $\pi = (4123)$ of the set $\{1, 2, 3, 4\}$.

\subsection{Extension of the Burgers hierarchy}\label{ExtBurgers}
Theorem \ref{KP-B} shows the relation between the KP equation and the
Burgers hierarchy for the function $w$ defined in terms of the
solution $f$ of the linear system (\ref{heat}) via the Cole-Hopf transformation,
\begin{equation}\label{CH}
w=\partial_x\,\ln f,\qquad {\rm or}\quad f^{(1)}=wf,
\end{equation}
where $f^{(1)}=\partial_xf$. Furthermore, Proposition \ref{Mm11sol} illustrates
the $(M-1,1)$-soliton solutions obtained from the special choices for the functions
$f$ and $w$.
In order to construct more general type of soliton solutions of the KP equation,
we consider an extension of the Cole-Hopf transformation (\ref{CH}) to the case
where the function $f$ satisfies an $N$-th order linear equation,
\begin{equation}\label{higherCH}
f^{(N)}=w_1f^{(N-1)}+\cdots+w_{N-1}f^{(1)} + w_Nf,
\end{equation}
where $f^{(j)}:=\partial_x^j f$. The coefficient functions $(w_1,w_2,\ldots,w_N)$ 
can be constructed from $N$ linearly independent solutions of 
(\ref{higherCH}), and their time evolutions with respect to $t_n,\, n>1$
are determined by the compatibility conditions of (\ref{higherCH}) with the linear system 
(\ref{heat}). For example, when $N=2$, the compatibility condition
$\partial_y(f_{xx})=\partial^2_x(f_y)$ with $y=t_2$  gives 
$\partial_y(w_1f^{(1)}+w_2f)=\partial_x^2(w_1f^{(1)}+w_2f)$. Then by
equating the coefficients of $f^{(1)}$ and $f$ in the last expression,
leads to the following equations for $(w_1,w_2)$, 
\begin{align*}
\partial_{y}w_1&=2w_1w_{1,x}+w_{1,xx}+2w_{2,x},\\
\partial_{y}w_2&=2w_2w_{1,x}+w_{2,xx}.
\end{align*}
Notice that if $w_2=0$, this system is reduced to the Burgers equation.
In general, the variables $(w_1,w_2,\ldots,w_N)$ satisfy an {\it $N$-component} Burgers 
equation in the $y$-variable while the higher flows with respect to 
the ``times" $t_n$ for $n=3,4,\ldots$ are the symmetries of
this coupled equation, thus forming an $N$-component Burgers hierarchy (this will be 
reformulated in terms of the Sato theory in the next section, also see \cite{Ha:87}).

We regard (\ref{higherCH}) as an $N$-th order {\it ordinary} differential equation,
and consider a fundamental set of $N$ solutions denoted by $\{f_1,f_2,\ldots,f_N\}$,
which satisfy
\[
f_i^{(N)}=w_1f_i^{(N-1)}+\cdots + w_Nf_i,\qquad i=1,2,\ldots, N.
\]
The above form a linear algebraic system which can be solved for the coefficients
$\{w_1,w_2,\ldots,w_N\}$ by applying Cramer's formula. In particular we have
\begin{align*}
w_1&=\frac{1}{{\rm Wr}(f_1,\ldots,f_N)}\,\left|\begin{matrix}  
f_1&f_1^{(1)}&\cdots&f_1^{(N-2)}&f_1^{(N)}\\
f_2&f_2^{(1)}&\cdots&f_2^{(N-2)}&f_2^{(N)}\\
\vdots&\vdots&\ddots&\vdots&\vdots \\
f_{N}&f_N^{(1)}&\cdots&f_N^{(N-2)}&f_N^{(N)}
\end{matrix}\right| \\
&=\frac{\partial}{\partial x}\ln {\rm Wr}(f_1,\ldots,f_N).
\end{align*}
where ${\rm Wr}(f_1,\ldots,f_N)$ is the Wronskian determinant of $\{f_1,\ldots,f_N\}$.
For $N=1$, this is just the Cole-Hopf transformation (\ref{CH}) where
$w_1=w$ solves the pKP equation. In what follows, we show that this is also 
true for the general case. That is, the function $w_1 = \partial_x\ln {\rm Wr}(f_1,\ldots,f_N)$ 
solves the pKP equation, and $u:=2w_{1,x}=2\partial_x^2\ln {\rm Wr}(f_1,\ldots,f_N)$ 
solves the KP equation. The Wr$(f_1,\ldots,f_N)$ is an example of the {\it $\tau$-function }
for the KP equation,
which plays a very important role of the KP theory (see Section \ref{S:tau}).
We now state the well-known theorem that leads to the Wronskian formulation of the
multi-soliton solution for the KP equation.
\begin{Theorem}\label{PHrelation}
Let the $\tau$-function be given by the Wronskian determinant,
\begin{equation}\label{KP-tau}
\tau(x,y,t)=\left|\begin{matrix}  
f_1&f_1^{(1)}&\cdots&f_1^{(N-1)}\\
f_2&f_2^{(1)}&\cdots&f_2^{(N-1)}\\
\vdots&\vdots&\ddots&\vdots \\
f_{N}&f_N^{(1)}&\cdots&f_N^{(N-1)}
\end{matrix}\right|\,,
\end{equation}
where $\{f_1,f_2,\ldots,f_N\}$ are linearly independent functions satisfying
$\partial_yf_j=f_j^{(2)}$ and $\partial_tf_j=f_j^{(3)}$ with $f_j^{(n)} := \partial_x^nf_j$
for $j=1,\ldots,N$.
Then $u(x,y,t) = 2\partial_x^2\ln \tau(x,y,t)$ satisfies the KP equation (\ref{kp}).
\end{Theorem}
We prove the theorem by employing an identity called the Pl\"ucker 
relation (see for example \cite{H:04}), satisfied by the maximal minors of an $N \times M$ matrix 
\[
\varPhi :=\left(\begin{matrix}
f_1 & f_1^{(1)} & \cdots & f_1^{(M-1)}\\
f_2& f_2^{(1)} & \cdots & f_2^{(M-1)}\\
\vdots& \vdots & \vdots & \vdots  \\
f_N& f_N^{(1)} &\cdots &f_N^{(M-1)}
\end{matrix}\right) \,.
\]
with $N < M$. A maximal minor $\varphi(l_1, l_2, \cdots, l_N)$ is the determinant of the
submatrix formed by $N$ columns $\varPhi_i = (f_1^{(i-1)}, f_2^{(i-1)}, \cdots, f_N^{(i-1)})^T$
of $\varPhi$ where $ i \in \{l_1, l_2, \cdots, l_N\}$, i.e.
\[
\varphi(l_1,l_2,\ldots,l_N)={\rm det}\left[\varPhi_{l_1},\varPhi_{l_2},\ldots,\varPhi_{l_N}\right].
\]
\begin{Lemma}\label{Prelation}
The set of maximal minors $\varphi[i,j]:=\varphi(1,2,\cdots,N-2,N-2+i,N-2+j)$ of the matrix 
$\varPhi$ defined above, with $1\le i<j\le 4$, satisfy the identity,
\begin{equation}\label{PLphi}
\varphi[1,2]\,\varphi[3,4]-\varphi[1,3]\,\varphi[2,4]+\varphi[1,4]\,\varphi[2,3]=0.
\end{equation}
\end{Lemma}
\begin{Proof}
Consider the following $2N\times 2N$ determinant expressed
in terms of the columns $\varPhi_i$ of the matrix $\varPhi$,
\[{\rm det}\left[
\begin{matrix}
\varPhi_1 &\cdots & \varPhi_{N-1} & \varPhi_1& \cdots & \varPhi_{N-2}&\varPhi_{N}&\varPhi_{N+1}&\varPhi_{N+2}\\
 0  &\cdots & 0  & \varPhi_1& \cdots & \varPhi_{N-2}&\varPhi_{N}&\varPhi_{N+1}&\varPhi_{N+2} \end{matrix}\right] 
\,.
\]
This determinant is obviously zero. Then, Laplace expansion of the first
determinant in terms of $N \times N$ minors yields the desired identity.
\end{Proof}
The three-term identity (\ref{PLphi}) is one of the Pl\"ucker relations which
play an important role in revealing the
geometric structure underlying the KP equation. This will be discussed
in Section~\ref{Grassmann}, where we will introduce the general form
of the Pl\"ucker relations. We now prove Theorem \ref{PHrelation} (the following proof is
the same as that in \cite{H:04}):
\begin{Proof} 
Substituting the relation $u = 2 \partial_x^2(\ln \tau)$ into the KP equation (\ref{kp}),  
integrating twice with respect to $x$ and setting the integrations constants
to zero, yields the following equation in $\tau$,
\begin{align}\label{taukp}
4(\tau\tau_{xt}-\tau_x\tau_t)-3\tau_{xx}^2-\tau\tau_{xxxx}+
4\tau_x\tau_{xxx}+3\tau_y^2-3\tau\tau_{yy}=0.
\end{align}
The $\tau$-function is given by the maximal minor, $\tau=\varphi(1,2,\ldots,N)=:\varphi[1,2]$, 
using the notation of Lemma \ref{Prelation}. Suitable combinations of
the derivatives of $\tau$ also correspond to other maximal minors of the matrix $\varPhi$.
In particular, we have \begin{align*}
\varphi[1,3]&=\tau_x,\qquad \varphi[1,4]=\sf{1}{2}(\tau_{xx}+\tau_y),\qquad
\varphi[2,3]=\sf{1}{2}(\tau_{xx}-\tau_y),\\
\varphi[2,4]&=\sf{1}{3}(\tau_{xxx}-\tau_t),\qquad
\varphi[3,4]=\sf{1}{12}(\tau_{xxxx}+3\tau_{yy}-4\tau_{xt})
\end{align*}
Substituting the above expressions into the Pl\"ucker relation (\ref{PLphi}) in 
Lemma \ref{Prelation}, we have
\begin{align*}
0 = &\varphi[1,2]\,\varphi[3,4]-\varphi[1,3]\,\varphi[2,4]+\varphi[1,4]\,\varphi[2,3]\\
=~&-\sf{1}{12}\left(\tau(4\tau_{xt}-\tau_{xxxx}-3\tau_{yy})+
4\tau_x(\tau_{xxx}-\tau_t)+3(\tau_y+\tau_{xx})(\tau_y-\tau_{xx})\right)
\end{align*}
which is equivalent to (\ref{taukp}). 
\end{Proof}
Theorem \ref{PHrelation} provides a direct {\it linearization} scheme for the 
KP equation 
in the sense that a large class of solutions whose $\tau$-function is given by 
the Wronskian determinant (\ref{KP-tau}) can be constructed from the solutions
of the linear equation (\ref{higherCH}) together with the evolution equations
$f_y = f_{xx}$ and $f_t = f_{xxx}$. Note that since $f(x,y,t)$ admits a rather
general integral representation (see Section 2.2), this linearization scheme gives
rise to different classes of solution besides the soliton solutions, by proper
choice of measure and contour of integration in the $k$-plane. It is a remarkable fact
that all solutions constructed in this way identically satisfy the KP equation by
virtue of the Pl\"ucker relation (\ref{PLphi}) which is an algebraic identity.
However, it should be emphasized that the significance of Theorem \ref{PHrelation} is
much deeper than that is presented here. In the original scope of Sato theory,
the differential operator in (\ref{higherCH}) is in fact a reduction ($N$-truncation)
of a pseudo-differential operator of infinite order (see Section 3) all of whose 
coefficients are determined via a single holomorphic function namely, the Sato $\tau$-function. 
The Wronskian determinant 
(\ref{KP-tau}) is a representation of the Sato $\tau$-function in the $N$-truncated case.
It is in this (infinite-dimensional) setting of the full Sato theory that the KP equation
expressed in terms of the Sato $\tau$-function as in (\ref{taukp}) can itself be
interpreted as the Pl\"ucker relation (\ref{PLphi}) in a suitable sense (see also Remark \ref{GrassmannSato}).

\begin{Remark}
The equation (\ref{taukp}) is often rewritten in the Hirota bi-linear form
\begin{equation*}
%\label{kpb}
P(D_x,D_y,D_t)\tau\cdot\tau:=(4D_xD_t-D_x^4-3D_y^2)\tau\cdot\tau=0, 
\end{equation*}
where $D_z$ with $z=x,y$ or $t$ is the usual Hirota derivative defined by
\[
D^n_{z}f\cdot g:=\frac{\partial^n}{\partial \epsilon^n}
\left[f(z+\epsilon)g(z-\epsilon)\right]\Big|_{\epsilon=0} \,.
\]
Note here that the soliton dispersion relation (\ref{Sdispersion}) can be obtained from the Hirota bi-linear form,
i.e. $P(2K_{[i,j]}^x,2K_{[i,j]}^y,2\Omega_{[i,j]})=0$.  Thus the dispersion relation
has a direct connection to the Pl\"ucker relation (\ref{PLphi}).

In his famous book \cite{H:04}, Hirota expresses the minors $\phi[i,j]$ with the elegant diagrams called the Maya diagrams introduced by Sato (which is also
equivalent to the Young diagrams). Explicit proofs of the solution formulas for other integrable systems
written in the Hirota bi-linear forms are also given in terms of the Maya diagrams.
\end{Remark}
\begin{Example} Consider the case with $N=2$ and $M=3$:
The linearly independent functions $f_1$ and $f_2$ are expressed as, 
\[
f_i=\sum_{j=1}^3a_{ij}E_j\,,\qquad i=1,2, 
\]
with $E_j(x,y,t)=\exp(k_jx+k_j^2y+k_j^3t)$ for $j=1,2,3$.
In this case the $\tau$-function in (\ref{KP-tau}) can be explicitly given by 
\begin{align*}
\tau=&\left|\begin{matrix} f_1 &f_1^{(1)}\\f_2&f_2^{(1)}\end{matrix}\right|=
\left|\begin{pmatrix} a_{11}&a_{12}&a_{13}\\a_{21}&a_{22}&a_{23}\end{pmatrix}
\begin{pmatrix}E_1 &k_1E_1\\E_2&k_2E_2\\E_3&k_3E_3\end{pmatrix}\right|\\
=&\left|\begin{array}{cc}a_{11}&a_{12}\\
a_{21}&a_{22}\end{array}\right|E(1,2)+
\left|\begin{array}{cc}a_{11}&a_{13}\\a_{21}&a_{23}\end{array}\right|E(1,3)+
\left|\begin{array}{cc}a_{12}&a_{13}\\a_{22}&a_{23}\end{array}\right|E(2,3)\,,
\end{align*}
where we have used  the Binet-Cauchy theorem and $E(i,j)={\rm Wr}(E_i,E_j)=(k_j-k_i)E_iE_j$. 
Let us investigate 
a concrete situation where $(k_1,k_2,k_3)=(-\sf{5}{4},-\sf{1}{4},\sf{3}{4})$, and
\begin{equation*}\left(
\begin{array}{ccc}
a_{11}&a_{12}&a_{13}\\
a_{21}&a_{22}&a_{23}
\end{array}\right)=\left(
\begin{array}{ccc}
1&0&-b\\
0&1&a
\end{array}\right),
\end{equation*}
with positive constants $a$ and $b$.
The $\tau$-function is then given by
\[
\tau=E(1,2)+aE(1,3)+bE(2,3).
\]
In order to carry out the asymptotic analysis in this case one needs to 
consider the sum of two $\eta_j(c)$, i.e. $\eta_{i,j}=\eta_i+\eta_j$ for 
$1\le i<j\le 3$. This can still be done using Figure \ref{fig:1},
but a more effective way is described in Section 5 (see the graph of $\eta(k,c)$
in Figure \ref{fig:eta24} and equations (\ref{etakc}), (\ref{dominant})).

 For $y\gg 0$, the transitions of the dominant exponentials are given by following scheme:
$$
E(1,2)\longrightarrow E(1,3) \longrightarrow E(2,3),
$$ 
as $c$ varies from large positive (i.e. $x\to-\infty$) to large negative values (i.e. $x\to\infty$). 
The boundary between the regions with the dominant exponentials $E(1,2)$ and $E(1,3)$
defines the $[2,3]$-soliton solution since here the $\tau$-function can be approximated as
\begin{align*}
\tau&\approx E(1,2)+aE(1,3)=(k_2-k_1)E_1\left(E_2+a\frac{k_3-k_1}{k_2-k_1}\,E_3\right)\\
&=2(k_2-k_1)E_1e^{\sf{1}{2}(\theta_2+\theta_3-\theta_{23})}\cosh\sf{1}{2}(\theta_2-\theta_3+\theta_{23}),
\end{align*}
so that we have
\[
u=2\partial_x^2\ln \tau \approx \sf{1}{2}(k_2-k_3)^2 
\mathrm{sech}^2\sf{1}{2}(\theta_2-\theta_3+\theta_{23})\,,
\]
where $\theta_{23}$ is given by
\[
\theta_{23}=\ln\frac{k_2-k_1}{k_3-k_1}-\ln a \qquad {\rm i.e.}\qquad a=\frac{k_2-k_1}{k_3-k_1}e^{-\theta_{23}}.
\]
Note here that since the logarithm of the coefficient of $\cosh$-function in $\tau$ above  
is linear in $x$, it does not contribute to the solution $u$.
The constant $a$ can be used to determine the location of this soliton. A similar computation
as above near the transition boundary of the dominant exponentials $E(1,3)$ and $E(2,3)$
yields
\begin{align*}
\tau &\approx 2(k_3-k_1)a E_3e^{\sf{1}{2}(\theta_1+\theta_2-\theta_{12})}\cosh
\sf{1}{2}(\theta_1-\theta_2+\theta_{12}),\\
u &\approx \sf{1}{2}(k_1-k_2)^2 \mathrm{sech}^2\sf{1}{2}(\theta_1-\theta_2+\theta_{12})\,,
\end{align*}
where $\theta_{12}$ is given by
\[
\theta_{12}=\ln\frac{k_3-k_1}{k_3-k_2}-\ln\frac{b}{a}\qquad{\rm i.e.}\qquad b=\frac{k_2-k_1}{k_3-k_2}e^{-\theta_{12}}.
\]
This gives $[1,2]$-soliton and its location is determined by the constant $b$.

%%%%%%%%%%%%%%%%%%%%%%%%%%%%%%%%%%%
\begin{figure}
\begin{center}
\includegraphics[height=4.5cm]{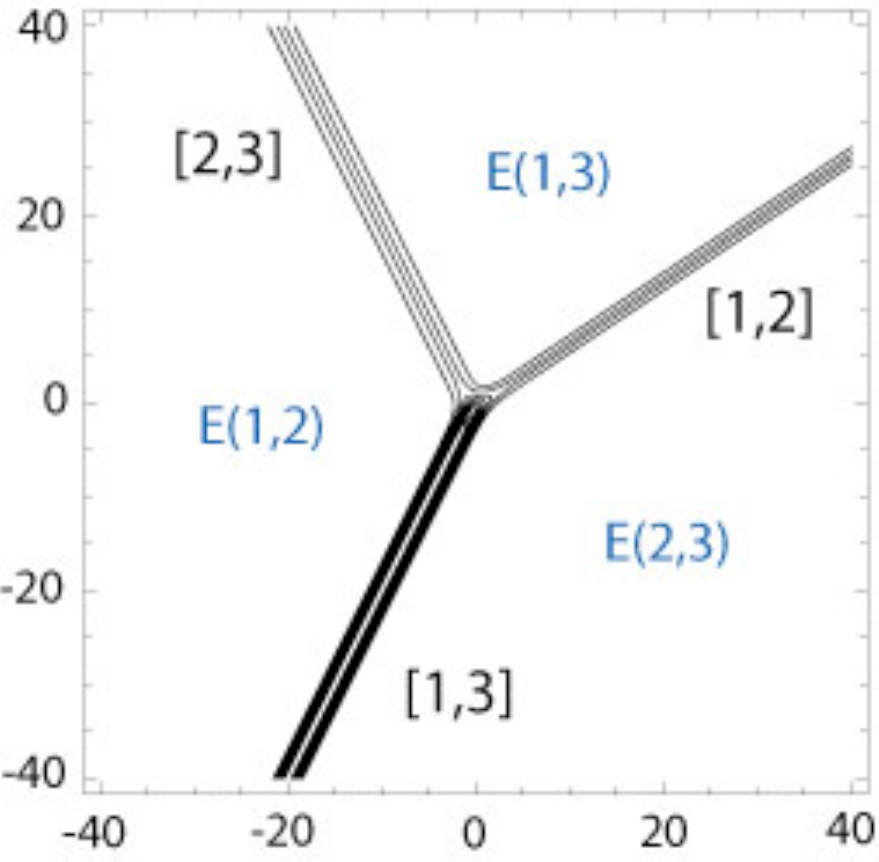}
\end{center}
\caption{A $(1,2)$-soliton solution. 
The $E(i,j)$ on each region indicates the dominant exponential term.
The $k$-parameters are the same as those in Figure \ref{fig:1}.
The parameters in the $A$-matrix are chosen as $a=\frac{1}{2}$ and $b=1$,
so that three line-solitons meet at the origin at $t=0$.\label{fig:3}}
\end{figure}
%%%%%%%%%%%%%%%%%%%%%%%%%%%%%%%%%%%

For $y\ll 0$,
there is only one transition, namely
$$
E(2,3) \longrightarrow E(1,2),
$$
as $c$ varies from large positive value (i.e. $x\to\infty$) to large negative 
value (i.e. $x\to-\infty$). 
In this case, a $[1,3]$-soliton is formed for $y\ll 0$ at the boundary of the
dominant exponentials $E(2,3)$ and $E(1,2)$.  The contour plot
of the line-soliton solution is shown in Figure \ref{fig:3}. 
Notice that this figure can be obtained from Figure \ref{fig:2} by changing $(x,y)\to(-x,-y)$.
This solution can be represented by the chord diagram corresponding to the permutation 
$\pi=(231)$ shown below. 
%%%%%%%%%%%%%%%%%%%%%%%%%%%
\begin{figure}[h]
\begin{center}
\includegraphics[height=1.8cm]{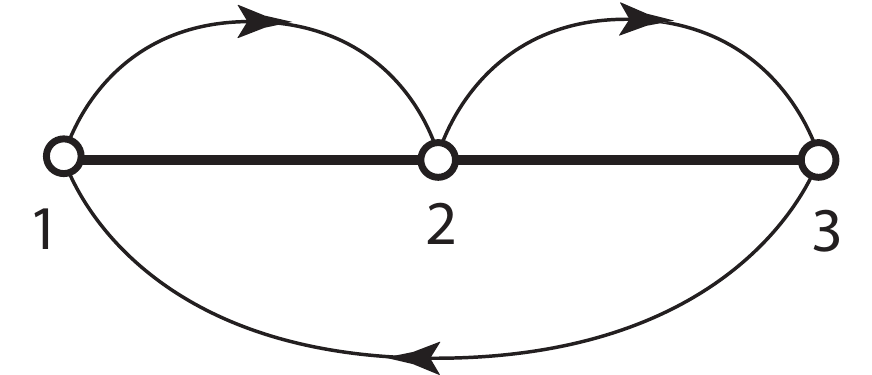}
\end{center}
\end{figure}
%%%%%%%%%%%%%%%%%%%%%%%%%%%%%%%%
Note that this diagram is the {\it $\pi$-rotation} of the chord diagram in Example \ref{ex:12soliton}
whose permutation $\pi=(312)$ is the inverse of $\pi=(231)$.

\end{Example}

%%%%%%%%%%%%%%%%%%%%%%%%%%%%%%%%%%%%%%%%%

\section{Sato theory of the KP hierarchy}\label{C:Sato}
In this section we review briefly some basic facts from the Sato theory of the KP hierarchy
(see also \cite{D:91, MJD:00}).
We first review the Lax formalism for an infinite set of evolution equations
whose compatibility conditions give rise to the flows corresponding to the KP hierarchy.
The KP equation is the first nontrivial member of this hierarchy
associated with the time variables $t_1 := x, \, t_2:=y$ and $t_3:=t$.
The main purpose is to highlight the basic framework of integrability
underlying the KP theory. 
Then we show that the results of the previous section are direct consequences
of the Sato theory. In particular, we emphasize the importance of the $\tau$-function
and explain the central role of the $\tau$-function in the KP hierarchy.

\subsection{Lax formulation of the KP equation}
Let $L$ be a first order pseudo-differential symbol,
$$L=\partial+u_1+u_2\partial^{-1}+u_3\partial^{-2}+\cdots$$
where the coefficients $u_i=u_i({\bf t})$ depend on infinitely many variables
${\bf t}=(t_1,t_2,t_3,\cdots)$. The symbol $\partial := \partial_x$ is a differential
operator whereas $\partial^{-1}$ is a formal integration, 
satisfying $\partial \partial^{-1}=\partial^{-1}\partial=1$. The operation of $\partial^{\nu}$
with $\nu\in\mathbb{Z}$ is given by the generalized Leibnitz rule 
$$\partial^\nu f=\sum_{j\geq 0}\left(\begin{array}{c}\nu\\j\end{array}\right)
\partial_x^j(f)\partial^{\nu-j}.$$
For example, we have
$\partial f = f_x + f \partial$ and 
$\partial^{-1}f = f\partial^{-1}-f_x\partial^{-2}+f_{xx}\partial^{-3}- \cdots$ (the
latter expression follows from the usual integration by parts formula).
Here we define the {\it weights} for the functions $u_i$ and $\partial^{\nu}$ as
\[
{\rm wt}(u_i)=i,\qquad {\rm wt}(\partial^{\nu})=\nu.
\]
So the $L$ has a homogeneous weight one. We also remark that the term $u_1$ in $L$ 
can be eliminated by a gauge transformation $g$ such that $u_1=-g^{-1}\partial_x(g)$, i.e.
\[
L\longrightarrow g^{-1}Lg=\partial +\tilde u_2\partial^{-1}+\tilde u_3\partial^{-2}+\cdots.
\]
(This can be extended so that we can eliminate {\it all} $u_j$ with an appropriate gauge operator $g=W$,
which will be discussed in Section \ref{Satoform}.)
We will henceforth consider $L$ without the $u_1$ term,
\begin{equation}\label{L}
L=\partial+u_2\partial^{-1}+u_3\partial^{-2}+\cdots \,.
\end{equation}

The Lax form of the KP hierarchy is defined by the infinite set of nonlinear equations
\begin{equation}\label{Lax}
\partial_{t_n}L=[B_n, L]\qquad {\rm with}\quad B_n=(L^n)_{\geq0}\quad n=1,2,\ldots,
\end{equation}
where $(L^n)_{\ge 0}$ represents the polynomial part of $L^n$ in $\partial$, 
i.e. $B_n$ is a differential operator of order $n$,
and $[B_n, L] := B_nL-LB_n$ is the usual commutator of operators. 
Since $[B_n, L]=[L^n-(L^n)_{<0}, L]=[L, (L^n)_{<0}]$, each side of the equation (\ref{Lax}) 
is a pseudo-differential operators of order $\leq -1$. Here $(L^n)_{<0}$ is the
negative part of $L^n$ in $\partial$, and note $[\partial,(L^n)_{<0}]$ has
no polynomial part.  Thus for $n > 1$, each equation
in (\ref{Lax}) is a consistent but infinite system of coupled $(1+1)$-evolution equations
in $t_n$ and $x$, for the variables $\{u_i:i=2,3,\cdots\}$. The case $n=1$ yields the equations
$\partial_{t_1}u_i = \partial_xu_i$, so we identify $t_1$ with $x$.
The infinite system (\ref{Lax}) is compatible, as prescribed by the following theorem:
\begin{Theorem}\label{ZS}
The differential operators $B_n=(L^n)_{\ge0}$ satisfy
\begin{equation}
\partial_{t_m} B_n-\partial_{t_n} B_m+[B_n,B_m]=0.
\label{zs}
\end{equation}
Consequently, the flows defined by (\ref{Lax}) commute i.e., for any $n,m\ge 1$,
\[
\partial_{t_n}(\partial_{t_m}L) = \partial_{t_m}(\partial_{t_n}L).
\]
\end{Theorem}
\begin{Proof}
It follows from (\ref{Lax}) that $\partial_{t_m}(L^n) = [B_m,L^n]$.
Then using the decomposition $L^n=B_n+(L^n)_{<0}$, we have
\begin{align*}
\partial_{t_m}(L^n)-\partial_{t_n}(L^m) &= [B_m,L^n]-[B_n,L^m] \\
&= 
[B_m,B_n]-[(L^m)_{<0}, \, (L^n)_{<0}].
\end{align*}
The differential part ($\geq 0$) of the above equation gives (\ref{zs}).

To prove that the flows commute we compute using (\ref{Lax}) once again, and obtain
\begin{align*}
 \partial_{t_m}(\partial_{t_n}L) - \partial_{t_n}(\partial_{t_m}L) &= 
 [\partial_{t_m}B_n,L]+[B_n,\partial_{t_m}L]-[\partial_{t_n}B_m,L]-[B_m,\partial_{t_n}L]\\
&=[\partial_{t_m} B_n-\partial_{t_n} B_m,L]+[B_n,[B_m,L]]-[B_m,[B_n,L]]\,.
\end{align*}
Applying the Jacobi identity for commutators, the right hand side of the above
equations becomes $[\partial_{t_m} B_n-\partial_{t_n} B_m+[B_n,B_m],L]$, which
vanishes due to (\ref{zs}).
\end{Proof}
Equations (\ref{zs}) are called the Zakharov-Shabat equations for the KP hierarchy.
Note that for a given pair $(n,m)$ with $n > m$, (\ref{zs}) gives a closed system 
of $n-1$ equations for $u_2, u_3, \cdots, u_{n}$, in the variables $t_m, t_n$ and $x$.
For example, consider the case with $n=3$ and $m=2$, i.e. $B_2=(L^2)_{\geq 0}=\partial^2+2u_2$ and
$B_3=(L^3)_{\geq 0}=\partial^3+3u_2+3(u_{2,x}+u_3)$, then
the Zakharov-Shabat equation (\ref{zs}) for $B_2$ and $B_3$ gives the system
\begin{equation*}\left\{
\begin{aligned}
&\partial_{t_2}u_{2}=u_{2,xx}+2u_{3,x} \\
&2\partial_{t_3}u_{2} = 3(u_{2,x}+u_3)_{t_2} -(u_{2,xx}-3u_{3,x} + 3u_2^2)_x  
\end{aligned}\right.
\end{equation*}
After setting $t_2=y, t_3=t$, and eliminating $u_3$ from the system, 
$u=2u_2$ satisfies the KP equation (\ref{kp}).

We also remark that the KP hierarchy (\ref{Lax}) is given by the compatibility of the linear system
\begin{equation}\label{eigen}
\left\{\begin{array}{ccl}
L\phi&=&k\phi\,,\\[0.5ex]
\partial_{t_n}\phi&=&B_n\phi\,,\qquad n=1,2,\ldots\,,
\end{array}\right.
\end{equation}
where $k\in\mathbb{C}$, the spectral parameter, and $\phi(\mathbf{t} ;k)$ is the 
eigenfunction of the KP hierarchy. The compatibility among the second set of equations
gives the Zakharov-Shabat equations.

\subsection{The dressing transformation}\label{Satoform}
As we mentioned in the previous section, the Lax
operator $L$ can be gauge-transformed into the trivial operator $\partial$,
i.e.
\begin{equation}\label{dressing}
L \longrightarrow \partial=W^{-1}LW,
\end{equation}
where the operator of the gauge transformation is defined by
\begin{equation}\label{Woperator}
W = 1-w_1\partial^{-1}-w_2\partial^{-2}-w_3\partial^{-3}+\cdots \,.
\end{equation}
The coefficient functions $w_i$ are related to the
coefficients $u_j$ of $L$ via the relation $LW = W\partial$, and which yields
\begin{align*}
 u_2 & = w_{1,x}\,, \quad u_3 = w_{2,x}+w_1w_{1,x}\,, \quad
u_4  = w_{3,x}+(w_1w_2)_x-w_{1,x}^2+w_1^2w_{1,x}\,,\quad \cdots\\
u_{j+1}& = w_{j,x} + F_{j+1}(w_1, w_2, \ldots, w_{j-1})\,, \quad\cdots \,,
\end{align*}
where $F_{j+1}$ are differential polynomials of weight $j+1$ (note ${\rm wt}(w_j)=j$).
 Thus the $w_j$ can be considered as the
primary variables whose $x$-derivatives determine the KP variables as illustrated
by the KP and pKP equations mentioned in Section \ref{B-KPrelation}. The evolutions of
the $w_j$ with respect to the time variables $t_n$ can be prescribed in a consistent
fashion by requiring that the dressing operator $W$ satisfy the following system of equations, 
\begin{equation}\label{Sato}
\partial_{t_n} W=B_nW-W\partial^n  \qquad {\rm for} \quad n=1,2,\ldots \,, 
\end{equation}
where $B_n$ is now given by $B_n= (W\partial^nW^{-1})_{\geq0}$. Notice that this expression for $B_n$
as a differential operator
is a consequence of the equations, $[(\partial_{t_n}W)W^{-1}]_{\ge 0}=0$.
Equation (\ref{Sato}) is sometimes referred to as the Sato equation.

The following Theorem asserts that the KP hierarchy is generated by the 
dressing (or conjugation)
of the trivial commutation relation $[\partial_{t_n}-\partial^n, \partial] = 0$
by the operator $W$. Because of this result, the (inverse) gauge transformation, $\partial\to L$, is called the dressing 
transformation for the KP hierarchy.
\begin{Theorem}\label{W}
If the operator $W$ satisfies the Sato equation (\ref{Sato}),
then the operator $L=W\partial W^{-1}$ satisfies the Lax equation (\ref{Lax})
for the KP hierarchy, and the operators $B_n =(W\partial^nW^{-1})_{\ge0}= (L^n)_{\geq 0}$ satisfies the 
Zakharov-Shabat equations (\ref{zs}).
\end{Theorem}
\begin{Proof} First, a direct calculation using $L=W\partial W^{-1}$ and
$L^n=W\partial^nW^{-1}=B_n+(L^n)_{< 0}$ shows that the operator
\begin{align*}
W(\partial_{t_n}-\partial^n)W^{-1} &= \partial_{t_n}-(\partial_{t_n}W)W^{-1}-W\partial^nW^{-1}\\
&=\partial_{t_n}-((\partial_{t_n}W)+W\partial^{n})W^{-1}
= \partial_{t_n} - B_n \,, 
\end{align*}
where the last equality is due to (\ref{Sato}). Then equations (\ref{Lax}) and
(\ref{zs}) follow from the commutator relations
\begin{align*}
0 &= W[\partial_{t_n}-\partial^n, \partial]W^{-1} =
\partial_{t_n}L - [B_n, L] \,, \\
0 &= W[\partial_{t_n}-\partial^n, \partial_{t_m}-\partial^m]W^{-1} =
[\partial_{t_n}-B_n, \partial_{t_m}-B_m] \,.
\end{align*}
\end{Proof}
It follows from Theorem \ref{W} that the flows defined by the Sato 
equation (\ref{Sato}) are commutative, i.e. they satisfy the compatibility
condition $\partial_{t_m}(\partial_{t_n}W) = \partial_{t_n}(\partial_{t_m}W)$.
Indeed, the compatibility condition for (\ref{Sato}) is equivalent to  
$[\partial_{t_n}+(L^n)_{<0}, \partial_{t_m}+(L^m)_{<0}] = 0$. Using $L^n = B_n+(L^n)_{<0}$, 
the commutator term on the left hand side can be decomposed as 
$[\partial_{t_n} - B_n, \partial_{t_m} - B_m] +
[\partial_{t_n}-B_n, L^m] - [\partial_{t_m}-B_m, L^n]\,,$
which vanish due to Theorem (\ref{W}). Finally we note that the KP linear system
(\ref{eigen}) is obtained by the dressing action: $\phi = W\phi_0$ where the
vacuum eigenfunction $\phi_0$ satisfies the {\it bare} linear system
\begin{equation}\label{vac}
\left\{\begin{array}{ccl}
\partial \phi_0 &=& k \phi_0\,,  \\[0.5ex]
 \partial_{t_n}\phi_0 &=&  \partial^n \phi_0=k^n\phi_0\,,
\qquad n=1,2,\cdots\,.
\end{array}\right.
\end{equation}
This equation together with
the Sato equation (\ref{Sato}) form the basic ingredients of the dressing procedure.
We will use the bare eigenfunction in the normalized form,
\begin{equation}\label{bareF}
\phi_0({\bf t};k)=e^{\theta({\bf t};k)} \qquad 
{\rm with}\quad \theta({\bf t};k)=\sum_{n=1}^{\infty}k^nt_n.
\end{equation}

\subsection{The $\tau$-function}\label{S:tau}
The $\tau$-function (\ref{KP-tau}) introduced in Section 2.3 plays a 
fundamental role in Sato theory for the KP hierarchy. In this subsection
we demonstrate explicitly how the $\tau$-function is related to the dressing
operator $W$ which satisfies the Sato equation (\ref{Sato}).
We restrict our discussion in this paper to a finite truncation of the 
infinite order pseudo-differential operator $W$ for simplicity only, while 
capturing the flavor of the general theory. Let us then consider the dressing operator
\[
W=1-w_1\partial^{-1}-w_2\partial^{-2}-\cdots -w_N\partial^{-N}\,,
\]
and define the differential operator,
\begin{equation*}
W_N:=W\partial^N
=\partial^N-w_1\partial^{N-1}-w_2\partial^{N-2}-\cdots-w_N~~.
\end{equation*}
Since $W$ satisfies the Sato equation, it is immediate that
$W_N$ also satisfies 
\begin{equation*}
\frac{\partial W_N}{\partial t_n}=B_nW_N-W_N\partial^n\qquad
B_n=(W\partial^n W^{-1})_{\geq 0}~~.
\end{equation*}
The following Proposition establishes the compatibility conditions leading to
the extended Burgers hierarchy introduced in Section \ref{C:KP}.
\begin{Proposition} 
The $N$-th order differential equation $W_Nf=0$ is invariant
under any flow of the linear heat hierarchy (\ref{heat}), $\{\partial_{t_n}f=\partial_x^n f :
n=1,2,\ldots\}$.
\label{WN}
\end{Proposition}
\begin{Proof}
It suffices to show that $\partial_{t_n}(W_Nf)=0$.
Then the desired result follows from the uniqueness of the differential equation:
\begin{eqnarray*}
\partial_{t_n}(W_Nf)&=&\partial_{t_n}(W_N)f+W_N\partial_{t_n}f\\
&=&(B_nW_N-W_N\partial^n)f+W_N\partial_{t_n}f\\
&=&B_n(W_Nf)+W_N\left(\partial_{t_n}f-\partial_x^nf\right)=0.
\end{eqnarray*}
\end{Proof}
Proposition \ref{WN} provides the compatible system considered in Section \ref{ExtBurgers}
\begin{eqnarray}\left\{
\begin{aligned}
&W_Nf=0,\\
&\partial_{t_n}f=\partial_x^nf, \qquad n=1,2,\ldots.
\end{aligned}\right.\label{EBurgers}
\end{eqnarray}
Furthermore, a set $\{f_j:j=1,2,\ldots,N\}$ of linearly independent solutions
of $W_Nf=0$, i.e.
\[
f^{(N)}=w_1f^{(N-1)}+w_2f^{(N-2)}+\cdots +w_{N-1}f^{(1)}+w_N f\,,
\]
can be employed to explicitly construct the coefficient functions $w_i$
of the dressing operator $W$ as follows
\begin{equation}
w_i=\frac{(-1)^{i+1}}{\tau}\left|\begin{matrix}
f_1 & \cdots &f_1^{(N-i-1)}& f_1^{(N-i+1)} & \cdots & f_1^{(N)} \\
f_2 & \cdots &f_2^{(N-i-1)}& f_2^{(N-i+1)} & \cdots & f_2^{(N)} \\
\vdots &       &\vdots &\vdots &           &\vdots \\
f_N & \cdots &f_N^{(N-i-1)}& f_N^{(N-i+1)} & \cdots & f_N^{(N)} \\
\end{matrix}\right|\,,
\label{wi}
\end{equation}
where $\tau={\rm Wr}(f_1,\ldots,f_N)$. Note that since the $t_n$-dependence of the
$w_i$ is given via the evolution equations 
$\partial_{t_n}f_j= \partial^n_xf_j$ for $j =1,2, \ldots, N$, one also 
has an explicit solution of the Sato equation (\ref{Sato}).

The formula (\ref{wi}) for the coefficients $w_i$ can be exploited to obtain
an elegant expression for the KP eigenfunction $\phi$
via the $\tau$-function (see for example \cite{D:91, MJD:00}).
\begin{Lemma}\label{wavefunction}
The eigenfunction $\phi$ of the linear system (\ref{eigen}) can be expressed as
\[
\phi({\bf t};k)=\frac{\tau({\bf t}-[k^{-1}])}{\tau({\bf t})}\phi_0({\bf t};k)\,,
\]
where $\phi_0=e^{\theta({\bf t};k)}$ with $\theta({\bf t};k)=\sum_{n=1}^{\infty}k^nt_n$, and
\[
\left({\bf t}-[k^{-1}]\right):=\left(t_1-\frac{1}{k},~t_2-\frac{1}{2k^2},~t_3-\frac{1}{3k^3},~\ldots\right).
\]
\end{Lemma}
We provide a simple proof of this well-known formula to preserve
the self-contained style of the present article (see also \cite{D:91}).
\begin{Proof}
First note that using (\ref{wi}), the eigenfunction $\phi$ can be written as
\begin{align*}
\phi&= W\phi_0=\left(1-\frac{w_1}{k}-\frac{w_2}{k^2}-\cdots - \frac{w_N}{k^N}\right)\phi_0\\
&=\frac{1}{\tau}\left|\begin{matrix}
f_1  & f_1^{(1)} & \cdots & f_1^{(N)} \\
\vdots &\vdots &\ddots & \vdots\\
f_N & f_N^{(1)} & \cdots & f_N^{(N)}\\
k^{-N}&k^{-N+1}&\cdots & 1
\end{matrix}\right|\,.
\end{align*}
Using the elementary column operations, the determinant in the numerator of the above expression can be re-written as
\[
\frac{(-1)^N}{k^N}\left|\left(f_i^{(j)}-kf_i^{(j-1)}\right)_{1\le i,j\le N}\right|\,.
\]
From the integral representation of the functions $f_i$,
\[
f_i({\bf t})=\int_C e^{\theta({\bf t};\lambda)}\rho_i(\lambda)\,d\lambda 
\qquad{\rm for}\quad i=1,2,\ldots,N\,,
\]
each matrix element in this determinant is given by
\begin{align*}
f_i^{(j)}({\bf t})-kf_i^{(j-1)}({\bf t})&=
-k\int_C\lambda^{j-1}\left(1-\frac{\lambda}{k}\right)e^{\theta({\bf t};\lambda)}\,\rho_i(\lambda)\,d\lambda\\
&=-k\int_C\lambda^{j-1}e^{-\sum_{n=1}^{\infty}\frac{\lambda^n}{nk^n}}e^{\theta({\bf t}; \lambda)}
\rho_i(\lambda)\,d\lambda\\
&=-ke^{-\sum_{n=1}^{\infty}\frac{1}{nk^n}\partial_{t_n}}f_i^{(j-1)}({\bf t})=
-kf_i^{(j-1)}\left({\bf t}-[k^{-1}]\right)\,,
\end{align*}
where we have used $\ln(1-\sf{\lambda}{k})=-\sum_{n=1}^{\infty}\sf{\lambda^n}{nk^n}$. 
This completes the proof.
\end{Proof} 

One should note that the expression for $\phi$ in Lemma \ref{wavefunction}
also holds in the general case for the full untruncated version of the
operator $W$.  Expanding this formula with respect to $k$, we have an explicit formula for 
$w_i$'s in terms of the $\tau$-function, i.e.
\[
w_i=-\frac{1}{\tau}\,S_i(-\tilde{\partial})\tau,
\]
where $\tilde{\partial}:=(\partial_{t_1},\sf{1}{2}\partial_{t_2},\sf{1}{3}\partial_{t_3},\ldots)$ and
$S_n({\bf z})$'s are the complete homogeneous symmetric functions 
(or sometimes referred to as the elementary Schur polynomials) defined by
\begin{equation}\label{Sformula}
\exp\left(\sum_{n=1}^{\infty}k^nz_n\right)=\sum_{n=0}^{\infty}S_n({\bf z})k^n\,, \qquad
{\bf z}=(z_1,z_2,\ldots)\,.
\end{equation}
The explicit expressions of these polynomials are given by
\begin{gather*}
S_0({\bf z}) = 1, \quad S_1({\bf z}) = z_1, \quad S_2({\bf z}) = z_2 + \frac{z_1^2}{2}, \quad 
S_3({\bf z}) = z_3 + z_1z_2 + \frac{z_1^3}{6}, \quad \ldots \quad , \\
S_n({\bf z})=
\sum_{l_1+2l_2+\cdots+nl_n=n}\frac{z_1^{l_1}z_2^{l_2}\cdots z_n^{l_n}}{l_1!l_2!\cdots l_n!}\,.
\end{gather*}
Note that for an $N$-truncated operator $W$ (i.e. $w_n=0$ if $n>N$) the $\tau$-function 
satisfies the constraints
\[
S_n(-\tilde{\partial})\tau=0\qquad{\rm for}\quad n>N.
\]

Lemma \ref{wavefunction} will be now used to derive a set of first integrals of the KP equation
that will prove to be useful in our classification of the line-soliton solutions
in Section 5.
The integrability of the KP equation may be demonstrated by the existence of an infinite number
of conservation laws in the form,
\[
\partial_{t_n} h_j=\partial_x g_{j,n}\,,
\]
for some conserved densities $h_j$ and the corresponding conserved fluxes $g_{j,n}$.
These functions are differential polynomials of $u_i$'s in the Lax operator $L$, and
they can be found as follows:
Differentiating the quantity $\phi^{-1}\partial_x\phi$ with respect to $t_n$ and using
the evolution equation $\partial_{t_n}\phi=B_n\phi$, we first derive the conservation law,
\begin{equation}\label{CL}
\partial_{t_n}\left(\phi^{-1}\partial_x\phi\right)=\partial_x\left(\phi^{-1}B_n\phi\right).
\end{equation}
Next we invert (\ref{L}) using the generalized Leibnitz rule 
to express the differential operator $\partial$ in $\partial_x\phi$ in terms of $L$ as
\[
\partial=L-v_2L^{-1}-v_3L^{-2}-\cdots,
\]
where $v_j$'s are the {\it differential polynomials} of $u_i$'s. Then the conserved 
density $\phi^{-1}\partial_x\phi$ can be written as an infinite series after using
$L^n\phi = k^n\phi, \, n \in \mathbb{Z}$,
\[
\phi^{-1}\partial_x\phi=k-\frac{v_2}{k}-\frac{v_3}{k^2}-\cdots,
\]
Each function $v_j$ is a conserved density of the KP hierarchy; the first few are given by
\begin{align*}
v_2&=u_2,\qquad v_3= u_3,\qquad v_4=u_4+u_2^2,\qquad v_5=u_5-3u_2u_3+u_2u_{2,x},\quad\cdots.
\end{align*}
Note that these are nonlocal in the KP solution, $u=2u_2$
(except $v_2$). However, they can be expressed simply as certain differential
polynomials of the variable $w_1 = \partial_x(\ln \tau)$. Namely, we have
\begin{align*}
\phi^{-1}\partial_x\phi &=\partial_x\ln\phi=
\partial_x\left[\theta({\bf t};k)+\ln\tau({\bf t}-[k^{-1}])-\ln\tau({\bf t})\right]\\
&=k+\partial_x\left[\exp\left({-\sum_{n=1}^{\infty}\frac{1}{nk^n}\partial_{t_n}}\right)-1\right]
\ln\tau({\bf t})=k+\sum_{n=1}^{\infty}\frac{1}{k^n}\,\partial_xS_n(-\tilde{\partial})\,\ln\tau\,,
\end{align*}
which leads to
\begin{equation}\label{CDtau}
v_{n+1}=-\partial_x S_n(-\tilde{\partial})\,\ln\tau = -S_n(-\tilde{\partial})\,w_1 \,.
\end{equation}
For example, we have
\begin{align*}
v_2&=\partial^2_x\ln\tau,\quad v_3=\sf{1}{2}\partial_x\left(\partial_{t_2}-\partial_x^2\right)\ln\tau,
\quad v_4=\sf{1}{3}\partial_x\left( \partial_{t_3}-\sf{3}{2}\partial_x\partial_{t_2}+
\sf{1}{2}\partial_x^3\right)\ln\tau,
\cdots\qquad \\ 
v_{n+1}&=\sf{1}{n}\partial_x\left(\partial_{t_n} +( {\rm h.o.d.})\right)\,\ln\tau,\qquad \cdots.
\end{align*}
where (h.o.d) indicates the terms including higher powers of the derivatives. Moreover,
if the solutions $u_i$'s
of the KP hierarchy decrease rapidly to zero as $|x|\to\infty$, one can define the integrals $C_n$ by
\begin{equation}\label{Integral}
C_n:=\int_{-\infty}^{\infty}\,v_{n+1}(x,\ldots)\,dx\qquad {\rm for}\quad n=1,2,\dots.
\end{equation}
In particular,
if the $\tau$-function gives one line-soliton of $[i,j]$-type, i.e. $\tau({\bf t})=E_i({\bf t})+aE_j({\bf t})$ with $E_i({\bf t})=\exp(\theta({\bf t};k_i))$, then from (\ref{CDtau}) the integrals take
\begin{equation}\label{AVofCD}
C_n=-S_n(-\tilde{\partial})\ln \tau\Big|_{x=-\infty}^{x=\infty}=\frac{1}{n}\,\left(k_j^n-k_i^n\right).
\end{equation}
Notice here that  if $k_i<k_j$, $\tau\approx E_i=e^{\theta({\bf t};k_i)}$ for $x\ll0$ and $\tau\approx aE_j
=ae^{\theta({\bf t};k_j)}$ for $x\gg 0$.
In general, if there are $N$ line-solitons of $[i_l,j_l]$-type for $l=1,\ldots,N$,
then we have
\[
C_n=\sum_{l=1}^N\frac{1}{n}\left(k^n_{j_l}-k^n_{i_l}\right),
\]
(see Corollary \ref{pairing} below for the proof).
This expression is similar to the $N$-soliton solutions of the KdV equation (see for example \cite{ZF:71}),
but for the KP equation, the $C_n$'s are also independent of $y$. This fact will be used in 
Proposition \ref{ij} below.

\begin{Remark}
An alternative set of the conserved densities can be found by observing the following
tautological equations (see p.99 in \cite{D:91}),
\[
\partial_{t_m}(\partial_x\partial_{t_n}\ln\tau)=\partial_x(\partial_{t_m}\partial_{t_n}\ln\tau).
\]
That is, the conserved densities obtained from those equations are given by
\[
\tilde{v}_{n+1}:=\frac{1}{n}\partial_x\partial_n\,\ln\tau.
\]
However one can show that the integrals $\tilde{C}_n=\int_{-\infty}^{\infty}\tilde{v}_{n+1}\,dx=\int_{-\infty}^{\infty}v_{n+1}=C_n$ for all $n$. 
\end{Remark}

In Section \ref{C:Classification}, we will discuss the classification problem of the solutions
obtained from the $\tau$-function with finite dimensional solutions of the $f$-equation,
and will show that the classification is completely characterized by the asymptotic behavior of 
the $\tau$-function.
Before getting into the classification problem, in the next Section, we give a brief review on the
real Grassmann manifold for the background information necessary for the problem.

%%%%%%%%%%%%%%%%%%%%%%%%%%%%%%%%%%%%%%%%%%%%%%%%%%%%%%%%%%%%%%%%%%%
\section{Introduction to the Grassmannian Gr$(N,M)$}\label{Grassmann}
The main purpose of this Section is to explain a mathematical
background of soliton solutions of the KP equation which is provided by
the real Grassmannian Gr$(N,M)$, the set of $N$-dimensional subspaces
of $\mathbb{R}^M$. This is a finite 
dimensional version of the universal Grassmannian introduced by Sato in 1981
\cite{S:81}.

\subsection{Grassmannian Gr$(N,M)$}
Let $\{f_1, f_2, \ldots, f_N\}$ be a set of $N$ linearly 
independent vectors in $\mathbb{R}^M$. Then the set forms a basis for $V$,
\[
V = {\rm Span}_{\mathbb{R}} \{f_1, f_2, \ldots , f_N \}\,, \quad {\rm and}
\quad {\rm Gr}(N,M) = \{V : V \subset \mathbb{R}^M \}\,.
\]
Let us begin with some simple examples:
\begin{Example} The Grassmannian Gr$(1,M)$: Suppose $M=2$, then Gr$(1,2)$ is the set of all
lines through the origin in the plane. Thus, any two non-zero vectors in 
$\mathbb{R}^2$ are on the same line if they are proportional, and they
determine the {\it same} point of $\mathrm{Gr}(1,2)$. Since every point
$(x_1, x_2) \neq (0, 0)$ of $\mathbb{R}^2$ lies on some line $l$ through 
the origin, we can represent $l$, which is also a {\it point} on $\mathrm{Gr}(1,2)$,
as 
\begin{equation*}
{\rm Gr}(1,2) \ni l = (x_1:x_2) := \{c(x_1, x_2)\,|\,c \neq 0,\,(x_1, x_2) \in 
\mathbb{R}^2\setminus\{(0,0)\} \}\,.
\end{equation*}
The line $(x_1:x_2)$ is called the homogeneous coordinate for Gr$(1,2)$
which has then the structure of a one-dimensional manifold. This manifold
admits the decomposition,
\[
{\rm Gr}(1,2)=\{(1:a):a\in\mathbb{R}\}\cup\{(0:1)\}\cong\mathbb{R}\cup\{\infty\},
\]
which is the projective line $\mathbb{R}P^1$,
and it can be identified with the unit  circle $S^1$
(i.e. $a\in\mathbb{R}$ in the decomposition parametrizes the slope of each line, and
the $y$-axis is identified as $(0:1)$).

The set of all lines passing through the origin of $\mathbb{R}^M$ is
represented by the Grassmannian Gr$(1,M)$ which can be identified with the 
$M-1$ dimensional projective space $\mathbb{R}\mathrm{P}^{M-1}$. Homogeneous
coordinates and the decomposition of Gr$(1,M)$ can be described in the same way 
as in the previous example.
In this case, one has the well-known Schubert decomposition given by 
$$ {\rm Gr}(1,M)=\bigcup_{j=1}^M F_j \quad {\rm with} \quad
F_j = \{(0:\ldots:0:1:a_{1}: \ldots:a_{M-j}): a_k\in\mathbb{R}\} \,,
$$ 
where each {\it cell} $F_j$ has the structure of a $(M-j)$-dimensional vector space;
the last one being $F_M=(0:\ldots:0:1)$ which represents a point (the $0$-dimensional cell)
in Gr$(1,M)$.
\end{Example}

We now consider a representation for the general case of Gr$(N,M)$ which plays
a key role in the classification of the line-soliton solutions of the KP equation as we
show in the next section. With respect to a basis $\{E_1, E_2, \ldots, E_M\}$ of 
$\mathbb{R}^M$, the basis vectors $f_i$ of the $N$-dimensional subspace $V$ can be 
expressed as
\begin{equation}
f_i=\sum_{j=1}^M a_{ij}E_j\,, \qquad{\rm for}\quad i=1,2,\ldots,N\,,
\label{Amatrix}
\end{equation}
where $A := (a_{ij})$ is a $N \times M$ matrix. Note that the rank of
$A$ is $N$ since the $\{f_1,f_2,\ldots,f_N\}$ are linearly independent.
A different choice of basis for $V$ will induce
row operations on the original matrix:\, $A \to BA$ with $B \in {\rm GL}(N)$.
Consequently, any matrix $A$ can be canonically chosen in the {\it reduced row 
echelon form} (RREF) by taking appropriate $B\in {\rm GL}(N)$.
This canonical form of $A$ has a distinguished set of {\it pivot} columns
labeled by $\mathcal{I} = \{i_1,i_2,\ldots,i_N\}$ with $1 \leq i_1 < i_2 < \ldots < i_N \leq M$
such that the submatrix $A_{\mathcal{I}}$ formed by the column set 
$\mathcal{I}$ is the identity matrix $I_N$.
For any given $A$, its unique RREF provides a coordinate for a point of Gr$(N,M)$.
Moreover, the set of all such $A$ matrices in RREF, and with a fixed pivot set $\mathcal{I}$
forms an affine coordinate chart or {\it cell} $W_\mathcal{I}$ which (similar to Gr$(1,M)$) 
gives the Schubert decomposition of the Grassmannian  
\begin{equation}
\mathrm{Gr}(N,M)=\bigcup_{\mathcal{I}\subset\{1,\ldots,M\}} W_\mathcal{I}\,.
\label{schubert}
\end{equation}
Each cell $W_\mathcal{I}$ is called a Schubert cell. For example, 
if $\mathcal{I}=\{1,2,\ldots,N\}$, then the Schubert cell $W_\mathcal{I}$ contains all $A$ matrices 
whose RREF is given by
\begin{equation*}\left(
\begin{array}{ccccccc}
1&0&\cdots&0&*&\cdots&*\\
0&1&\cdots&0&*&\cdots&*\\
\vdots&\vdots&\ddots&\vdots&\vdots&\vdots&\vdots\\
0&0&\cdots&1&*&\cdots&*
\end{array}\right)
\end{equation*}
where the $N(M-N)$ entries of the right-hand block are arbitrary. This particular
Scubert cell $W_\mathcal{I}$ is often refered to as the {\it top} cell of the Grassmannian 
$\mathrm{Gr}(N,M)$, and has the maximum number of free parameters marked by $*$. 
It follows from this that the dimension of $\mathrm{Gr}(N,M)$ is $N(M-N)$.
A straightforward counting of the maximum number of free parameters for an $A$
in RREF with a given pivot set $\mathcal{I} = \{i_1,\ldots,i_N\}$ shows that 
the dimension of the cell $W_\mathcal{I}$ is given by
$${\rm dim}\, W_\mathcal{I} =N(M-N)+\frac{1}{2}N(N+1)-(i_1+i_2+\cdots+i_N)\,.$$
\begin{Example} For Gr$(2,4)$, the Schubert decomposition (\ref{schubert}) is given by
\begin{equation*}
\mathrm{Gr}(2,4)=\bigcup_{1\le i<j\le 4}W_{\{i,j\}}\,.
\end{equation*}
There are six cells $W_{\{i,j\}}$ with $\mathrm{dim}\,W_{\{i,j\}}=7-(i+j)$, which are given by
\begin{eqnarray*}
& {\rm (a)}\,\, W_{\{1,2\}} = \left\{\begin{pmatrix} 1&0&*&*\\0&1&*&* \end{pmatrix}\right\} \quad \qquad  
& {\rm (b)}\,\, W_{\{1,3\}} = \left\{\begin{pmatrix} 1&*&0&*\\0&0&1&* \end{pmatrix} \right\}\\
& {\rm (c)}\,\, W_{\{1,4\}} = \left\{\begin{pmatrix} 1&*&*&0\\0&0&0&1 \end{pmatrix}\right\} \quad \qquad  
& {\rm (d)}\,\, W_{\{2,3\}} = \left\{\begin{pmatrix} 0&1&0&*\\0&0&1&* \end{pmatrix}\right\} \\
& {\rm (e)}\,\, W_{\{2,4\}} = \left\{\begin{pmatrix} 0&1&*&0\\0&0&0&1 \end{pmatrix}\right\} \quad \qquad  
& {\rm (f)}\,\, W_{\{3,4\}} = \left\{\begin{pmatrix} 0&0&1&0\\0&0&0&1 \end{pmatrix} \right\}
\end{eqnarray*}
\end{Example}

\subsection{The Pl\"ucker embedding and the $\tau$-function}
We now introduce a coordinate system called the Pl\"ucker coordinates for
the Grassmannian Gr$(N,M)$ which is related directly to the Wronskian form of the $\tau$-function (\ref{KP-tau}) in Theorem \ref{PHrelation}. Given any basis $\{f_1,\ldots,f_N\}$ of an $N$-dimensional
subspace $V \subset \mathbb{R}^M$, the exterior product space $\bigwedge^NV$ is a 
one-dimensional subspace $\bigwedge^NV \subset \bigwedge^N\mathbb{R}^M$ spanned by 
the wedge product $\nu_f:=f_1\wedge f_2\wedge\cdots\wedge f_N$.
If $\{g_1,\ldots,g_N\}$ is another basis for $V$ with $B\in {\rm GL}(N)$ so that 
$(g_1,\ldots,g_N)=(f_1,\ldots,f_N)B$, then $ \nu_g =\det(B)\nu_f$.
Consequently, $\bigwedge^NV$ is an element of $P(\bigwedge^N\mathbb{R}^M)$,
the projectivization of $\bigwedge^N\mathbb{R}^M$.
The Pl\"ucker embedding is a map from Gr$(N,M)$ to $P(\bigwedge^N\mathbb{R}^M)$,
that sends $V \subset \mathbb{R}^M$ to $\bigwedge^NV \subset \bigwedge^N\mathbb{R}^M$.
In terms of the $A$-matrix in (\ref{Amatrix}), the explicit form of this map
is given by
\begin{equation}\label{Pembedding} 
\nu_f = f_1\wedge f_2\wedge\cdots\wedge f_N = \sum_{1\leq i_1<\cdots<i_N\leq M} \hspace{-0.2 in}
\xi(i_1,i_2,\cdots,i_N)E_{i_1}\wedge E_{i_2}\wedge\cdots\wedge E_{i_N} \,,
\end{equation}
where $\xi(i_1,i_2,\cdots,i_N)$ are the maximal minors of $A$ formed by its
columns $A_i, \, i \in \{i_1,i_2,\ldots,i_N\}$. Recall that the maximal minors
were introduced in Section 1.3 (following Theorem \ref{PHrelation}).
The minors in the set $\{\xi(i_1,\cdots,i_N):1\le i_1<\cdots<i_N\le M\}$ are defined 
up to a scale factor, and are called the Pl\"ucker coordinates for the Grassmannian 
$\mathrm{Gr}(N,M)$. These coordinates are not independent, but satisfy the identities 
\begin{equation}\label{plucker}
\sum^{N+1}_{i=1}(-1)^{i-1}\xi(\alpha_1,\cdots,\alpha_{N-1},\beta_i)
\xi(\beta_1,\cdots,\beta_{i-1},\beta_{i+1},\cdots,\beta_{N+1})=0,
\end{equation}
called the Pl\"ucker relations which hold because the exterior form
$\nu_f = f_1\wedge \cdots\wedge f_N$ is totally {\it decomposable}, i.e.,
$v \wedge \nu_f = 0, \, \forall \, v \in V$. The Pl\"ucker relations can be
derived using elementary linear algebra from the Laplace expansion
of the following $2N \times 2N$ determinant formed by the columns $A_i$ of
the matrix $A$, similar to Lemma \ref{Prelation},
\begin{equation*}
\left|\begin{matrix}
A_{\alpha_1} &\cdots & A_{\alpha_{N-1}}  & A_{\beta_1}& \cdots & A_{\beta_{N+1}}\\
 0  &\cdots & 0  & A_{\beta_1}& \cdots & A_{\beta_{N+1}} \\
 \end{matrix}\right| = 0 \,.
\end{equation*}
The Pl\"ucker coordinates, modulo the
Pl\"ucker relations, give the correct dimension of Gr$(N,M)$ which
is typically less than the dimension of $P(\bigwedge^N\mathbb{R}^M)$.
\begin{Example} For Gr$(2,4)$, the Pl\"{u}cker coordinates are given by the
maximal minors,
\[
\xi(1,2), \quad \xi(1,3), \quad \xi(1,4), \quad\xi(2,3), \quad\xi(2,4), \quad\xi(3,4).
\]
Taking $\alpha_1=1$, $(\beta_1,\beta_2,\beta_3)=(2,3,4)$ in (\ref{plucker}) gives the
only Pl\"{u}cker relation in this case,
$$\xi(1,2)\xi(3,4)-\xi(1,3)\xi(2,4)+\xi(1,4)\xi(2,3)=0,$$
which is the same as (\ref{PLphi}).
Since $\dim(\bigwedge^2\mathbb{R}^4) = 6$, and the projectivization 
gives dim($P(\bigwedge^2\mathbb{R}^4)) = 6-1=5$. Then with one Pl\"ucker relation, 
the dimension of Gr$(2,4)$ turns out be 4, which is consistent with the
dimension of the top cell $W_{\{1,2\}}$ as shown in Example 3.2 (case (a)).
\end{Example}

We next show that  the $\tau$-function defined by 
the Wronskian determinant in (\ref{KP-tau}) can be identified as a point on the
Grassmannian Gr$(N,M)$. First note that the set of functions,
\[
\left\{E_j=e^{\theta_j}=\exp\left(\sum_{n=1}^{\infty}k_j^nt_n\right) :j=1,2,\ldots,M\right\}\,,
\]
with distinct real parameters $k_j$, gives a linearly independent set because 
 $\mathrm{Wr}(E_1,E_2,\cdots,E_M) \neq 0$.  Then an $N$-dimensional subspace 
$V \subset \mathbb{R}^M$ is given by 
$\mathrm{Span}_{\mathbb{R}}\{f_1,f_2,\ldots,f_N\}$ with $N$ linearly independent solutions 
$\{f_1,f_2,\ldots,f_N\}$ of $f_y=f_{xx}$ and $f_t=f_{xxx}$
as in Theorem \ref{PHrelation}. The coefficients of $E_j$ in (\ref{Amatrix})
form a $N \times M$ matrix $A$ of rank $N$, whose row-space is isomorphic
to $V$ so that $A$ represents a point of Gr$(N,M)$. One can then see that 
the $\tau$-function 
$\tau = \mathrm{Wr}(f_1,f_2,\ldots,f_N)$
gives the Pl\"ucker embedding Gr$(N,M) \to {P}(\bigwedge^N\mathbb{R}^M)$ defined in (\ref{Pembedding}):
Indeed expanding the Wronskian determinant (\ref{KP-tau}) by Binet-Cauchy formula,
we have
\begin{equation}\label{tauexpansion}
\tau=\mathrm{Wr}(f_1,f_2,\ldots,f_N)
=\sum_{1\le i_1<\cdots<i_N\le M} \hspace{-0.2 in}
\xi(i_1,i_2,\ldots,i_N)E(i_1,i_2,\ldots,i_N)\,,
\end{equation}
where $\xi(i_1,i_2,\ldots,i_N)$ are the Pl\"ucker coordinates given by the maximal minors of 
the $A$-matrix, and $E(i_1,i_2,\ldots,i_N)={\rm Wr}(E_{i_1},E_{i_2},\ldots,E_{i_N})$.
Note here that $E(i_1,i_2,\ldots,i_N)$ can be
identified as $E_{i_1}\wedge E_{i_2}\wedge\cdots E_{i_N}$, and forms a basis 
for $\bigwedge^N\mathbb{R}^M$, i.e.
\[
{\rm Span}_\mathbb{R}\left\{E(i_1,i_2,\ldots,i_N):1\le i_1<i_2<\cdots<i_N\le M\right\}\cong 
\bigwedge^N\mathbb{R}^M\,.
\]
 Note here that the sum
$k_{i_1}+k_{i_2}+\ldots+k_{i_N}$ should be distinct for distinct sets
$\{i_1,i_2,\ldots,i_N\}$ in order for the $\{E(i_1,i_2,\ldots,i_N)\}$ 
to be independent.

\begin{Remark}\label{GrassmannSato}
We just mention here that an infinite dimensional Grassmann manifold appears when
we consider the case where the functions $\{f_1,\ldots,f_N\}$ are analytic.
In this case, one can expand
$f_i({\bf t})$ in terms of the elementary Schur polynomials defined in (\ref{Sformula}), 
\begin{align*}
f_i({\bf t})&=\int_Ce^{\theta(\mathbf{t};k)}\,\rho_i(k)\,dk=
\sum_{n=1}^{\infty}a_{i,n}S_{n-1}({\bf t}),
\end{align*}
where $\theta(\mathbf{t};k)=\sum_{n=1}^{\infty}k^nt_n$, and $a_{i,n}$ are given by the moment integrals,
\[
a_{i,n}=\int_Ck^n\,\rho_i(k)\,dk\qquad {\rm for}\qquad
\left\{\begin{array}{lll}
1\le i\le N,\\
1\le n.
\end{array}\right.
\]
Thus the $A$-matrix has the size $N\times \infty$, and each $A$-matrix represents a point
on an infinite dimensional Grassmann manifold Gr$(N,\infty)$. The $\tau$-function then has the 
well-known expansion form with the Schur polynomials,
\begin{equation}\label{Satotau}
\tau(\mathbf{t})=\sum_{1\le i_1<i_2<\cdots<i_N}\xi(i_1,i_2,\ldots,i_N)S_{\{i_1,i_2,\ldots,i_N\}}({\bf t}),
\end{equation}
where $S_{\{i_1,\ldots,i_N\}}({\bf t})$ is defined by
\[
S_{\{i_1,i_2,\ldots,i_N\}}({\bf t})={\rm Wr}(S_{i_1-1},S_{i_2-1},\ldots,S_{i_N-1})(\mathbf{t}).
\]
The index set $\{i_1,i_2,\dots,i_N\}$ is commonly expressed by a Young diagram
(see for example \cite{MJD:00}). Then the main result of the Sato theory is that
the function defined by (\ref{Satotau}) is the $\tau$-function, which 
 represents a point on the Grassmann manifold \cite{S:81}.
\end{Remark}

In the following section, we will show that the line-soliton solutions
 arise from the $\tau$-function given by (\ref{tauexpansion}). The Grassmannian
formulation of this $\tau$-function provides a geometric characterization of
the solutions by identifying the different solution classes with distinct Grassmann cells.
Of particular physical interests are those solutions which are regular
in $x, y$ and $t$. Such solutions are realized if the associated $\tau$-function has 
non-negative Pl\"ucker coordinates, that is, all  $\xi(i_1,i_2,\ldots,i_N) \geq 0$ in (\ref{tauexpansion}).
The matrices $A$ having this property are called {\it totally non-negative} (TNN) matrices;
they parametrize the totally non-negative Grassmannian Gr$^+(N,M) \subset \mathrm{Gr}(N,M)$
(see also \cite{P:06}).

%%%%%%%%%%%%%%%%%%%%%%%%%%%%%%%%%%%%%%%%%%
\section{Classification of soliton solutions}\label{C:Classification}
In this section, we present a classification scheme of the line-soliton solutions
based on the asymptotic behavior of the $\tau$-function 
(\ref{tauexpansion}). It turns out that the classification problem is closely
related to the characterization of the totally nonnegative Grassmannian cells
of Gr$(N,M)$ recently studied by Postnikov et al (see for example \cite{P:06, W:05}).

\subsection{Asymptotic line-solitons}
Let us recall that the $\tau$-function is given by the sum of exponential terms,
\begin{equation}\label{tauexp}
\tau({\bf t})=
\sum_{1\le m_1<\cdots<m_N\le M}\xi(m_1,\ldots,m_N)E(m_1,\ldots,m_N)({\bf t}),
\end{equation}
where $E(m_1,\ldots,m_N)={\rm Wr}(E_{m_1},\ldots,E_{m_N})$ with $E_m({\bf t})=e^{\theta({\bf t};k_m)}=\exp(\sum_{n=1}^{\infty}k_m^nt_n)$.
We then impose the following conditions:
\begin{itemize}
\item[(i)] The $k$-parameters are
ordered as $k_1 < k_2 < \ldots < k_M$, and the sums $k_i+k_j$ are all distinct (this implies that
all $[i,j]$-solitons have different slopes).
\item[(ii)] The $A$-matrix is TNN, that is, all its maximal minors
$\xi(m_1,m_2,\dots,m_N)$ are non-negative (i.e. the $\tau$-function is 
positive definite for all $x,y,t$, hence the solution $u=2\partial_x^2(\ln\tau)$ is non-singular).
\end{itemize}

The asymptotic spatial structure of the solution $u(x,y,t)$ is determined
from the consideration of which exponential term $E(m_1,\ldots,m_N)$ in the $\tau$-function
dominates in different regions of the $xy$-plane for large $|y|$.
 For example, if only one 
exponential term $E(m_1,\dots,m_N)$ in the $\tau$-function is dominant 
in a certain region, then the solution $u = 2 \partial_x^2(\ln\tau)$ remains exponentially small 
at all points in the interior of any given dominant region, but is localized at 
the {\it boundaries} of two distinct regions where a balance exists between  
two dominant exponentials in the $\tau$-function (\ref{tauexp}).

Furthermore, it is also possible to deduce the following result from the
asymptotic analysis of the $\tau$-function:
\begin{Proposition}
The dominant exponentials of the $\tau$-function in adjacent
regions of the $xy$-plane are of the form $E(i,m_2,\dots,m_N)$
and $E(j,m_2,\dots,m_N)$. That is, the exponentials contain
$N-1$~common phases and differ by only one phase. 
\label{ij}
\end{Proposition}
We prove the Proposition by applying the following Lemma.
\begin{Lemma}\label{uniqueness}
Let $\{s_n(x_1,x_2,\ldots,x_N):n=1,\ldots,N\}$ be the set of $N$ power sums
in $N$ real variables defined by  
\[
s_n:=x_1^n+x_2^n+\cdots+x_{N}^n, \qquad n=1,2,\ldots,N.
\]
Suppose the set of $N$ equations 
$$  s_n(x_1,x_2, \ldots, x_N) = c_n \,, \qquad n=1,\ldots,N\,,$$ 
admit solutions $(x_1,\ldots,x_{N})$ for appropriate values of the constants $c_n$, 
then the solution set is unique up to the permutation of $N$ elements $\{x_1, \ldots, x_N\}$.
%Consequently, if the elements are distinct then (by renaming the variables, if necessary)
%there is only one solution of the form $(x_1,\ldots,x_{N})$ with $x_1<\ldots<x_N$.
\end{Lemma}
\begin{Proof}
The proof follows from the Newton's identities relating the power sums $\{s_n\}$
and the elementary symmetric polynomials $\{\sigma_n\}$ which are defined as
\[
\sigma_n=\sum_{1\le i_1<\cdots<i_n\le N}x_{i_1}x_{i_2}\cdots x_{i_n}\qquad n=1,\ldots,N\,.
\]
For $n=1,2,\ldots,N$, the Newton's identities are recursively given by
$$
s_n + a_1s_{n-1} + \ldots + a_{n-1}s_1 + na_n = 0\,, \qquad 
a_k = (-1)^k \sigma_k \,. 
$$
Then the solutions of $s_n=c_n, \,\, n=1,\ldots,N$, which exist by assumption, 
correspond the set of $N$ roots $x_1, \ldots, x_N$ of the $N$-th degree monic polynomial,
\[
x^{N}+a_1x^{N-1}+a_2x^{N-2}+\cdots+a_{N}=0,
\]
where the coefficients $\{a_n\}$ are uniquely determined by $\{c_n\}$ using Newton's identity.
Obviously, the set of roots $(x_1,\ldots,x_n)$ is unique up to permutations. 
\end{Proof}
Now we prove Proposition \ref{ij}:
\begin{Proof}
Let $E(i_1,\ldots,i_N)$ and $E(j_1,\ldots,j_N)$ be the two dominant exponentials 
in adjacent regions, then near the boundary of the two regions, the $\tau$-function can be
approximately given by 
\[
\tau\approx \xi_1E(i_1,\ldots,i_N)+\xi_2E(j_1,\ldots,j_N),
\]
where $\xi_1$ and $\xi_2$ are the corresponding minors of the $A$-matrix.
The remaining terms in the $\tau$-function are exponentially small in comparison with those dominant
terms.
Then the solution $u=2\partial_x^2(\ln\tau)$ locally has the one-soliton 
form as in (\ref{OneHsoliton}), i.e.
\[
u\approx \mathcal{A}\sech^2\left(\sum_{n=1}^{\infty}\mathcal{K}_nt_n+\Theta^0\right)\,,
\]
where $\mathcal{A}=\sf{1}{2}(Q-P)^2$ and $\mathcal{K}_n=\sf{1}{2}(Q^n-P^n)$. Moreover, $P$
and $Q$ satisfy
\begin{equation*}
Q^n-P^n=(k_{j_1}^n+\cdots+k_{j_N}^n) -(k_{i_1}^n+\cdots+k_{i_N}^n)\qquad n=1,2,\ldots\,.
\end{equation*}
Here it is assumed (without any loss of genericity) that
$k_{j_1}+\cdots+k_{j_N}>k_{i_1}+\cdots+k_{i_N}$ so that $P < Q$.
Let us rewrite the first $N+1$ of the above equations as follows
\begin{equation}\label{PQ}
P^n+k_{j_1}^n+\cdots+k_{j_N}^n=Q^n+k_{i_1}^n+\cdots+k_{i_N}^n := c_n\,,
\end{equation}
where $c_n$'s are values of each sum. Then $\{P,k_{j_1},\ldots,k_{j_N}\}$ and 
$\{Q,k_{i_1},\ldots,k_{i_N}\}$ are solutions to the 
system of equations $x_1^n + \ldots x_{N+1}^n = c_n, \, n=1,\ldots,N+1$.
Hence, from Lemma \ref{uniqueness} these two sets must be the
same up to permutation of their elements.
Now recall that the sets $\{k_{i_n}\}$ and $\{k_{j_n}\}$ are distinct and $P \neq Q$.
Then the one-to-one correspondence between the sets $\{P,k_{j_1},\ldots,k_{j_N}\}$ 
and $\{Q,k_{i_1},\ldots,k_{i_N}\}$ implies that $P$ must be one of $\{k_{i_n}\}$, and 
$Q$ must be one of $\{k_{j_n}\}$. Then the remaining $N-1$ elements of the set
$\{k_{i_n}\}$ must be the same as the remaining $N-1$ elements of the set $\{k_{j_n}\}$,
proving the Proposition.
\end{Proof}

 As an immediate consequence of Proposition \ref{ij},
the asymptotic behavior of the KP solution is given by 
\begin{equation}
u(x,y,t) \approx \sf12(k_j-k_i)^2 \mathrm{sech}^2\sf12(\theta_j-\theta_i+\theta_{ij}) \,,
\label{uasymp}
\end{equation}
in the neighborhood of the line $x+(k_i+k_j)y$=constant, which forms the boundary 
between the regions of dominant exponentials $E(i,m_2,\ldots,m_N)$ and $E(j,m_2,\ldots,m_N)$.
Equation (\ref{uasymp}) defines an {\it asymptotic} $[i,j]$-soliton as a result of those 
two dominant exponentials. Then the condition (i) below (\ref{tauexp}) implies that
those asymptotic solitons have the different slopes, so that they all separate asymptotically
for $|y|\gg 0$.
 
\begin{figure}
\begin{center}
\includegraphics[height=4.3cm]{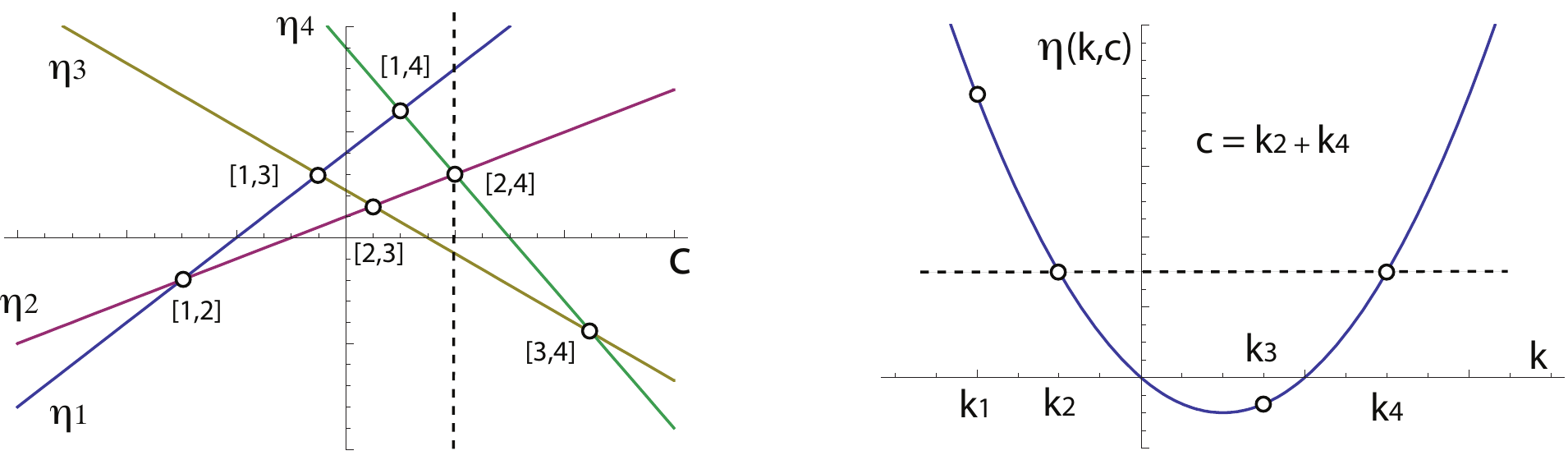}
\caption{The left figure shows $\eta_j(c)=k_j(k_j-c)$ for $j=1,\ldots,4$. 
Each $[i,j]$ at the intersection point of $\eta_i(c)=\eta_j(c)$ indicates the exchange $i\leftrightarrow j$.
We assume that there is at most one intersection point for each $c$ (genericity condition for $k_j$'s).
The right one illustrates the $\eta$ as the function of $k$ with $c=k_2+k_3$, that is, the
$\eta$ along the dotted line in the left figure passing through the intersection point $\eta_2(c)=\eta_4(c)$.
This figure shows the order $\eta_3<\eta_2=\eta_4<\eta_1$.\label{fig:eta24}}
\end{center}
\end{figure}

Next we discuss how to determine which exponential term in a given $\tau$-function
is actually dominant in the direction of particular line $x=-cy$ for $y \to \pm\infty$.
The basic idea is the same as in Example \ref{ex:12soliton}.
Along the line $x=-cy$, each exponential term $E(m_1,m_2,\ldots,m_N)$ has the form,
\[
C\exp\left(\sum_{n=1}^N\eta(k_{m_n},c)\,y+\theta_{m_n}^0(t)\right),
\]
where $C$ and $\theta_{m_n}^0(t)$ are constants for fixed $t$, and $\eta(k,c)$ is defined by
(cf. (\ref{eta})),
\begin{equation}\label{etakc}
\eta(k,c)=k(k-c)\qquad{\rm and}\qquad \eta_j(c)=k_j(k_j-c)\,.
\end{equation}
Then for $y\gg 0$ (or $\ll 0$), we look for the dominant (or least) sum of $\eta(k_{m_n},c)$
for each $c$. First note that we have
\[
\eta_i(c)=\eta_j(c) \qquad {\rm if}\qquad c=k_i+k_j,
\]
that is, in the direction of $[i,j]$-soliton, those terms are in balance. 
Since the $k$-parameters are ordered as $k_1<k_2<\cdots<k_M$, we have 
the following dominance relation among the other $\eta_m(c)$'s along $c=k_i+k_j$,
\begin{equation}\label{dominant}
\left\{\begin{array}{lll}
\eta_i=\eta_j<\eta_m \quad {\rm if}\quad 1\le m<i~{\rm or}~ j<m\le M,\\[1.0ex]
\eta_i=\eta_j>\eta_m\quad {\rm if}\quad i<m<j.
\end{array}\right.
\end{equation}
In order to find the dominant sum, the graph of $\eta(k,c)$ is particularly useful.
Figure \ref{fig:eta24} illustrates the case with $M=4$.
%%%%%%%%%%%%%%%%%%%%%%%%%%%%%%%%%
%%%%%%%%%%%%%%%%%%%%%%%%%
We demonstrate how to identify the asymptotic line-solitons from a given 
$\tau$-function using Lemma \ref{ij} and the relations (\ref{dominant}).
\begin{Example}
We consider the $A$-matrix given by
\[
 A = \begin{pmatrix} 1 & 1 & 0 & 0 \\ 0 & 0 & 1& 1 \end{pmatrix}\,.
 \]
Then from (\ref{tauexp}) the $\tau$-function has the form,
\[
\tau=E(1,3)+E(1,4)+E(2,3)+E(2,4).
\]
Since Proposition \ref{ij} implies that  the line-solitons are localized along the
phase transition lines $x+cy=$constant with $c=k_i+k_j$, we look for dominant exponential terms
in the $\tau$-function along those directions. 

For $y\gg0$, the $c$-values along a
line $x = -cy$ decrease as we sweep clockwise from negative to positive
$x$-axis. So, we start with the largest value $c = k_3+k_4$, and try to
determine if the line $[3,4]$ actually corresponds to a line-soliton.
The relation (\ref{dominant})
gives $\eta_1 , \eta_2 > \eta_3=\eta_4$, along the
line $[3,4]$. This implies that $\eta_1+\eta_2$ is the most dominant 
combinations along this line. However, since the corresponding
exponential term $E(1,2)$ is {\it absent} in the $\tau$-function, it turns out that
$E(1,3)$ and $E(1,4)$ are the two dominant exponentials 
in the $\tau$-function along the line $[3,4]$ satisfying the condition of Lemma \ref{ij}. 
Therefore, there is a $[3,4]$-soliton. An almost identical
argument implies that $[1,2]$ is another line-soliton. But for example,
along the line $[1,3]$, the following relations hold:\, 
$\eta_4 > \eta_1=\eta_3,\, \eta_2 < \eta_1=\eta_3$. This implies that
$\eta_1+\eta_4$ and $\eta_3+\eta_4$ are the two dominant combinations.
But $E(1,4)$ is the only dominant exponential term along the line $[1,3]$, since $E(3,4)$ is absent from the
$\tau$-function.
Hence the solution $u \approx 0$.  In this fashion by checking
along the lines $[i,j]$ for all possible $(i,j)$ pairs, we conclude that $[1,2]$ and
$[3,4]$ are the only two asymptotic line-solitons as $y\gg 0$.  

For $y\ll 0$, we now look for the least sum $\eta_i+\eta_j$
which implies the dominant exponential $E(i,j)$.
We continue to sweep clockwise from positive towards negative $x$-axis 
with the $c$-values still decreasing. Again, we start with $c = k_3+k_4$, then we have
the relation $\eta_1, \eta_2 > \eta_3=\eta_4$. But since $E(3,4)$   
is absent from the $\tau$-function, $E(2,3)$ and $E(2,4)$
provide the dominant balance leading to the line-soliton $[3,4]$.
Considering $c= k_2+k_4$, we have
$\eta_3<\eta_2=\eta_1, \eta_4$, which implies
that $E(2,3)$ is the only dominant exponential in the $\tau$-function,
hence $u \approx 0$ along the line $[2,4]$. Proceeding similarly as in the case for $y\gg 0$ 
leads to the conclusion that $[1,2]$ and
$[3,4]$ are the only two asymptotic line-solitons for $y\ll 0$.
\label{1234soliton}
\end{Example}
 From the above example, it is important to notice the fact that whether or not 
the dominant (or least) combinations inferred from (\ref{dominant})
are actually {\it present} in the given $\tau$-function ultimately decides
if a line $[i,j]$ corresponds to a line-soliton. In turn, this depends on the
coefficients of the exponential terms given by the maximal minors of
the $A$-matrix. Therefore, it suffices to specify only the $A$-matrix rather than the
whole $\tau$-function in order determine the asymptotic line-solitons. The next 
example illustrates this point.

\begin{Example}\label{3421soliton}
Consider the $2 \times 4$ matrix,
\begin{equation*}\label{ExA}
A=\begin{pmatrix}
1 & 0 & 0 &-\\
0 & 1 & + & +
\end{pmatrix}\,,
\end{equation*}
where some of the non-zero entries are indicated by their signs which
ensure that all non-zero maximal minors of $A$ are positive.
In this case, there are five nonzero minors,
\[
\xi(1,2),\quad \xi(1,3),\quad\xi(1,4),\quad\xi(2,4),\quad\xi(3,4), 
\]
and one missing $\xi(2,3)$.

Now look for the asymptotic solitons as $y\gg 0$. First we can immediately 
see that $[3,4]$-soliton is impossible since the dominant exponents along the 
line $[3,4]$ from (\ref{dominant}) is given by $\eta_1+\eta_2$. 
Since $\xi(1,2) \neq 0$, $\tau(x,y,t) \approx E(1,2)$ implying
that $u \approx 0$ along $[3,4]$. For the same reason, $[1,4]$ and
$[1,2]$-solitons are also impossible. Let us then check the $[2,4]$-soliton. 
From (\ref{dominant}), $\eta_3<\eta_2=\eta_4<\eta_1$, 
and since $\xi(1,2)\ne0, \, \xi(1,4)\ne0$, the $\tau$-function has the dominant phase
balance $\tau \approx E(1,2) + \xi(1,4)E(1,4)$ along $[2,4]$.
Therefore $[2,4]$ corresponds to an asymptotic line-soliton as $y \to \infty$.
Moreover, the $[1,3]$-soliton also exists for similar reasons.
Thus, we have two asymptotic line-solitons for $y\gg 0$.

We next look for the asymptotic solitons for $y\ll 0$.  
It is easy to see that $[1,2]$ and $[3,4]$-solitons are impossible.
Then consider the $[1,3]$-soliton. In this case, (\ref{dominant}) implies
that $\eta_2<\eta_1=\eta_3<\eta_4$.
Thus the two exponents $E(1,2)$ and $E(2,3)$ are dominant for $y\ll0$.
However, $\xi(2,3)=0$ implies that this is impossible. So $[1,3]$-soliton does not exist
as $y\ll 0$. For similar reasons, $[2,4]$-soliton is also impossible.
Now check $[1,4]$-soliton. In this case, we have 
$\eta_2, \eta_3<\eta_1=\eta_4$ from (\ref{dominant}) (see also
Figure \ref{fig:eta24}). Since $\xi(2,3)=0$, the dominant exponent $E(2,3)$ is {\it not} 
present in the $\tau$-function, but there does exist a dominant balance with
$E(1,3)$ and $E(3,4)$.  Thus there is a dominant
phase transition along the line $[1,4]$.  A similar argument applies for the
transition line $[2,3]$ which corresponds the other line-soliton as $y \ll 0$.
In Figure \ref{fig:3421},  we show the soliton solution given by this matrix.
%%%%%%%%%%%%%%%%%%%%%%%%%%%%%
\begin{figure}[t]
\centering
\includegraphics[scale=0.56]{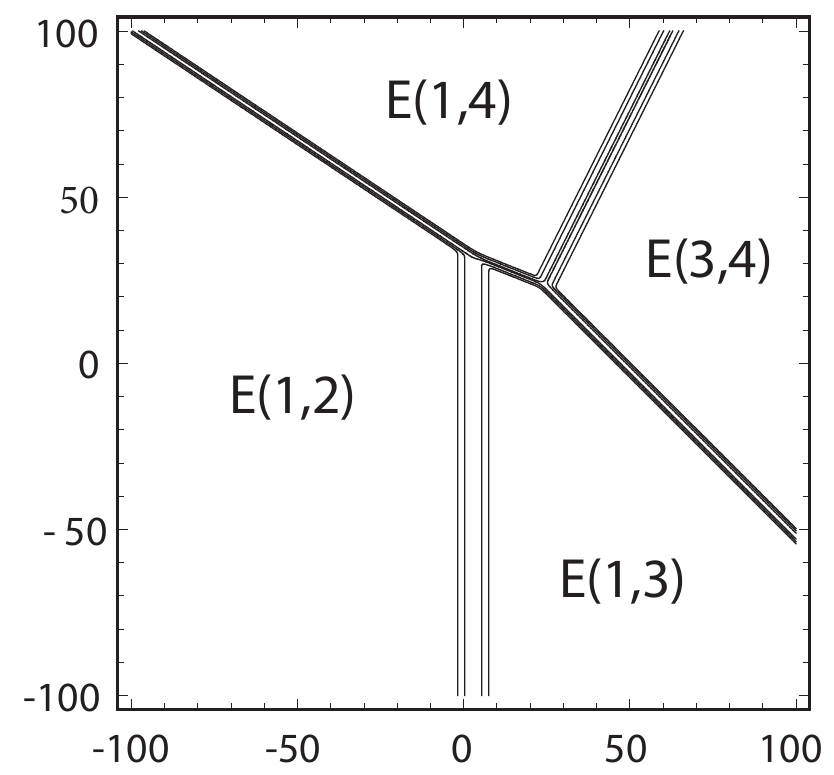}  \hskip 1.5cm
\includegraphics[scale=0.56]{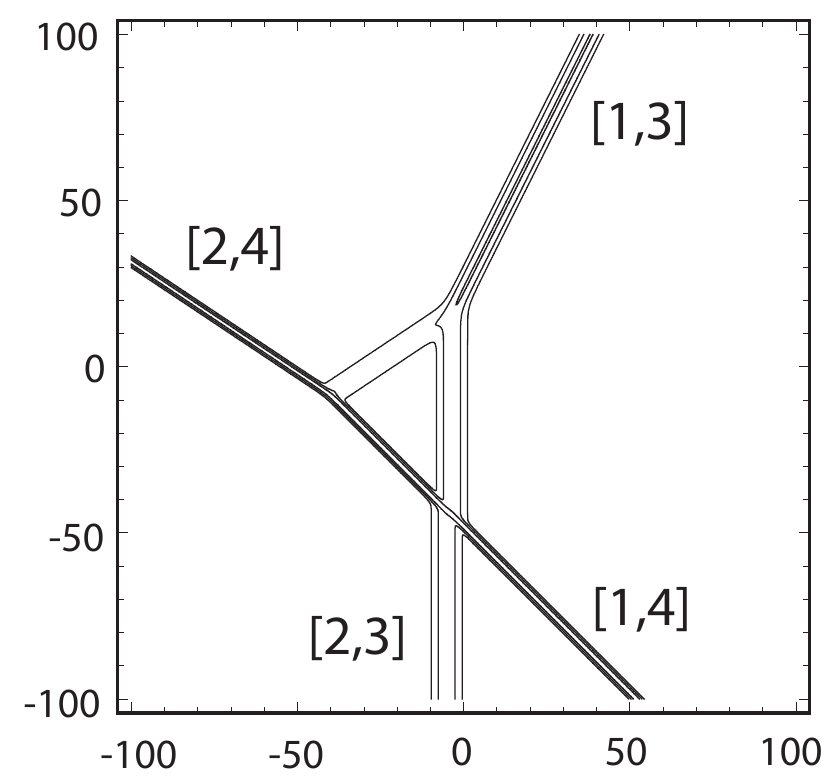} \\
\caption{A $(2,2)$-soliton solution. The left left figure is
at $t=-16$ and the right one at $t=16$. Each $E(i,j)$ for $1\le i<j\le 4$ in the left figure
shows the dominant exponential in the region, and
the boundaries of those regions give the line-solitons.
 The parameters $(k_1,\ldots,k_4)$ are
chosen as $(-1,-\sf{1}{2},\sf{1}{2},2)$.\label{fig:3421}}
\end{figure}
%%%%%%%%%%%%%%%%%%%%%%%%%%%%%%%%
\end{Example}

\subsection{Characterization of the line-solitons}
We saw in the Examples \ref{1234soliton} and \ref{3421soliton} that the $A$-matrix
plays a key role in the identification of the asymptotic line-solitons since
they determine if particular dominant phase combinations are in fact present in
the given $\tau$-function. Therefore, in order to obtain a complete characterization 
of the asymptotic line-solitons it is necessary to consider the structure
of the $N \times M$ coefficient matrix $A$. For the remainder of this article, we will 
consider the matrix~$A$ to be in RREF, and we will also assume that $A$ is 
{\it irreducible} as defined below:
\begin{Definition} An $N\times M$ matrix $A$ is irreducible if each 
column of $A$ contains at least one nonzero element, or each row contains at least
one nonzero element other than the pivot once $A$ is in RREF.
\end{Definition}
The reason for this assumption is the following: if an $N\times M$ matrix $A$ is 
{\it not} irreducible, then the solution $u=2\partial_x^2(\ln\tau)$ can be obtained from
a $\tau$-function with a matrix $\tilde{A}$ of smaller size than the original one
(that is, the size of $A$ is reducible). If a column of $A$ is identically zero,
then it is clear that we can re-express the functions $f_i$ in terms
of a $N \times (M-1)$ coefficient matrix $\tilde{A}$ obtained from $A$ by deleting its 
zero column. Or, suppose that an $N \times M$ matrix $A$ in RREF has a row whose 
elements are all zero except for the pivot, then it can be deduced from 
(\ref{tauexpansion}) that the corresponding $\tau$-function gives the same KP 
solution $u$ which can be obtained from a another $\tau$-function associated with a 
$(N-1) \times (M-1)$ matrix $\tilde{A}$.  For example, consider the matrix,
\begin{equation*}
A=\left(
\begin{array}{cccc}
1&a_{12}&a_{13}& 0\\
0&0&0&1
\end{array}\right), 
\end{equation*}
which  yields $f_1=E_1+a_{12}E_2+a_{13}E_3$ and $f_2=E_4$.
Then $\tau= \mathrm{Wr}(f_1,f_2)= E_4(k_4f_1-f_{1,x})$. Factoring out the
exponential $E_4$, we find that both $\tau$ and $\tilde{\tau}=k_4f_1-f_{1,x}$
give the same KP solution $u$. Furthermore, note that 
$\tilde{\tau}=\tilde{f} =(E_1,E_2,E_3)\tilde{A}^T$, where $\tilde{A}$ is the $1\times 3$ matrix,
\[
\tilde{A}=(\tilde{a}_{11},\tilde{a}_{12},\tilde{a}_{13})=(k_4-k_1,\,k_4-k_2a_{12},\,k_4-k_3a_{13})\,.
\]

\begin{figure}
\begin{center}
\includegraphics[height=5.8cm]{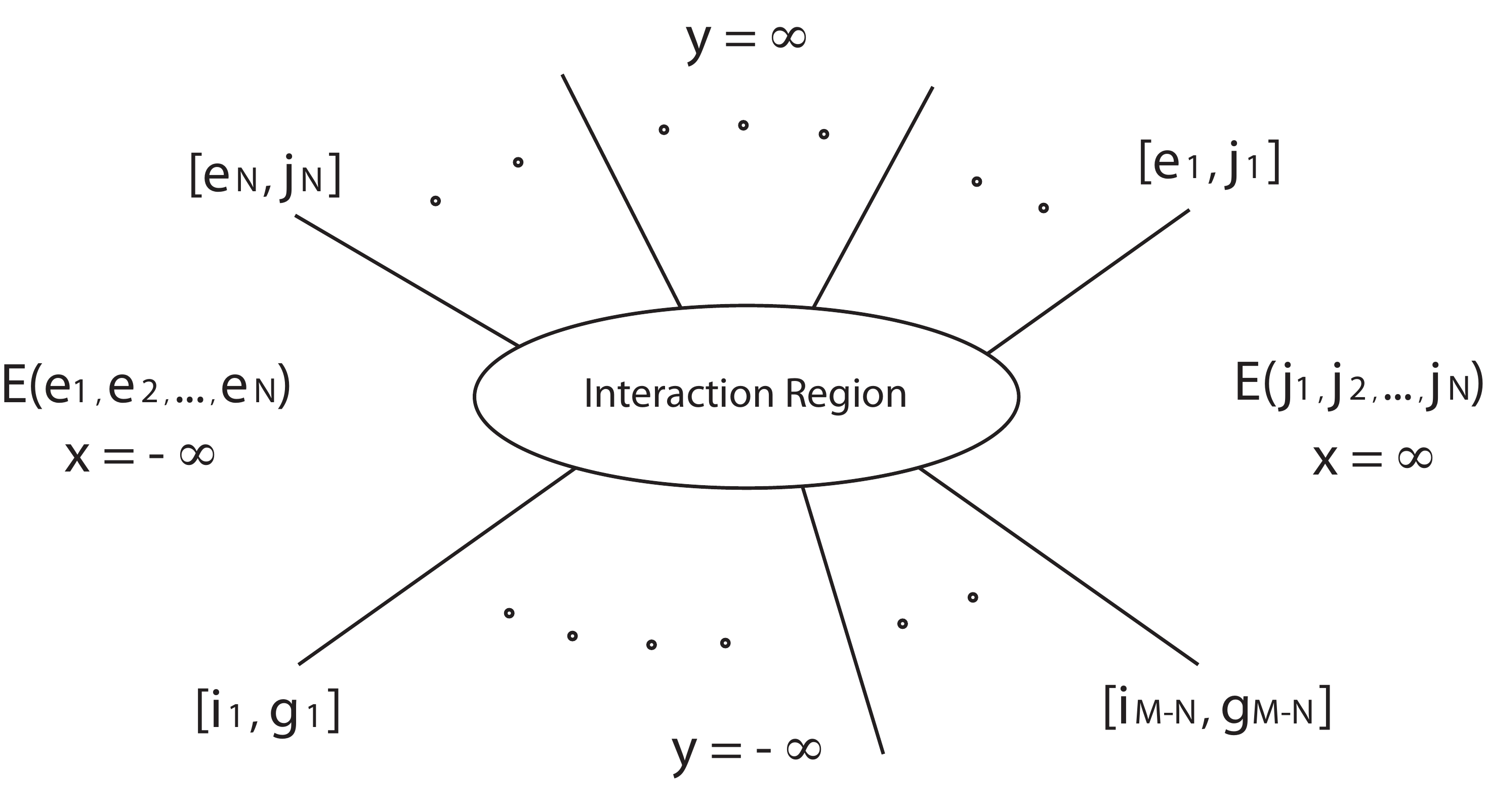}
\caption{$(N_-,N_+)$-soliton solution. The asymptotic line-solitons
are denoted by their index pairs $[e_n,j_n]$ and $[i_m,g_m]$. The sets $\{e_1,e_2,\ldots,e_{N}\}$ and
$\{g_1,g_2,\ldots,g_{M-N}\}$ indicate pivot and non-pivot indices, respectively.
Here $N_-=M-N$ and $N_+=N$ for the $\tau$-function on Gr$(N,M)$, and
$E(\cdot,\ldots,\cdot)$ represents the dominant exponential in that region.\label{NMsoliton}}
\end{center}
\end{figure}

\begin{figure}
\begin{center}
\includegraphics[height=7cm]{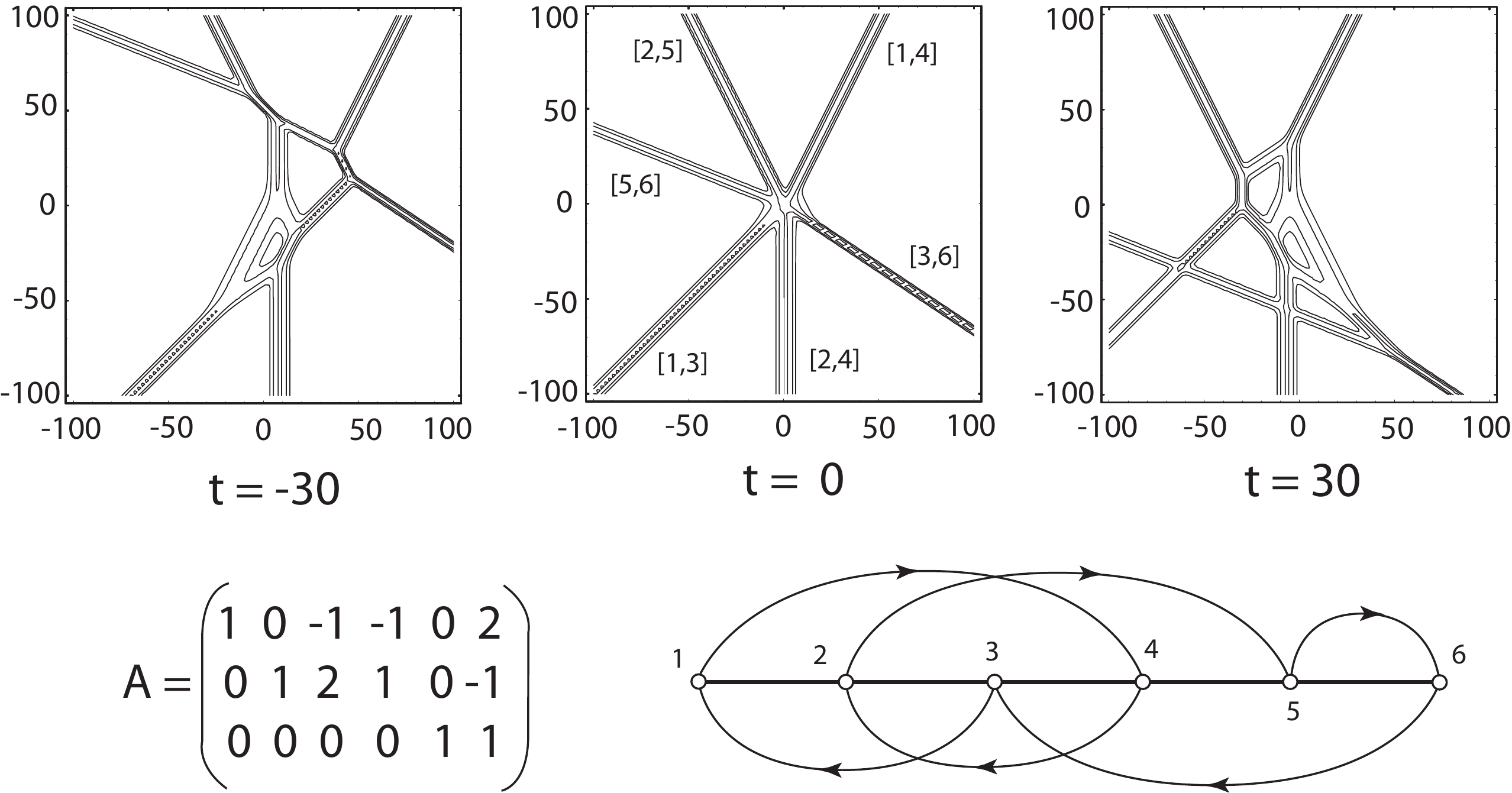}
\caption{The time evolution of a $(3,3)$-soliton solution. The permutation of this solution
is $\pi=(451263)$. The $k$-parameters are chosen as $(k_1,k_2,\ldots,k_6)=(-1,-\sf{1}{2},0,\sf{1}{2},1,\sf{3}{2})$. The dominant exponential for $x\ll 0$ is $E(1,2,5)$, and each dominant exponential
is obtained through the derangement representing the solution, e.g. after crossing $[5,6]$-soliton
in the clockwise direction,
the dominant exponential becomes $E(1,2,6)$. That is, $[5,6]$-soliton is given by
the balance of those two exponentials. \label{fig:33soliton}}
\end{center}
\end{figure}

We now present a classification scheme of the line-soliton solutions
by identifying the asymptotic line-solitons as $y \to\pm \infty$. We denote a
line-soliton solution by $(N_-,N_+)$-soliton whose asymptotic form
consists of $N_-$ line-solitons as $y\to-\infty$ and $N_+$ line-solitons 
for $y\to\infty$ in the $xy$-plane as shown in Figure \ref{NMsoliton}.
%%%%%%%%%%%%%%%%%%%%%%%%%%%
%%%%%%%%%%%%%%%%%%%%%%%%%%%%%%%%%
The following Proposition provides a general result characterizing 
the asymptotic line-solitons of
the $(N_-,N_+)$-soliton solutions.
\begin{Proposition}
Let $\{e_1,e_2,\ldots,e_N\}$ be the pivot indices, and let $\{g_1,g_2,\ldots,g_{M-N}\}$ be 
the non-pivot indices of an irreducible and totally non-negative $N\times M$ matrix $A$. 
Then the soliton 
solution generated by the $\tau$-function in (\ref{tauexp}) with the matrix $A$ has
\begin{itemize}
\item[(a)] $N$ asymptotic line-solitons as $y\to\infty$, each 
defined uniquely by the line $[e_n, j_n]$ for some $j_n$, and
\item[(b)] $M-N$ asymptotic line-solitons as $y\to-\infty$, each
defined uniquely by the line $[i_m, g_m]$ for some $i_m$.
\end{itemize}
\label{engn}
\end{Proposition}
\begin{Proof}
Let $[i,j]$ denote an asymptotic line-soliton for $y \gg 0$, and
let it correspond to the dominant exponentials $E(i,m_2,\ldots,m_N)$ and 
$E(j,m_2,\ldots,m_N)$ in (\ref{tauexp}). Hence in particular, the coefficient
minor $\xi(i,m_2,\ldots,m_N) \neq 0$. Assume that $A_i$ is {\it not} a pivot column.
Then $A_i$ must be spanned by the pivot columns of $A$ to its left since $A$ is 
in RREF. That is, $A_i = \sum_{r=1}^nc_rA_{e_r}$, where $e_n < i$. Therefore,
$$ \xi(i,m_2,\ldots,m_N) = \sum_{r=1}^nc_r\xi(e_r, m_2,\ldots,m_N) \neq 0\,.$$
Suppose $c_s$ is the coefficient of the first nonzero term in the above sum, then the
minor $\xi(e_s, m_2,\ldots,m_N) \neq 0$, implying that the corresponding exponential
$E(e_s, m_2,\ldots,m_N)$ is {\it present} in the $\tau$-function.
Furthermore, from (\ref{dominant}) $\eta_{e_s} > \eta_i$ along $[i,j]$, so that
$E(e_s, m_2,\ldots,m_N) > E(i,m_2,\ldots,m_N)$. But this is impossible
as $E(i,m_2,\ldots,m_N)$ is the dominant exponential in the $\tau$-function along $[i,j]$.
Therefore, $A_i$ must be a pivot column, proving part (a) of the Theorem.

Now suppose $[i,j]$ is an asymptotic line-soliton for $y \ll 0$ and
that it corresponds to the dominant exponentials $E(i,m_2,\ldots,m_N)$ 
and $E(j,m_2,\ldots,m_N)$ in (\ref{tauexp}). Then the nonzero minor 
$\xi(i,m_2,\ldots,m_N) \neq 0$ which implies that the column set 
$B = \{A_i, A_{m_2}, \ldots, A_{m_N}\}$ forms a basis for $\mathbb{R}^N$. 
In particular, $B$ spans the column $A_j$, i.e., $A_j = cA_i + \sum_{r=2}^Nc_rA_{m_r}$.
Then,
$$ \xi(i,m_2,\ldots,m_{r-1},j,m_{r+1}, \ldots,m_N) = 
c_r\xi(i, m_2,\ldots,m_N)\,, \qquad r=2,\ldots,N \,.$$
If $A_j$ is a pivot column, then there is at least one index $s$  such that $m_s>j$.
Then it follows from (\ref{dominant}) that 
$\eta_{m_s} > \eta_j$ along the line $[i,j]$, and for $y \ll 0$, we have
$ E(i,m_2,\ldots,m_{s-1},j,m_{s+1}, \ldots,m_N) \gg E(i,m_2,\ldots,m_N)$.
But this contradicts with $E(i,m_2,\ldots,m_N)$ being dominant.

To prove uniqueness of each index pair, first consider $y>0$, and assume that 
there are {\it two} line-solitons $[e_n, j_n]$ and $[e_n, j'_n]$ with a given 
pivot index $e_n$, and where $j_n > j'_n$.
These line-solitons are localized along the lines $x+(k_{e_n}+k_{j_n})y =$ constant
and $x+(k_{e_n}+k_{j'_n})y = $ constant with $k_{e_n}+k_{j_n} > k_{e_n}+k_{j'_n}$.
Notice from Proposition \ref{ij} that the phase $\theta_{e_n}$
is replaced by the phase $\theta_{j_n}$ in the exponential term which
is dominant immediately to the right of the $[e_n, j_n]$-soliton.
On the other hand, $\theta_{e_n}$ must be present in the exponential term which
is dominant immediately to the left of the $[e_n, j'_n]$-soliton.
Consequently, there must be an intermediate line-soliton $[e_m, e_n]$
for some pivot index $e_m < e_n$, localized along the line $x+(k_{e_m}+k_{e_n})y =$ constant,
which must satisfy $k_{e_m}+k_{e_n} > k_{e_n}+k_{j'_n}$.
But the latter inequality is impossible due to ordering of the indices $e_m < e_n < j'_n$
and the ordering $k_r < k_s$ for $r<s$, giving the required contradiction.
Hence for a given pivot index $e_n$ there is a unique line-soliton $[e_n, j_n]$ as 
$y \to \infty$. The uniqueness proof for the $[i_m, g_m]$-solitons for $m=1,\ldots, M-N$,
and as $y \to -\infty$, is similar.
\end{Proof}
The unique index pairings in Proposition \ref{engn} have a combinatorial interpretation.
Let $[M] := \{1,2,\ldots,M\}$ be the integer set and recall that
$\{e_1,\ldots,e_N\} \cup \{g_1,\ldots,g_{M-N}\}$ is a disjoint partition
of $[M]$. Define the pairing map $\pi: [M] \to [M]$ according to parts (a) and (b) 
of Proposition \ref{engn} as follows:
\begin{equation}
\left\{\begin{array}{lll}
\pi(e_n) &= j_n\,,\quad & n=1,2,\ldots,N\,, \\[1.0ex]
\pi(g_m) &= i_m\,, \quad &m=1,2,\ldots,M-N\,,
\end{array}\right.
\label{pi}
\end{equation}
where $e_n$ and $g_m$ are respectively, the pivot and non-pivot indices of
the $A$-matrix. Then 
as a consequence of Proposition \ref{engn} we have the following.
\begin{Corollary}
The map $\pi: [M] \to [M]$ is a bijection. That is, $\pi \in \mathcal{S}_M$, where
$\mathcal{S}_M$ is the group of permutations for the index set $[M]$.
\label{pairing}
\end{Corollary}
\begin{Proof}
We prove this by finding an explicit relation among those parameters $\{k_{i_m},k_{j_n}\}$ and
$\{k_{e_n},k_{g_m}\}$:
For this purpose, we consider the integrals $C_n$ in (\ref{Integral}) for the $(M-N,N)$-soliton 
solutions obtained in Proposition \ref{engn}. Here we recall that
\[
C_n=\int_{-\infty}^{\infty}v_{n+1}(x,y,\cdots)\,dx, \qquad n=1,2,\ldots,M,
\]
where $v_{n+1}=-\partial_xS_n(-\tilde{\partial})\ln\tau$ (see (\ref{CDtau})).
Note that for $|y|\gg0$ the $\tau$-function is asymptotically equal to a 
single exponential function in each of the $M$ asymptotic
sectors in the $xy$-plane (see Figure \ref{NMsoliton}).
That is, in each asymptotic sector the $\tau$-function has the form
$$\tau ~\approx ~\xi(m_1,\ldots,m_N)E(m_1,\ldots,m_N)\,,$$
and all other terms are exponentially small in comparison with this dominant exponential. Then
we have
$$ \ln \tau~ \approx  ~\sum_{j=1}^N\left(\sum_{n=1}^\infty k_{m_j}^nt_n\right) + c\,,$$
where $c$ is a constant determined by $\xi(m_1,\ldots,m_N)$ and the $k$-parameters.
We now calculate $C_n$ using Proposition \ref{engn} for $y\gg0$ and $y\ll 0$.

For $y\gg0$, there are $N+1$ asymptotic regions, and the boundary of adjacent regions
is given by the line $[e_n,j_n]$ for $n=1,\ldots,N$. Recall from Proposition
\ref{ij} that the dominant exponential terms in adjacent regions differ by only 
one phase. We now
 integrate $v_{n+1}$ along a horizontal line which passes through
those asymptotic regions, by using the above expression for $\ln \tau$
and the fact that $S_n(-\tilde{\partial})= -\sf{1}{n}\partial_{t_n} + (\rm{h.o.d.})$.
Since the contribution of the exponentially small terms vanish as $y \to \infty$,
we then obtain
\[
C_n =  \frac{1}{n}\sum_{r=1}^N\left(k_{j_r}^n-k_{e_r}^n\right) \,, \qquad
n = 1,2,\ldots\,,
\]
where each term in the sum is the contribution to the integral from the 
pair of dominant exponentials in adjacent asymptotic regions for $y\gg0$.  

For $y\ll0$, we follow the similar arguments as above and integrate $v_{n+1}$ along
a line across the $M-N+1$ asymptotic regions separated by the lines $[i_m,g_m]$ for
$m=1,\ldots,M-N$. We then obtain another
expression for the integral $C_n$, namely
\[
C_n = \frac{1}{n}\sum_{s=1}^{M-N}\left(k_{g_s}^n-k_{i_s}^n\right) \,, \qquad
n=1,2,\ldots\,.
\]

Since $C_n$ do not depend on $y$, the two expressions obtained for $y\gg0$ and
$y\ll 0$ must be the same, that is,  we obtain the relations,
$$
C_n=\frac{1}{n}\sum_{r=1}^N\left(k_{j_r}^n-k_{e_r}^n\right) =
\frac{1}{n}\sum_{s=1}^{M-N}\left(k_{g_s}^n-k_{i_s}^n\right)\,.
$$
Rearranging the terms, we have for $n=1,2,\ldots,M$
\begin{align*}
\sum_{j=1}^Mk_j^n&=\sum_{r=1}^Nk_{e_r}^n+\sum_{s=1}^{M-N}k_{g_s}^n\\
&=\sum_{r=1}^Nk_{j_r}^n+\sum_{s=1}^{M-N}k_{i_s}^n.
\end{align*} 
Then Lemma \ref{uniqueness} implies that there is a one-to-one correspondence between
two sets $\{k_{e_r},k_{g_s}\}$ and $\{k_{j_r},k_{i_s}\}$. Furthermore, since 
$\{k_{e_r},k_{g_s}\}$ is a set of distinct elements, then so is $\{k_{j_r},k_{i_s}\}$.
This proves that $\pi$ is a bijection.
\end{Proof}
Note that the permutation $\pi$ defined by (\ref{pi}) has no fixed point because 
$\pi(e_n)=j_n > e_n,\, n=1,\ldots,N$ and $\pi(g_m) = i_m < g_m, \, m=1,\ldots,M-N$. 
Such permutations are called {\em derangements}. Moreover, $\pi$ have exactly $N$
{\it excedances} defined as follows: an element $l \in [M]$ is an
{\em excedance} of $\pi$ if $\pi(l) > l$. The excedance set of $\pi$ in (\ref{pi})
is the set of pivot indices $\{e_1, e_2, \ldots, e_N\}$. We can now summarize the 
results of Proposition \ref{engn} and Corollary \ref{pairing} as follows:
\begin{Theorem}
\label{derangement}
Let $A$ be an $N\times M$ irreducible matrix which gives a coordinate of
a non-negative cell of the Grassmannian Gr$(N,M)$. Then the
 $tau$-function (\ref{tauexp}) associated with the $A$-matrix 
generates an $(M-N,N)$-soliton solutions. The total $M$ asymptotic line-solitons
associated with each of these solutions induce a pairing map $\pi$ defined by
(\ref{pi}). Moreover, $\pi$ is a derangement of the index set $[M]$ with
$N$ excedances given by the pivot indices $\{e_1, e_2, \ldots, e_N\}$ of the $A$-matrix
in RREF.
\end{Theorem}
The derangements $\pi \in \mathcal{S}_M$ are represented by linear chord diagrams
with the arrows above the line pointing from $e_n$ to $j_n$ 
for $n=1,2,\ldots,N$, while arrows below the line point from $g_m$ to $i_m$ for 
$m=1,2,\ldots,M-N$. Figure \ref{fig:33soliton} illustrates the time evolution of a $(3,3)$-soliton
solution. The chord diagram shows all asymptotic line-solitons for $y\to\pm\infty$.
%%%%%%%%%%%%%%%%%%%%%%%%%%%
%%%%%%%%%%%%%%%%%%%%%%%%%%%%%%%%%

\section{$(2,2)$-soliton solutions}\label{C:2-2soliton}
\begin{figure}[t]
\centering
\includegraphics[scale=0.53]{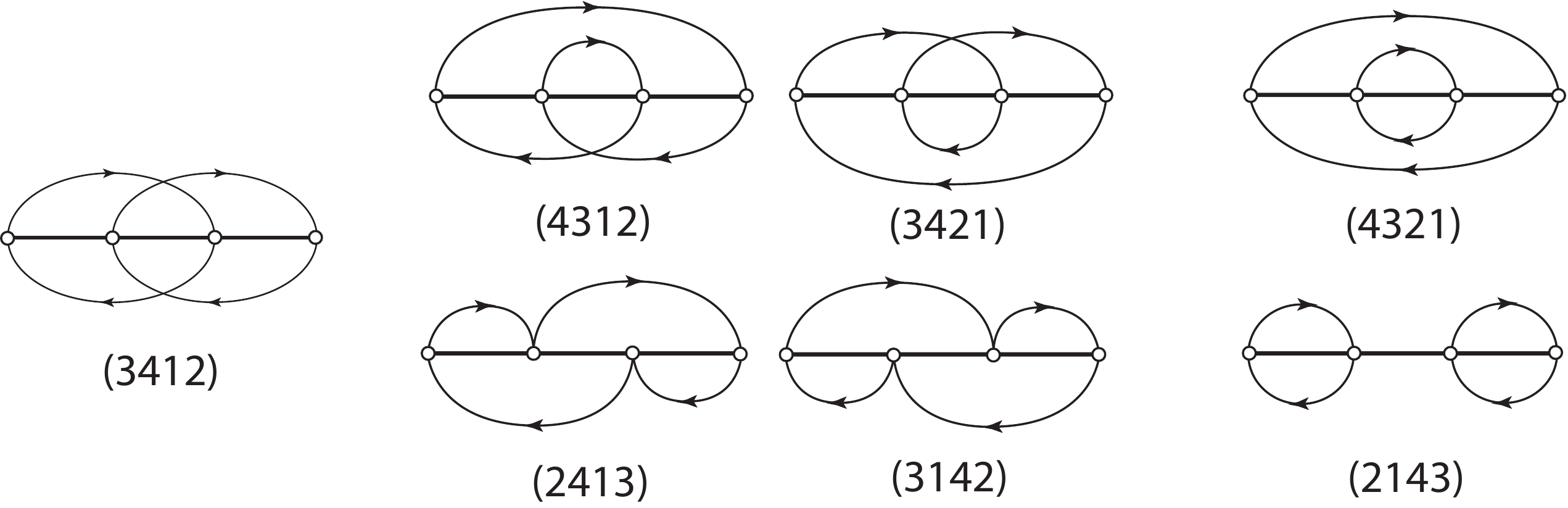} 
\caption{The chord diagrams for seven different types
of $(2,2)$-soliton solutions. Each diagram corresponds to a totally non-negative Grassmannian
cell in Gr$(2,4)$.
\label{fig:chords}}
\end{figure}

In this section we discuss all soliton solutions of the KP equation generated by the $2\times 4$ 
irreducible $A$-matrices with nonnegative minors. Theorem \ref{derangement} implies that each
of the soliton solutions consists of two asymptotic line-solitons as $y\to\pm\infty$, that is, we have
$(2,2)$-soliton solutions.  We outline below the classification scheme for the $(2,2)$-soliton solutions.
First we note that
there are only two types of irreducible matrices for the size $2\times 4$, i.e.
\[
\begin{pmatrix}
1 & 0 & -c & -d \\
0& 1 & a & b 
\end{pmatrix}\qquad {\rm and}\qquad 
\begin{pmatrix}
1& a & 0 & -c\\
0 & 0 &1 & b
\end{pmatrix}\,,
\]
The condition for nonnegative minors implies that the constants $a,b,c$ and $d$ must 
be non-negative. In the first case, one can easily see that $ad=0$ is impossible because
then $\xi(3,4)<0$ from the irreducibility. Then there are 5 cases for nonnegative minors with $ad\ne 0$, which are
(1) $ad-bc>0$, (2) $ad-bc=0$, (3) $b=0, c\ne0$, (4) $c=0, b\ne 0$, and (5) $b=c=0$.
For the second case, since $ab\ne 0$ due to irreducibility, we have only two cases:\,
(1) $c=0$, (2) $c\ne 0$.
Thus we have total seven different types of $A$-matrices, and each $A$-matrix gives 
a different $(2,2)$-soliton solution. 
We now summarize the results for all seven cases with $2\times 4$ irreducible and
totally non-negative $A$-matrices:

\begin{itemize}
\item[(a)] $\pi=(3412)$: This case gives {\it T-type} 2-soliton solution, and the asymptotic line-solitons are
$[1,3]$- and $ [2,4] $-types for $|y|\to\infty$. This type of solution was first obtained as the solution
of the {\it Toda} lattice hierarchy \cite{BK:03}, and this is why we call it ``T-type" (see also \cite{K:04}).
The $A$-matrix is given by
\[
A=\begin{pmatrix} 1&0&-c&-d\\0&1&a&b\end{pmatrix}\,,
\]
where $a,b,c,d>0$ are free parameters with $ad-bc>0$. This is the generic solution
on the maximum dimensional cell (four free parameters) of Gr$(2,4)$, and
contains all possible line-solitons in the process of interaction, i.e. $[i,j]$-solitons for
any $\{i,j\}$ pairs (see the next section for details).
\item[(b)] $\pi=(4312)$: The asymptotic line-solitons are given by
\begin{itemize}
\item[(i)] $[1,4]$- and $[2,3]$-solitons in $y\gg0$
\item[(ii)] $[1,3]$- and $[2,4]$-solitons in $y\ll0$.
\end{itemize}
The $A$-matrix is given by
\[
A=\begin{pmatrix} 1&0&-b&-c\\0&1&a&0\end{pmatrix}\,,
\]
where $a,b,c>0$ are free parameters. Note that two line-solitons for $y\ll 0$
are the same types as in the T-type solitons.  We also observe all possible
line-solitons including intermediate ones, but not for all $t$ unlike the T-type.
\item[(c)] $\pi=(3421)$: The asymptotic line-solitons are given by
\begin{itemize}
\item[(i)] $[1,3]$- and $[2,4]$-solitons in $y\gg 0$
\item[(ii)] $[1,4]$- and $[2,3]$-solitons in $y\ll 0$.
\end{itemize}
The $A$-matrix is given by
\[
A=\begin{pmatrix} 1&0&0&-c\\0&1&a&b\end{pmatrix}\,,
\]
where $a,b,c>0$ are positive free parameters.
This solution can be considered as a dual of the previous case (b), that is,
two sets of line-solitons for $y\gg 0$ and $y\ll 0$ are exchanged.
The example discussed in Example \ref{3421soliton} corresponds to this solution (see Figure
\ref{fig:3421}).

\item[(d)] $\pi=(2413)$: The asymptotic line-solitons are given by
\begin{itemize}
\item[(i)] $[1,2]$- and $[2,4]$-solitons in $y\gg 0$
\item[(ii)] $[1,3]$- and $[3,4]$-solitons in $y\ll0$.
\end{itemize}
The $A$-matrix is given by
\[
A=\begin{pmatrix} 1&0&-c&-d\\0&1&a&b\end{pmatrix}\,,
\]
where $a,b,c,d>0$ with $\xi(3,4)=ad-bc=0$. 

\item[(e)] $\pi=(3142)$: The asymptotic line-solitons are given by
\begin{itemize}
\item[(i)] $[1,3]$- and $[3,4]$-solitons in $y\gg 0$
\item[(ii)] $[1,2]$- and $[2,4]$-solitons in $y\ll 0$.
\end{itemize}
The $A$-matrix is given by
\[
A=\begin{pmatrix} 1&a&0&-c\\0&0&1&b\end{pmatrix}\,,
\]
where $a,b,c>0$. This solution is dual to the previous one (d) in the sense that the missing
minors are switched by $\xi(3,4)\leftrightarrow \xi(1,2)$. This case will be further discussed in the next subsection to
describe  a connection with the Mach reflection in shallow water waves.

\item[(f)] $\pi=(4321)$: This case gives  {\it P-type} 2-soliton solution, and the asymptotic line-solitons are
$[1,4]$- and $ [2,3] $-types for $|y|\to\infty$. This type of solutions fits better with the {\it physical}
assumption for the derivation of the KP equation, i.e. a quasi-two dimensionality with weak $y$-dependence.
This is why we call it ``P-type" (see \cite{K:04}). The situation is similar to the KdV case, for example,
two solitons must have different amplitudes, $A_{[1,4]}>A_{[2,3]}$.
The $A$-matrix is given by
\[
A=\begin{pmatrix} 1&0&0&-b\\0&1&a&0\end{pmatrix}\,.
\]
\item[(g)] $\pi=(2143)$: This case gives {\it O-type} 2-soliton solution, and the asymptotic line-solitons are
$[1,2]$- and $ [3,4] $-types for $|y|\to\infty$. The letter ``O" for this type is due to the fact that this solution
was  {\it originally} found to describe the two-soliton solution, see for example \cite{FN:83}, (see also \cite{K:04}). Interaction properties for solitons with equal amplitude has been discussed using this solution. However this solution becomes singular,
when those solitons are almost parallel to the $y$-axis; in this case, the order of the $k$-parameters
is reversed, that is, $k_2>k_3$ leading to $E(2,3)<0$).  This is contrary to the assumption of the 
quasi-two dimensionality for the KP equation \cite{M:77}. We will discuss this issue in the next section.
The $A$-matrix is given by
\[
A=\begin{pmatrix} 1&a&0&0\\0&0&1&b\end{pmatrix}\,.
\]
\end{itemize}
 In Figure \ref{fig:chords}, we list the chord diagrams for all those seven cases. One should note that
any derangement of $\mathcal{S}_4$ with exactly two excedances should be one of the graphs. 
This uniqueness in the general case has been used to count the number of totally non-negative
Grassmann cells \cite{P:06, W:05}. We also mention that these seven types of soliton solutions
have been found by applying recursive binary Darboux transformations 
(see Appendix in \cite{BPPP:01}).
%%%%%%%%%%%%%%%%%%%%%%%%%%%%%%%%%%%%%%%%%%
%%%%%%%%%%%%%%%%%%%%%%%%%%%%%%%%%%%

Now let us describe the details of some of the $(2,2)$-soliton solutions, which will be important for
an application of those solutions to shallow water problem discussed in the next Section.
In particular, we explain how the $A$-matrix uniquely determines 
the structure of the corresponding soliton solution such as the location of the 
solitons and their phase shifts.

\subsection{O-type soliton solutions}
\begin{figure}[t]
\centering
\includegraphics[scale=0.48]{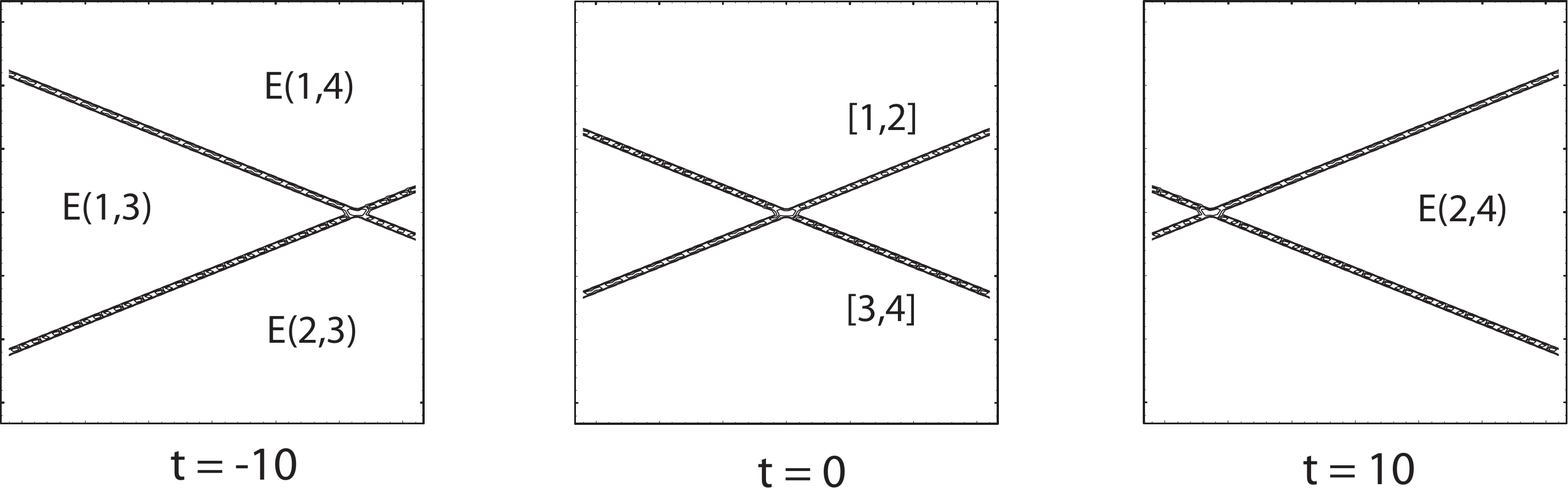} 
\caption{The time evolution of an O-type soliton solution. Each $E(i,j)$ indicates the dominant exponential in that region. The parameters are chosen as
$(k_1,k_2,k_3,k_4)=(-\frac{9}{4},-\frac{1}{4},\frac{1}{4},\frac{9}{4})$, so that both line-solitons have the same 
amplitude, $A_{[1,2]}=A_{[3,4]}=2$, and the directions of the wave-vectors are given by
$\Psi_{[3,4]}=-\Psi_{[1,2]}=\tan^{-1}\frac{5}{2}\approx 68.2^{\circ}$.
\label{fig:V4}}
\end{figure}

This is the original two-soliton solution, and  
the solutions correspond to the chord diagram of $\pi=(2143)$. A solution of this 
type consists of two full line-solitons of $[1,2]$ and $[3,4]$ (see Figure \ref{fig:V4}). 
Note here that they have phase shifts due to their collision.
Let us describe explicitly the structure of the solution of this type:
The $\tau$-function defined in (\ref{tauexp}) for this case is given by
\[
\tau=E(1,3)+bE(1,4)+aE(2,3)+abE(2,4)\,,
\]
where $a,b>0$ are the free parameters given in the $A$-matrix listed in the previous section.
As we will show that those two parameters can be used to fix the locations of 
those solitons, that is, they are determined
by the asymptotic data of the solution for large $|y|$.

For the later application of the solution, we assume that $[1,2]$-soliton has a ``positive'' $y$-component in the wave-vector (i.e. $\tan\Psi_{[1,2]}<0$), and $[3,4]$-soliton has a ``negative'' $y$-component, (i.e. $\tan\Psi_{[3,4]}>0$,
see Figure \cite{fig:1soliton}).
(Recall that any line-soliton has a ``negative'' $x$-component in the wave-vector.)
Then for the region with large positive $x$, we have $[1,2]$-soliton 
in $y>0$ and $[3,4]$-soliton in $y<0$. Those solitons are obtained by the balance 
between two exponential terms in the $\tau$-function: 

For $[1,2]$-soliton in $x>0$ (and $y\gg 0$), we have the dominant balance between $E(1,4)$ and $E(2,4)$.
Then the $\tau$-function can be written in the following form,
\begin{align*}
\tau&\approx bE(1,4)+abE(2,4)\\
&=2be^{\theta_4+\sf{1}{2}(\theta_1+\theta_2)}
\cosh\frac{1}{2}\left(\theta_1-\theta_2+\theta_{12}^{\pm}\right)\,,
\end{align*}
which leads to the $[1,2]$-soliton solution in the region near $\theta_1\approx \theta_2$ for large $x>0$,
\[
u=2{\partial_x^2}\ln\tau\approx 
\frac{1}{2}(k_2-k_1)^2\sech^2\frac{1}{2}\left(\theta_1-\theta_2+\theta_{12}^+\right).
\]
Here the phase shift $\theta_{12}^+$ ($+$ indicates $x>0$) is given by
\begin{equation}\label{Oshifta}
\theta_{12}^+=
\ln\frac{k_4-k_1}{k_4-k_2}-\ln a \qquad {\rm i.e.}\qquad a=\frac{k_4-k_1}{k_4-k_2}e^{-\theta_{12}^+}.
\end{equation}
The parameter $a$ determines the location of the $[1,2]$-soliton. For example, one can choose $a$ so that $\theta_{12}^+=0$, which
implies that the $[1,2]$-soliton is passing through the origin at $t=0$.

%%%%%%%%%%%%%%%%%%%%%%%%%%%%%%%%%%%%
%%%%%%%%%%%%%%%%%%%%%%%%%%%%%%%%%%%%%%%%

For $[3,4]$-soliton in $x>0$ (and $y\ll 0$), from the balance $\tau\approx aE(2,3)+abE(2,4)$, we have
\begin{equation*}
u\approx \frac{1}{2}(k_4-k_3)^2{\rm sech}^2\frac{1}{2}(\theta_3-\theta_4+\theta_{34}^+),
\end{equation*}
with the phase shift,
\begin{equation}\label{Oshiftb}
\theta_{34}^+=\ln\frac{k_3-k_2}{k_4-k_2}-\ln b \qquad {\rm i.e.}\qquad 
b=\frac{k_3-k_2}{k_4-k_2}e^{-\theta_{34}^+}.
\end{equation}
Thus the parameter $b$ determines the location of $[3,4]$-soliton, and
 one can choose appropriate $b$ so that $\theta_{34}^+=0$. Thus the parameters in the $A$-matrix
can be determined by the asymptotic data of the phase shifts $\theta_{12}^+$ and $\theta_{34}^+$ 
for $x\gg 0$ and $|y|\gg 0$.

Now calculating the phase shift for the $[3,4]$-soliton in $x<0$, one can find the total phase shift
along the line-soliton:
   From the balance between the terms with $E(1,3)$ and $bE(1,4)$,
and following the previous arguments,
we obtain
\[
\theta_{34}^-=\ln\frac{k_3-k_1}{k_4-k_1}-\ln b.
\]
It is interesting to note that the sum of the phase shifts for those solitons are conserved
in the $y$-direction for each $A$-matrix,
\begin{equation}\label{phaseconservation}
\theta_{12}^++\theta_{34}^-=\theta_{34}^++\theta_{12}^-=\ln\frac{k_3-k_1}{k_4-k_2}-\ln(ab).
\end{equation}

\begin{figure}[t!]
\centering
\includegraphics[scale=0.4]{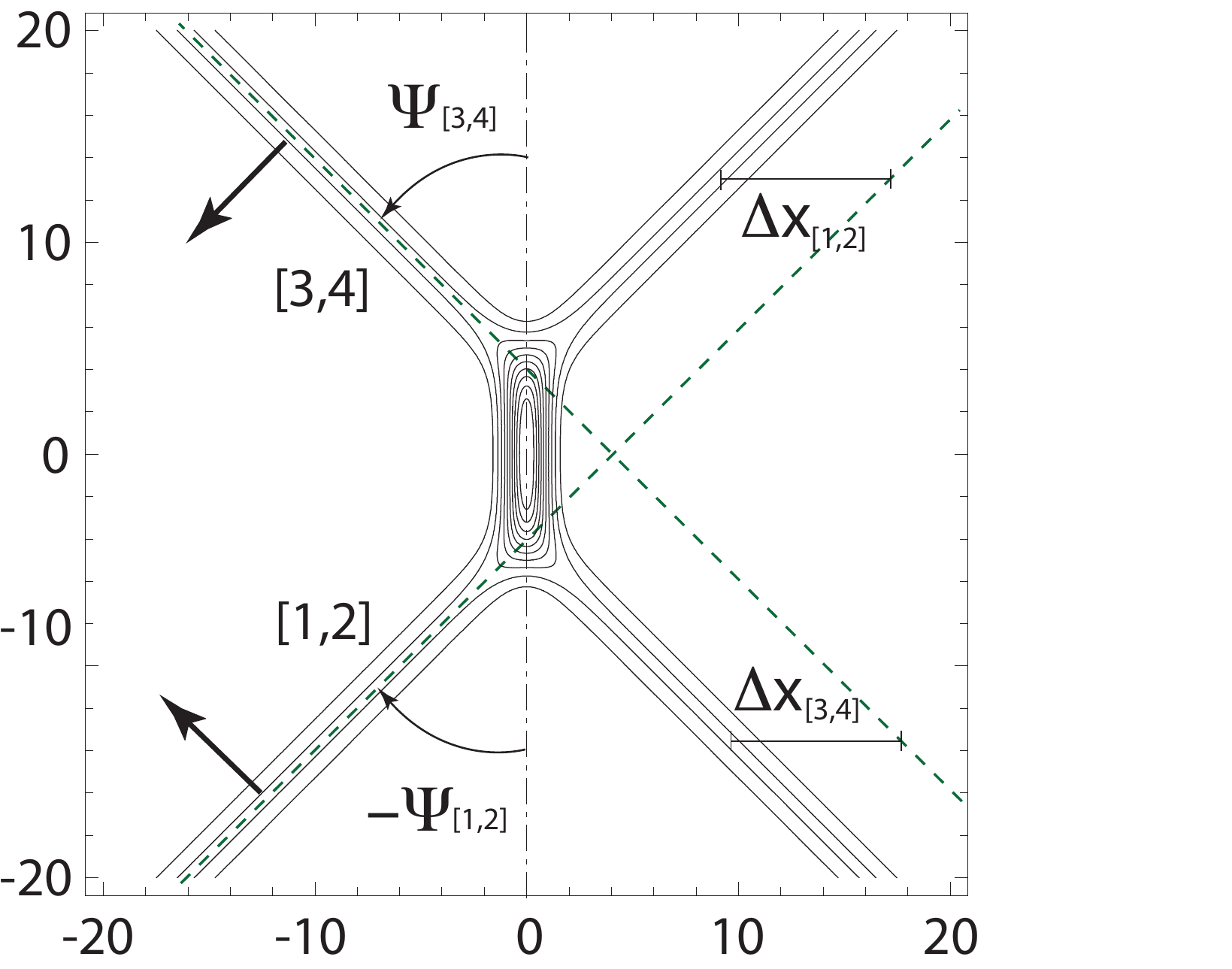} 
\caption{O-type interaction for two equal amplitude solitons.
The parameters $k_i$'s are taken by $(k_1,k_2,k_3,k_4)=(-1-10^{-4},-10^{-4},10^{-4},1+10^{-4})$,
which give $A_0=A_{[1,2]}=A_{[3,4]}=\frac{1}{2}$ and $\Psi_0=\tan\Psi_{[3,4]}=-\tan\Psi_{[1,2]}=1+2\times10^{-4}$ (i.e. $\Psi_{[3,4]}\approx
45.0057$).
The constants $a,b$ in the $A$-matrix are chosen so that the center of interaction
point is located at the origin (see (\ref{centerX})), and $u_{\rm max}=u(0,0,0)\approx 1.96$ (about four times
larger than each soliton amplitude $A_0$).\label{fig:Otype}}
\end{figure}

We measure the total phase shift for each soliton in the reference of the soliton in $x\ll0$,
that is, the total phase shift is defined by $\theta_{34}=\theta^+_{34}-\theta^-_{34}$ (recall that 
the phase shift is generated by interacting with other soliton, 
and the effect of the interaction propagates only in the ``positive'' $x$-direction).
We then obtain the well-known formula for the phase shift $\theta_{34}$ (see for example
\cite{H:04}),
\[
\theta_{34}=\ln\frac{(k_4-k_1)(k_3-k_2)}{(k_3-k_1)(k_4-k_2)}\,.
\]
Notice that this does not depend on the $A$-matrix. One can also see from 
(\ref{phaseconservation}) that
the total phase shift of $[1,2]$-soliton is the same as that of the $[3,4]$-soliton, i.e.
\[
\theta_{12}=\theta^+_{12}-\theta^-_{12}=\theta_{34}=\ln\Delta_{\rm O}\,,
\]
where we note
\[
\Delta_{\rm O}:=\frac{(k_3-k_2)(k_4-k_1)}{(k_4-k_2)(k_3-k_1)}=1-\frac{(k_2-k_1)(k_4-k_3)}{(k_4-k_2)(k_3-k_1)}<1.
\]
This implies that $\theta_{12}=\theta_{34}<0$, and each $[i,j]$-soliton shifts in $x$ with
\[
\Delta x_{[i,j]}=\frac{1}{k_j-k_i}\,\theta_{ij}.
\]
The negative phase shifts $\Delta x_{[1,2]}<0$ and $\Delta x_{[3,4]}<0$ indicate an {\it attractive} force
in the interaction, and
if the amplitudes are the same, i.e. $k_2-k_1=k_4-k_3$, then their phase shifts are the same,
i.e.  $\Delta x_{[1,2]}=\Delta x_{[3,4]}=\frac{1}{k_4-k_3}\theta_{34}<0$.
Figure \ref{fig:Otype} illustrates an O-type interaction of two solitons which
have the same amplitude, $A_{[1,2]}=A_{[3,4]}=A_0$,  and are symmetric with respect to the $y$-axis,
$\Psi_{[3,4]}=-\Psi_{[1,2]}=\Psi_0$.

%%%%%%%%%%%%%%%%%%%%%%%%%%%
%%%%%%%%%%%%%%%%%%%%%%%%%%%%%

O-type soliton solution has a steady X-shape with phase shifts in both line-solitons.
One can also find the formula of the maximum amplitude which occurs at the 
center of intersection point (center of the X-shape):  We place the soliton 
solution so that the origin $(0,0)$ is
the center of the $X$-shape at $t=0$. 
This implies that the sum of the phase shifts vanishes for
each soliton, i.e.
\begin{equation}\label{centerX}\left\{\begin{array}{lll}
\displaystyle{\theta^+_{12}+\theta^-_{12}=\ln\frac{(k_4-k_1)(k_3-k_1)}{(k_4-k_2)(k_3-k_2)}-2\ln a=0},\\[2.0ex]
\displaystyle{\theta^+_{34}+\theta^-_{34}=\ln\frac{(k_3-k_1)(k_3-k_2)}{(k_4-k_1)(k_4-k_2)}-2\ln b=0}.
\end{array}
\right.
\end{equation}
Those determine $a$ and $b$ in the $A$-matrix. Then the $\tau$-function becomes
\begin{align*}
\tau&=(k_3-k_1)\left(E_1E_3+E_2E_4 +\sqrt{\Delta_{\rm O}}\,(E_1E_4+E_2E_3)\right)\\
&\equiv E_1\left(E_3+\sqrt{\Delta_{\rm O}}\,E_4\right)+E_2\left(\sqrt{\Delta_{\rm O}}\,E_3+E_4\right)\\
&\equiv E_{12}^+\left(E_{34}^++\sqrt{\Delta_{\rm O}}\,E_{34}^-\right)+E_{12}^-\left(\sqrt{\Delta_{\rm O}}\,E_{34}^++E_{34}^-\right)\\
&\equiv \cosh \Theta^++\sqrt{\Delta_{\rm O}}\,\cosh\Theta^-,
\end{align*}
where $\alpha\equiv \beta$ implies that $\alpha$ and $\beta$ give the same solution $u$, and
$E_{ij}^{\pm}$ and $\Theta^{\pm}$ are defined by
\[
E_{ij}^{\pm}:=\exp\left(\pm\frac{\theta_i-\theta_j}{2}\right),\qquad
\Theta^{\pm}:=\frac{1}{2}\left[(\theta_1-\theta_2)\pm(\theta_3-\theta_4)\right].
\]
It is then not difficult to see that $u(x,y,0)$ attains the maximum at the origin, and we get
the maximum value $u_{\rm max}:=u(0,0,0)$,  which is
\begin{equation}\label{O-center}
\begin{array}{lll}
u_{\rm max}&=&\displaystyle{\frac{1}{2}\left((k_1-k_2)^2+(k_3-k_4)^2\right)+
\frac{1-\sqrt{\Delta_{\rm O}}}{1+\sqrt{\Delta_{\rm O}}}(k_1-k_2)(k_3-k_4)}\\[2.0ex]
&=& \displaystyle{A_{[1,2]}+A_{[3,4]}+
2\,\frac{1-\sqrt{\Delta_{\rm O}}}{1+\sqrt{\Delta_{\rm O}}}\sqrt{A_{[1,2]}A_{[3,4]}}}\,.
\end{array}
\end{equation}
(See also \cite{DSH:04, S:04}.)
Since $0<\Delta_{\rm O}<1$, we have the bound
\[
A_{[1,2]}+A_{[3,4]}<u_{\rm max}<\left(\sqrt{A_{[1,2]}}+
\sqrt{A_{[3,4]}}\right)^2\,.
\]

It is also interesting to note that the formula $\Delta_{\rm O}$ has critical cases at the values 
$k_1=k_2$ or $k_2=k_3$ or $k_3=k_4$. For the case with $k_1=k_2$ or $k_3=k_4$ (i.e. $\Delta_{\rm O}=1$),
one can see that one of the line-soliton becomes small, and the limit consists of just one-soliton
solution. On the other hand, for  the case $k_2=k_3$ (i.e. $\Delta_{\rm O}=0$), 
the $\tau$-function has only three terms, which corresponds to a solution showing
a Y-shape interaction (i.e. the phase shift
becomes infinity and the middle portion of the interaction stretches to infinity). 
This limit has been discussed in \cite{M:77, NR:77} as a resonant interaction of three 
waves to make Y-shape soliton. This limit gives a critical angle
between those solitons which can be found as follows:
First let us express each $k_j$ parameter in terms of the amplitude and the slope,
\begin{align*}
k_{1,2}&=\frac{1}{2}\left(\tan\Psi_{[1,2]}\mp \sqrt{2A_{[1,2]}}\right),\\
k_{3,4}&=\frac{1}{2}\left(\tan\Psi_{[3,4]}\mp\sqrt{2A_{[3,4]}}\right),
\end{align*}
where the angle $\Psi_{[i,j]}$ is measured in the counterclockwise direction from the $y$-axis
(see Figure \ref{fig:Otype}). In particular, we have
\[
\tan\Psi_{[1,2]}=-\sqrt{2A_{[1,2]}}+2k_2, \qquad \tan\Psi_{[3,4]}=\sqrt{2A_{[3,4]}}+2k_3.
\]
Without loss of generality, let us consider the special case when both solitons
are of equal amplitude and symmetric with respect the
$y$-axis i.e., $A_{[1,2]}=A_{[3,4]}=A_0$ and  $\Psi_{[3,4]}=-\Psi_{[1,2]}=\Psi_0>0$. 
This corresponds to setting $k_1 = -k_4$ and $k_2 = -k_3$. Then, for fixed
amplitude $A_0$, the angle $\Psi_0$ has a lower bound given by
\[
\tan\Psi_0 = \sqrt{2A_{0}}+2k_3 \geq \sqrt{2A_{0}} := \tan \Psi_c\,.
\]
The lower bound is achieved in the 
limit $k_2=k_3=0$, and the critical angle $\Psi_c$ is given by
\begin{equation}\label{critical}
\Psi_c=\tan^{-1}\sqrt{2A_0}\,.
\end{equation}
Note that this function is monotone increasing in $A_0$, and for $A_0=\frac{1}{2}$ the
critical angle occurs when the two solitons intersect perpendicularly. In this symmetric case,  from
(\ref{O-center}), the maximum amplitude given at the center of interaction can be calculated as
\begin{equation}\label{Omax}
u_{\rm max}=\frac{4A_0}{1+\sqrt{\Delta_{\rm O}}}, 
\qquad{\rm with}\quad \Delta_{\rm O}=1-\frac{2A_0}{\tan^2\Psi_0}.
\end{equation}
Thus, at the critical angle $\Psi_0=\Psi_c$ (i.e., $\Delta_{\rm O}=0$), 
we have $u_{\rm max}=4A_0$ (see also \cite{M:77,DSH:04,S:04}).

We also note that there are three other soliton solutions of $(2,2)$-type 
which have the same limit. These are the cases marked by $\pi=(3412), (2413)$ and 
$(3142)$, and their chord diagrams degenerate to the same graph of figure eight 
in the limit.  We will discuss these cases in the following sections
Since $(2413)$- and $(3142)$-types are dual each other, we only
discuss the $(3142)$-type.

One should note that if we use the form of the O-type solution even beyond 
the critical angle, i.e.  $k_3<k_2$, then the solution becomes singular (note that 
the sign of $E(2,3)$ changes). In earlier works, this was considered to be an 
obstacle for using the KP equation to describe an interaction of two line-solitons 
with a smaller angle.  On the contrary, the KP equation should give a better 
approximation to describe oblique interactions of solitons with smaller angles.
Thus one should expect to have explicit solutions of the KP equation describing
such phenomena. It turns out that the new types of $(2,2)$-soliton solutions 
discussed above can indeed serve as good models for describing line-soliton 
interactions of solitons with small angles. We will show in the next section how
these solutions are related 
to the {\it Mach reflections} discussed in \cite{M:77}.

\subsection{$(3142)$-type soliton solutions}
%%%%%%%%%%%%%%%%%%%%%%%%%%%%
\begin{figure}[t]
\centering
\includegraphics[scale=0.48]{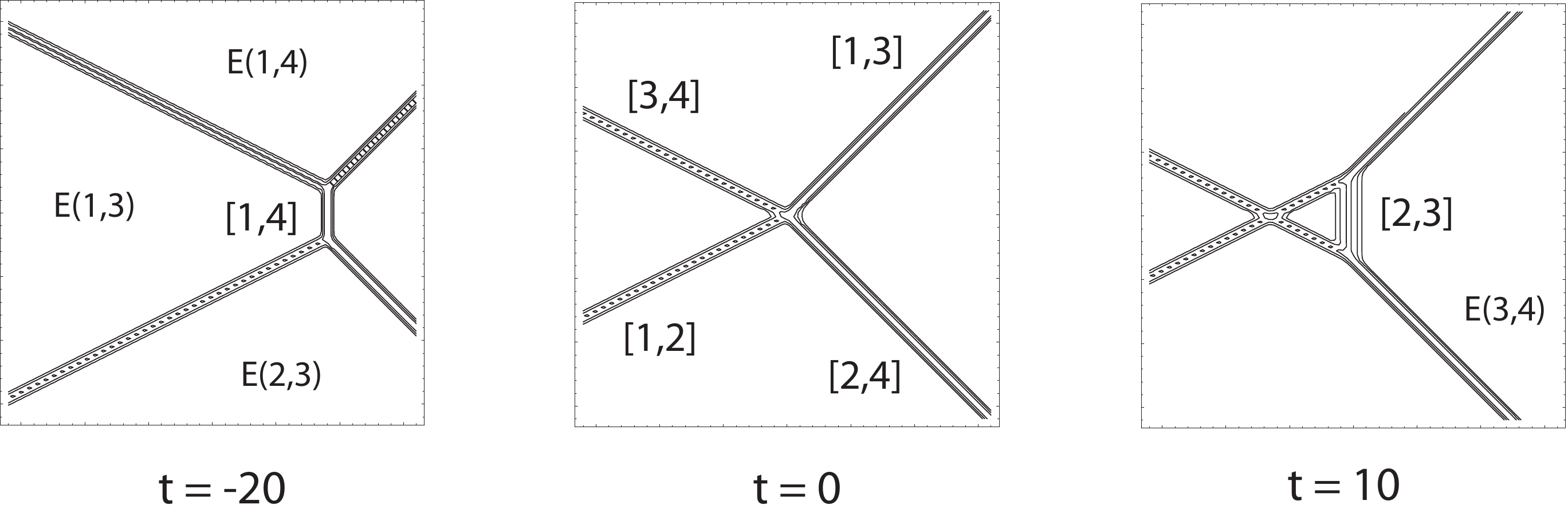} 
\caption{The time evolution of a $(3142)$-type soliton solution. The parameters are chosen as
$(k_1,k_2,k_3,k_4)=(-\frac{3}{2},-\frac{1}{2},\frac{1}{2},\frac{3}{2})$. The amplitudes of
those solitons are $A_{[1,3]}=A_{[2,4]}=2$ and $A_{[1,2]}=A_{[3,4]}=\sf{1}{2}$. The directions of the wave-vectors are $\Psi_{[2,4]}=-\Psi_{[1,3]}=45^{\circ}$. The critical angle is given by
$\Psi_c=\tan^{-1}\sqrt{2A_0}\approx 63.4^{\circ}$, which also gives $\Psi_c=\Psi_{[3,4]}=-\Psi_{[1,2]}$.
\label{fig:3142}}
\end{figure}
%%%%%%%%%%%%%%%%%%%%%%%%%%%%%%%
We consider a solution of this type which
 consists of two line-solitons for large positive $x$ and two other line-solitons for 
large negative $x$. We then assume that the slopes of two solitons in each region
have opposite signs, i.e. one in $y>0$ and other in $y<0$ (see Figure \ref{fig:3142}). 
The line-solitons for the $(3142)$-type solution can be determined from 
the balance between two appropriate exponential terms in its $\tau$-function which has the form,
\[
\tau=E(1,3)+bE(1,4)+aE(2,3)+abE(2,4)+cE(3,4)\,.
\]
The solution contains three free parameters $a,b$ and $c$, which can
be used to determine the locations of three (out of four) asymptotic line-solitons 
(e.g. two in $x\gg0$ and one in $x\ll0$). Thus, the parameters are completely 
determined from the asymptotic data on large $|y|$.

Let us first consider the line-solitons in $x\gg 0$: There are two line-solitons 
which are $[1,3]$-soliton in $y\gg0$ and $[2,4]$-soliton in $y\ll0$. The $[1,3]$-soliton 
is obtained by the balance between the exponential terms $bE(1,4)$ and $cE(3,4)$, 
and the $[2,4]$-soliton is by the balance between $aE(2,3)$ and $cE(3,4)$. Consequently,
the phase shifts of $[1,3]$- and $[2,4]$-solitons for $x\gg0$ are given by 
\begin{equation}\label{shift1324}
\theta_{13}^+=\ln\frac{k_4-k_1}{k_4-k_3}+\ln\frac{b}{c}\,, \qquad \quad
\theta_{24}^+=\ln\frac{k_3-k_2}{k_4-k_3}+\ln \frac{a}{c}\,.
\end{equation}
Choosing appropriate values for the ratios $a/c$ and $b/c$, one can set 
$\theta_{13}^+=\theta_{24}^+ =0$, thus fixing the locations of those line-solitons 
as passing through the origin at $t=0$.  Notice that there is one free parameter left
after this.

Now we consider the line-solitons in $x\ll0$: They are $[3,4]$-soliton in $y\gg0$ 
and $[1,2]$-soliton in $y\ll0$. The phase shifts are given respectively by
\begin{align*}
\theta_{34}^-=\ln\frac{k_3-k_1}{k_4-k_1}-\ln b,\qquad
\theta_{12}^-=\ln\frac{k_3-k_1}{k_3-k_2}-\ln a
\end{align*}

The four phase shifts and the three free parameters $a,b,c$ are related by
\begin{equation*}
a=\frac{k_3-k_1}{k_3-k_2}e^{-\theta_{12}^-}, \qquad b=\frac{k_3-k_1}{k_4-k_1}e^{-\theta_{34}^-},\qquad
c=\frac{k_3-k_1}{k_4-k_3}e^{-\theta_{13}^+-\theta_{34}^-},
\end{equation*}
with the conservation of total phase shifts for $y\to\pm\infty$ (as in the case of O-type,
see (\ref{phaseconservation})),
\[
\theta_{13}^++\theta_{34}^-=\theta_{24}^++\theta_{12}^-.
\]
We then define the parameter representing this conservation,
\begin{equation}\label{shiftc}
s:=\exp\left(-\theta_{13}^+-\theta_{34}^-\right),
\end{equation}
which leads to
\begin{equation}\label{shiftab}
a=\frac{k_3-k_1}{k_3-k_2}\,se^{\theta_{24}^+}, \qquad b=\frac{k_3-k_1}{k_4-k_1}\,se^{\theta_{13}^+},\qquad
c=\frac{k_3-k_1}{k_4-k_3}\,s.
\end{equation}
The $s$-parameter represents the relative locations of the intersection point of the
$[1,3]$- and $[3,4]$-solitons with the $x$-axis, in particular, 
$\theta_{13}^++\theta_{34}^- = 0$ when $s=1$. 
Thus the parameters $a,b$ and $c$ are related to  the locations of $[1,3]$-soliton (with $\theta_{13}^+$),
of $[2,4]$-soliton (with $\theta_{24}^+$), and the intersection point of $[1,3]$- and $[3,4]$-solitons
(with $s$).

%%%%%%%%%%%%%%%%%%%%%%%%%%%%%%%%%%%%%
\begin{figure}[t]
\centering
\includegraphics[scale=0.52]{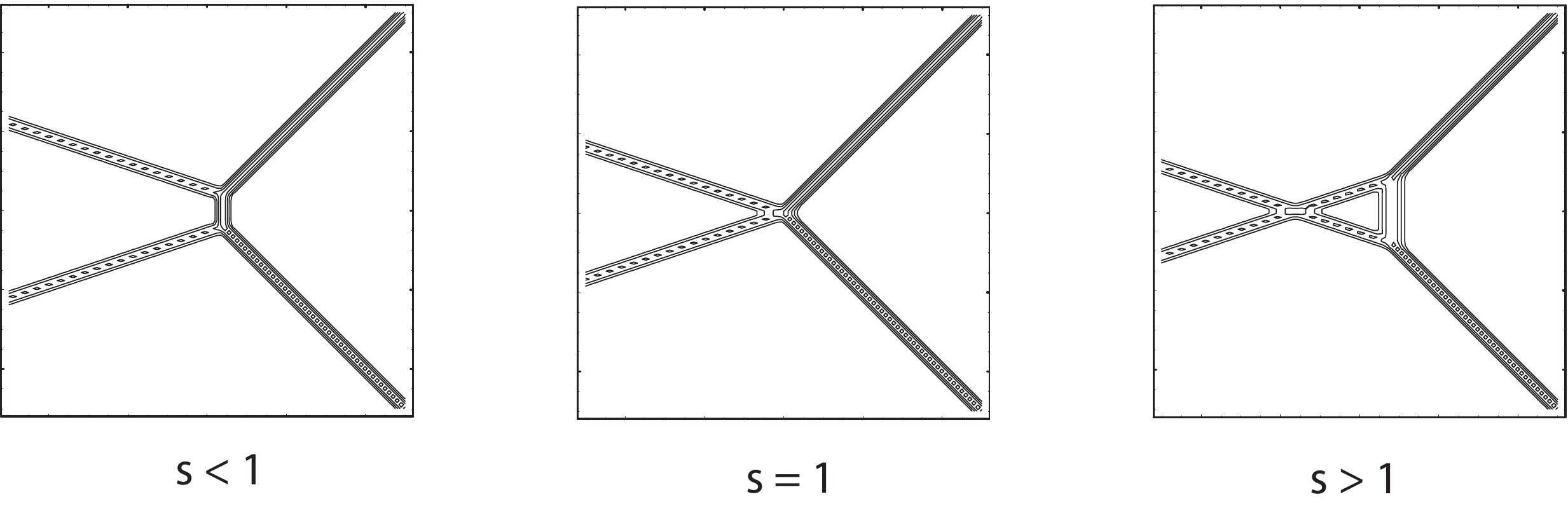} 
\caption{A $(3142)$-type soliton solution with the $s$-parameter.
 The $k$-parameters are taken as
$(k_1,k_2,k_3,k_4)=(-2,-1,1,2)$. The parameters in the $A$-matrix
are chosen as $a=\frac{3}{2}s, b=\frac{3}{4}s$ and $c=3s$, so that
$[1,3]$- and $[2,4]$-solitons meet at the origin, i.e. $\theta_{13}^+=\theta_{24}^+=0$
(see (\ref{shiftab})). 
Then at $s=1$, all the solitons
meet at the origin, i.e. the $s$-parameter shifts $[1,2]$- and $[3,4]$-solitons.
\label{fig:phase3142}}
\end{figure}
%%%%%%%%%%%%%%%%%%%%%%%%%%%%%%%%

Now we consider an example, in which the $[1,3]$- and $[2,4]$-solitons have the same amplitude
($A_{[1,3]}=A_{[2,4]}=A_0$) and they are symmetric with respect to the $x$-axis 
($\Psi_{[2,4]}=-\Psi_{[1,3]}=\Psi_0$). Then in terms of the $k$-parameters, we have
\[
k_3-k_1=k_4-k_2=\sqrt{2A_0}.
\]
Also the symmetry of the wave-vectors, i.e. $\Psi_{[2,4]}=\Psi_0=-\Psi_{[1,3]}$, gives
\[
k_2+k_4=-(k_1+k_3)=\tan\Psi_0.
\]
This implies that we have 
\begin{equation}\label{SymmetryinK}
k_4=-k_1>0,\qquad k_3=-k_2>0.
\end{equation} 
The angle $\Psi_0$ takes the value in $(0,\Psi_c)$, where the critical angle is given by
the condition $k_2=k_3=0$, i.e.
\[
\Psi_c=\tan^{-1}\sqrt{2A_0}<\frac{\pi}{2}.
\]
Notice that this formula is the same as that of the O-type soliton 
solution (see (\ref{critical})). In this sense, these two cases have the 
same limit but from opposite directions namely, $\Psi_0 \to \Psi_c$ from
above for the O-type, while $\Psi_0 \to \Psi_c$ from below for the $(3142)$-type.
This fact will be important for
the initial value problem discussed in Section \ref{IVP}.
     
 From (\ref{SymmetryinK}), one can easily deduce the following facts
for $[1,2]$- and $[3,4]$-solitons in $x<0$:
\begin{itemize}
\item[(a)] Those solitons have the same amplitude, i.e. 
\[
A_{[1,2]}=A_{[3,4]}=\frac{1}{2}(k_4-k_3)^2=\frac{1}{2}(k_4+k_2)^2=\frac{1}{2}\tan^2\Psi_0.
\]
Thus, if the $[1,3]$- and $[2,4]$-solitons in $x>0$ are close to the $y$-axis (i.e.
a small $\Psi_0$), then the amplitudes of the solitons in $x<0$ are small;
whereas at the critical angle
$\Psi_0=\Psi_c$, the solitons $[1,2]$ and $[3,4]$ in $x<0$ 
take the maximum amplitude $A_{[1,2]}=A_{[3,4]}=A_0$.
\item[(b)] The directions of the wave-vectors for the $[1,2]$ and $[3,4]$-solitons
are also symmetric, i.e.
\[
\tan\Psi_{[3,4]}=-\tan\Psi_{[1,2]}=k_3+k_4.
\]
Moreover, the symmetry (\ref{SymmetryinK}) implies that
$\tan\Psi_{[3,4]}=k_4-k_2=\sqrt{2A_{[2,4]}}=\sqrt{2A_0}$, so
\[
\Psi_{[3,4]}=\Psi_c=\tan^{-1}\sqrt{2A_0}.
\]
Thus the directions of the wave-vectors for the $[1,2]$ and $[3,4]$-solitons in $x<0$
depend only on the amplitude of the solitons in $x>0$ but not on their directions
(i.e., angle of their V-shape).
\end{itemize}

Let us choose the parameters in the $A$-matrix for the $(3142)$-soliton
solution appropriately, so that at $t=0$ all the solitons intersect at the 
origin (see Figure \ref{fig:3142}).
Then for $t>0$, the resonant interaction between $[1,3]$- and $[3,4]$-solitons (as well as
$[2,4]$- and $[1,2]$-solitons) generates an intermediate line-soliton (called ``stem" soliton)
which is $[1,4]$ soliton. The amplitude of this soliton is given by
\begin{equation}\label{stemA}
A_{[1,4]}=\frac{1}{2}(k_4-k_1)^2=2k_4^2=\frac{1}{2}\left(\sqrt{2A_0}+\tan\Psi_0\right)^2.
\end{equation}
Note here that at the critical angle $\Psi_0=\Psi_c$, the amplitude takes the maximum $A_{[1,4]}=4A_0$
(see \cite{TO:07, PTLO:05}).

For $t<0$, the resonant interaction between $[1,3]$- and $[1,2]$-solitons (as well as $[2,4]$- and
$[3,4]$-solitons) generates an intermediate line-soliton of $[2,3]$-soliton. 
The amplitude of $[2,3]$-soliton is given by
\[
A_{[2,3]}=\frac{1}{2}(k_3-k_2)^2=\frac{1}{2}\left(\sqrt{2A_0}-\tan\Psi_0\right)^2.
\]
Because of the symmetry (\ref{SymmetryinK}), both $[1,4]$- and $[2,3]$-solitons are
parallel to the $y$-axis, i.e. $\tan\Psi_{[1,4]}=\tan\Psi_{[2,3]}=0$.

\subsection{P-type soliton solutions}
%%%%%%%%%%%%%%%%%%%%%%%%%%%%%%%%%%%%
\begin{figure}[t]
\centering
\includegraphics[scale=0.48]{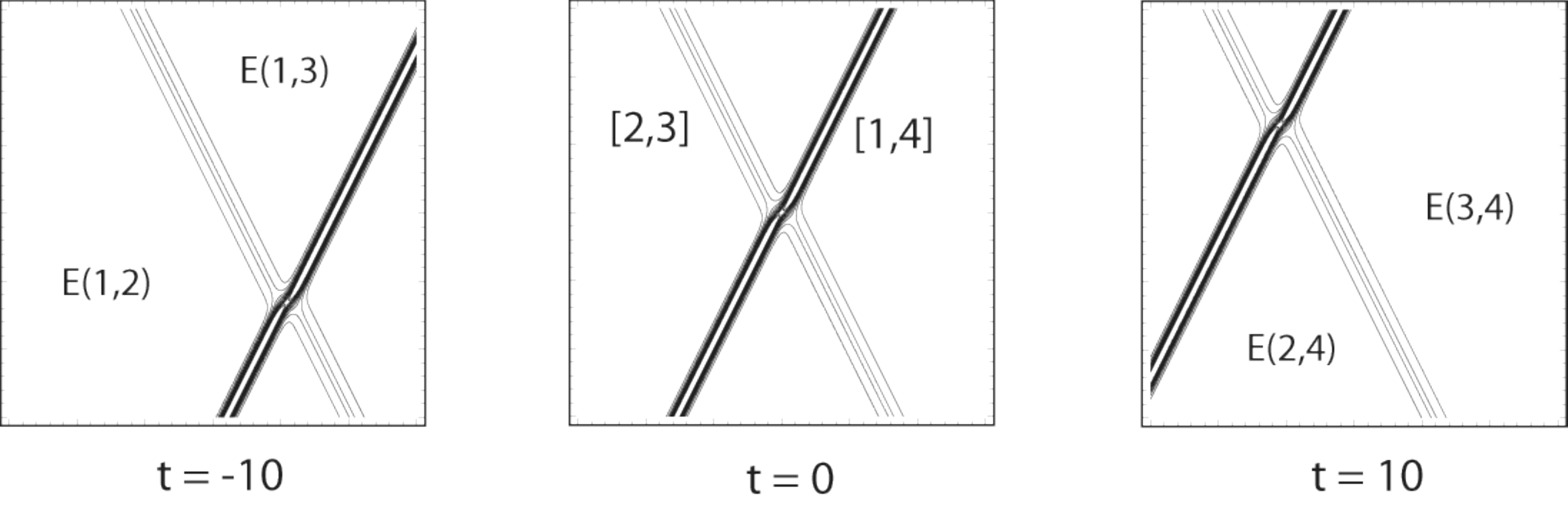} 
\caption{The time evolution of a P-type soliton solution. The parameters are chosen as
$(k_1,k_2,k_3,k_4)=(-\frac{3}{2},-\frac{1}{4},\frac{3}{4},1)$. 
The amplitudes are $A_{[1,4]}=\frac{25}{8}, A_{[2,3]}=\frac{1}{2}$. The directions of the wave-vectors are given by
$\Psi_{[2,3]}=-\Psi_{[1,4]}=\tan^{-1}\frac{1}{2}\approx 26.57^{\circ}$.
The parameters in the $A$-matrix are chosen as $a=\frac{5}{3}$ and $b=3$, so that
the center of the interaction point is located at the origin at $t=0$.
\label{fig:Ptype}}
\end{figure}
%%%%%%%%%%%%%%%%%%%%%%%%%%%%%%%%%%%%%%%%

This type of solution consists of two line-solitons, $[1,4]$ and $[2,3]$.
We again assume that $[1,4]$-soliton has positive slope and $[2,3]$-soliton has negative slope
(see Figure \ref{fig:Ptype}).

The phase shifts of  $[1,4]$- and $[2,3]$-soliton in $x\gg0$ are given by
\[
\theta_{14}^+=\ln\frac{k_3-k_1}{k_4-k_3}-\ln b,\qquad
\theta_{23}^+=\ln\frac{k_4-k_2}{k_4-k_3}-\ln a.
\]
Those parameters $a$ and $b$ in the $A$-matrix can be chosen, so that at $t=0$ those solitons
meet at the origin. 

The phase shifts of those solitons in $x\ll0$ are
\[
\theta_{14}^-=\ln\frac{k_2-k_1}{k_4-k_2}-\ln b,\qquad 
\theta_{23}^-=\ln\frac{k_2-k_1}{k_3-k_1}-\ln a.
\]
The total phase shift of $[1,4]$-soliton is given by
\[
\theta_{14}=\theta_{14}^+-\theta_{14}^-=\ln\frac{(k_4-k_2)(k_3-k_1)}{(k_2-k_1)(k_4-k_3)}>0\,.
\]
Again this value is the same as the shift for $[2,3]$-soliton, i.e. 
$\theta_{23}=\theta^+_{23}-\theta_{23}^- =\theta_{14}$. Because of the 
order $k_1<k_2<k_3<k_4$, the phase shift is always positive, i.e.
\begin{equation}\label{Pphase}
\Delta x_{[1,4]}=\frac{1}{k_4-k_1}\theta_{14}>0.
\end{equation}
This situation is similar to the case of the KdV solitons. Notice in particular
that the total phase shift has the opposite sign to that of the O-type soliton
solution \cite{BC:06}, which indicates a {\it repulsive} force in the interaction.

As in the case of O-type, one can find the amplitude at the center of interaction:
We calculate the sums of $\theta_{14}^{\pm}$ and $\theta_{23}^{\pm}$, and set
\[\left\{\begin{array}{lll}
\displaystyle{\theta_{14}^++\theta_{14}^-=\ln\frac{(k_2-k_1)(k_3-k_1)}{(k_4-k_2)(k_4-k_3)} -2\ln b=0},\\[2.0ex]
\displaystyle{\theta_{23}^++\theta_{23}^-=\ln\frac{(k_2-k_1)(k_4-k_2)}{(k_3-k_1)(k_4-k_3)}-2\ln a=0}.
\end{array}\right.
\]
Then following the same procedure as in the case of O-type, we get the amplitude at the center
of interaction which is located at the point of $x=y=t=0$,
\begin{equation}\label{P-center}
\begin{array}{lll}
u(0,0,0)&=&\displaystyle{\frac{1}{2}\left((k_4-k_1)^2+(k_3-k_2)^2\right)-
\frac{1-\sqrt{\Delta_{\rm P}}}{1+\sqrt{\Delta_{\rm P}}} (k_4-k_1)(k_3-k_2)}\\[2.0ex]
&=&\displaystyle{A_{[1,4]}+A_{[2,3]}-2\,\frac{1-\sqrt{\Delta_{\rm P}}}
{1+\sqrt{\Delta_{\rm P}}}\sqrt{A_{[1,4]}A_{[2,3]}}}\,,
\end{array}
\end{equation}
where $\Delta_{\rm P}$ is given by
\[
\Delta_{\rm P}=\frac{(k_2-k_1)(k_4-k_3)}{(k_3-k_1)(k_4-k_2)}\,.
\]
If we set $x=\sf{k_2-k_1}{k_4-k_1},\, y=\sf{k_4-k_3}{k_4-k_1}$, then we have
\begin{align*}
\frac{1-\sqrt{\Delta_{\rm P}}}{1+\sqrt{\Delta_{\rm P}}} = 
\frac{\sqrt{(1-x)(1-y)} - \sqrt{xy}}{\sqrt{(1-x)(1-y)}+\sqrt{xy}}=\frac{1-(x+y)}{(\sqrt{(1-x)(1-y)}+\sqrt{xy})^2}.
\end{align*}
Using the inequality (equivalent to $\sqrt{xy}\le \sf{1}{2}(x+y)$),
\[
\sqrt{(1-x)(1-y)}+\sqrt{xy} \leq 1.
\]
we have
\[
\frac{1-\sqrt{\Delta_{\rm P}}}{1+\sqrt{\Delta_{\rm P}}} \ge 1-(x+y)=\frac{k_3-k_2}{k_4-k_1}=\sqrt{\frac{A_{[2,3]}}{A_{[1,4]}}}.
\]
Then (\ref{P-center}) gives the following bound of the interaction amplitude,
\[
\left(\sqrt{A_{[1,4]}}-\sqrt{A_{[2,3]}}\,\right)^2<u(0,0,0)\le A_{[1,4]}-A_{[2,3]}\,,
\]
As a result, we find that the amplitude at the center of the amplitude is smaller
than the larger soliton-amplitude $A_{[1,4]}$ due to the P-type interaction
which acts as a repulsive force (see also the phase shift (\ref{Pphase})).

\subsection{T-type soliton solutions}
\begin{figure}[t!]
\centering
\includegraphics[scale=0.52]{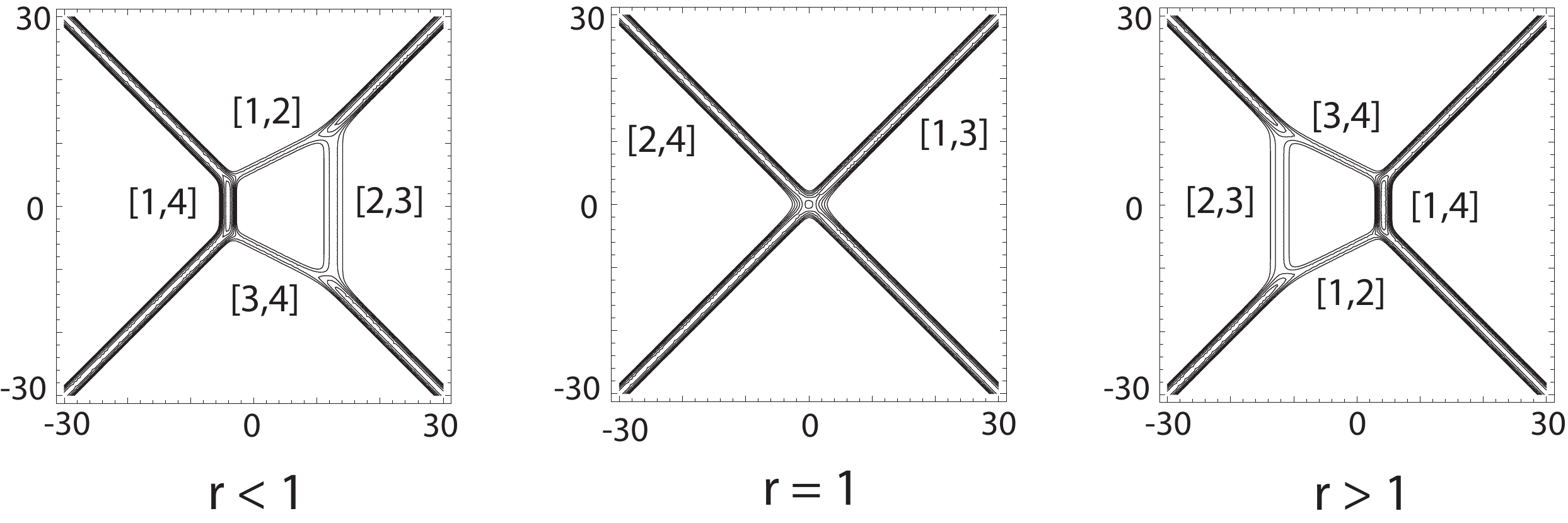} 
\caption{Two-soliton solution with T-type interaction. The parameters are chosen as
$(k_1,k_2,k_3,k_4)=(-\frac{3}{2},-\frac{1}{2},\frac{1}{2},\frac{3}{2})$ and $s=1$ (i.e. no phase
shifts). Two other elements $b,c$ in the $A$-matrix 
are chosen, so that two solitons intersect at the origin when $r=1$. 
Also notice that T-type soliton consists of $(2413)$-soliton in $x<0$ and
$(3142)$-soliton in $x>0$.
\label{fig:T}}
\end{figure}

There are four parameters in the $A$-matrix for T-type soliton
solution consisting of two asymptotic line-solitons,
$[1,3]$ and $[2,4]$. Here we explain that those parameters give the information of
the locations of those line-solitons, the phase shift and on-set of the opening of
a box (see Figure \ref{fig:T}). Thus three of those four parameters are determined
by the asymptotic data on large $|y|$, and we need an internal data for the other one.
%%%%%%%%%%%%%%%%%%%%%%%%%%%%%%%%%%%%%%%%%%
%%%%%%%%%%%%%%%%%%%%%%%%%%%%%%%%%%%%%%%%

Following the arguments in the previous section, one can find the phase shifts of the
line-solitons of $[1,3]$ and $[2,4]$:
For $[1,3]$-soliton in $x>0$ (and $y\gg0$), the phase shift is calculated as
\[
\theta_{13}^+=\ln\frac{k_4-k_1}{k_4-k_3}-\ln \frac{D}{b}\,,
\]
where $D=ad-bc=\xi(3,4)$. For the same soliton in $x<0$ (and $y \ll0$), we have
\[
\theta^-_{13}=\ln\frac{k_2-k_1}{k_3-k_2}-\ln c\,.
\]
So the total phase shift is given by
\[
\theta_{13}=\theta_{13}^+-\theta_{13}^-=\ln\frac{(k_4-k_1)(k_3-k_2)}{(k_2-k_1)(k_4-k_3)}-
\ln\frac{D}{bc}.
\]
Thus the shift depends on the $A$-matrix unlike the cases of O- and P-types, and it can take
any value.

For $[2,4]$-soliton in $x>0$ (and $y\ll0$), we have
\[
\theta_{24}^+=\ln\frac{k_3-k_2}{k_4-k_3}-\ln\frac{D}{c}.
\]
and for the same one in $x<0$ (and $y\gg0$), we have
\[
\theta_{24}^-=\ln\frac{k_2-k_1}{k_4-k_1}-\ln b.
\]
Note that the total phase shift $\theta_{24}=\theta_{24}^+-\theta_{24}^- $ is the 
same as that for $[1,3]$-soliton, i.e. the phase 
conservation along the $y$-axis $\theta_{13}^++\theta_{24}^-=\theta_{13}^-+\theta_{24}^+$
holds. 
Then as in (\ref{shiftc}) for the case of $(3142)$-type, we define the $s$-parameter,
\begin{equation}\label{sTtype}
s:=\exp(-\theta_{13}^+-\theta_{24}^-),
\end{equation}
which represents the intersection point of $[1,3]$- and $[2,4]$-soliton.
With the $s$-parameter, we have
\begin{equation}\label{bcD}
b=\frac{k_2-k_1}{k_4-k_1}\,se^{\theta_{13}^+},\qquad c=\frac{k_2-k_1}{k_3-k_2}\,se^{\theta_{24}^+},\qquad D=\frac{k_2-k_1}{k_4-k_3}\,s.
\end{equation}
Namely, the three parameters $b,c$ and $D=ad-bc$ determine the locations and the phase shift (i.e.
the intersection point of $[1,3]$- and $[2,4]$-solitons). 
One other parameter is then related to an on-set of a box
at the intersection point (see Figure \ref{fig:T}).

In order to characterize this parameter,
let us consider the intermediate solitons of $[1,4]$ and $[2,3]$.
First note that for $t\gg 0$, $[1,4]$-soliton appears as the dominant balance between
$E(1,2)$ and $E(2,4)$. This implies that the $\tau$-function can be written  by
\[
\tau\approx E(1,2)+dE(2,4)\equiv \cosh\frac{1}{2}(\theta_1-\theta_4+\theta_{14}^+)\,,
\]
where the super index ``+'' means $t>0$. Then the phase shift $\theta_{14}^+$ is given by
\[
\theta^+_{14}=\ln\frac{k_2-k_1}{k_4-k_2}-\ln\, d .
\]
Similarly one can get the phase shift $\theta_{14}^-$ for $t\ll 0$ as
\[
\theta^-_{14}=\ln\frac{k_3-k_1}{k_4-k_3}-\ln\frac{D}{a}.
\]
 Now consider the sum of $\theta_{14}^{\pm}$, i.e.
\begin{align*}
\theta_{14}^++\theta_{14}^-&=\ln\frac{(k_2-k_1)(k_3-k_1)}{(k_4-k_2)(k_4-k_3)}-\ln\frac{dD}{a}.
\end{align*}
Also, for the $[2,3]$-soliton, one can get
\begin{align*}
\theta_{23}^++\theta_{23}^-&=\ln\frac{(k_2-k_1)(k_4-k_2)}{(k_3-k_1)(k_4-k_3)}-\ln\frac{aD}{d}.
\end{align*}

  Now we introduce a parameter $r$ in the form,
\begin{equation}\label{r}
\frac{a}{d}=r\frac{k_4-k_2}{k_3-k_1}\,,
\end{equation}
so that we have
\begin{align*}
\theta_{14}^++\theta_{14}^-=\ln\,\frac{r}{s},\qquad \theta_{23}^++\theta_{23}^-=-\ln\,(rs).
\end{align*}
Suppose that at $t=0$, $[1,3]$- and $[2,4]$-solitons in $x>0$ are placed so that they meet at the origin,
that is, we choose $\theta_{13}^+=\theta_{24}^+=0$. Also if there is no phase shifts for those solitons,
i.e. $s=1$. then the sums become
\[
\theta_{14}^++\theta_{14}^-=\ln\,r=-(\theta_{23}^++\theta_{23}^-)\,.
\]
This implies that at $t=0$ if $r=1$, then the T-type soliton solution has an exact 
shape of ``X" without any opening of a box at the intersection point on the origin.
Moreover, at $t=0$ if $r>1$, then $[1,4]$-soliton appears in $x>0$ and $[2,3]$-soliton in $x<0$;
whereas if $0<r<1$, then $[1,4]$-soliton appears in $x<0$ and $[2,3]$-soliton in $x>0$.
Figure \ref{fig:T} illustrates those cases with $s=1$. The parameter $r$ determines the
exponential term that is dominant in the region inside the box. When
$r<1$, $E(2,4)$ is the dominant exponential term, and when $r>1$ the dominant
exponential is $E(1,3)$. One should note that the parameter $r$ cannot be
determined by the asymptotic data, that is, $r$ is considered as an ``internal" parameter.

\section{Numerical simulation with V-shape initial waves}\label{IVP}
In this section, we present some numerical simulations of the KP equation with
obliquely incident initial waves related to a physical situation (see for examples \cite{PTLO:05, TO:07, F:80}).
In these simulations, we change the sign of the time $t$, so that now the solitons move in
the positive $x$-direction (this seems to be a convention in the area of fluid mechanics).
A complete numerical study of the initial value problem will be reported in a future 
communication \cite{KK:09}.

\begin{figure}[t!]
\centering
\includegraphics[scale=0.4]{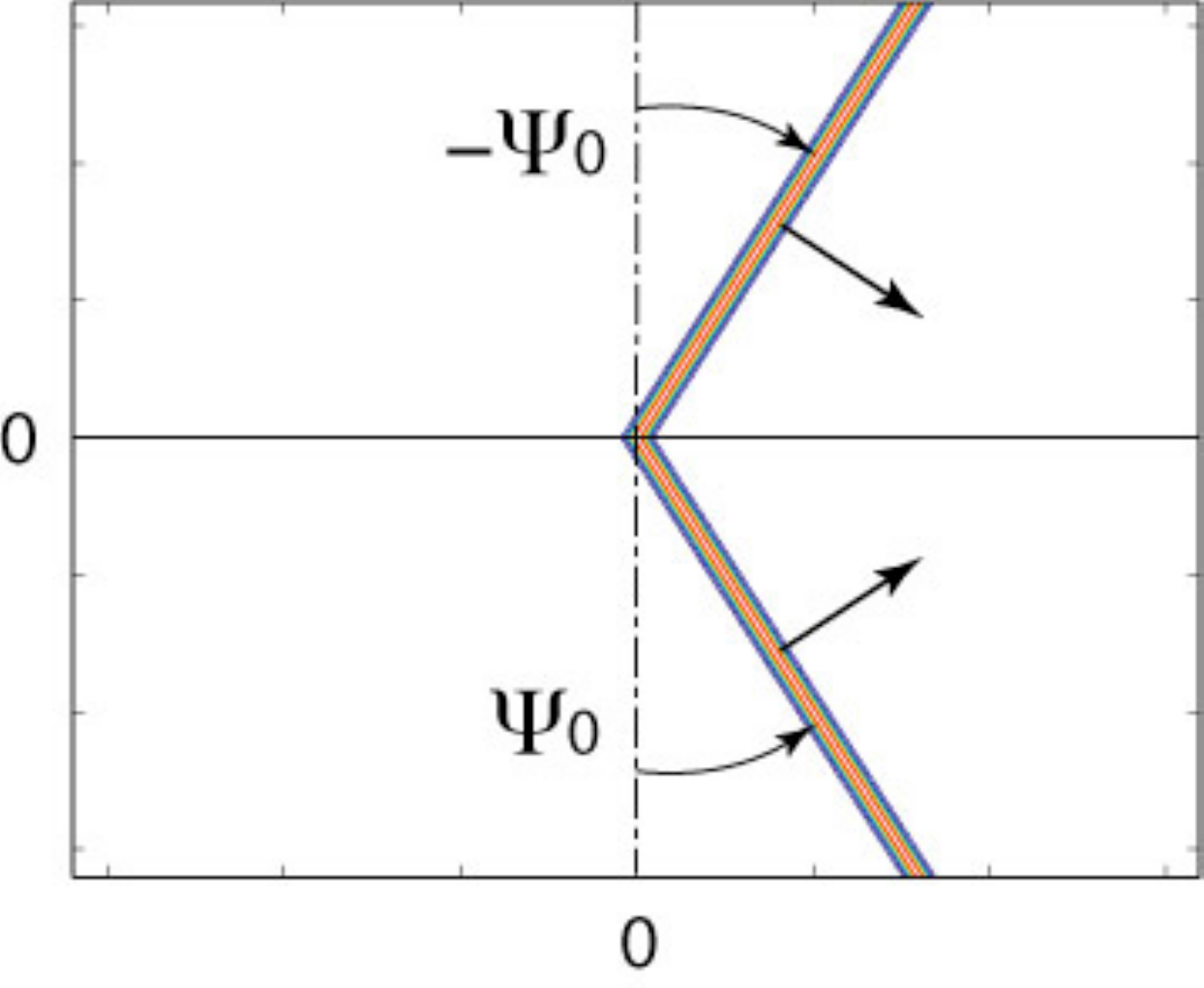} 
\caption{Initial data with V-shape wave. Each line of the V-shape is locally a line-soliton solution, $u(x,y,0)=A_0\sech^2\sqrt{\sf{1}{2}A_0}(x+c_0y)$ with $c_0=\tan(\pm\Psi_0)$.
We set those line-solitons to meet at the origin.  In our numerical simulation, we reverse the time, i.e. $t\to-t$,
so that the solitons are propagate in the positive $x$-direction, and small radiations
propagate in the negative $x$-direction.
\label{fig:IV}}
\end{figure}

We consider the initial data given in the shape of ``V'',
\begin{equation}\label{initialdata}
u(x,y,0)=A_0\sech^2 \sqrt{\sf{1}{2}A_0}\left(x-c_0|y|\right)\,,
\end{equation}
where $c_0=\tan\Psi_0>0$ with the angle measured from the $y$-axis (see Figure \ref{fig:IV}).
Note here that two semi-infinite line-solitons are propagating toward each other,
so that they interact strongly at the corner of the V-shape.
At the boundaries $y=\pm y_{\rm max}$ of the numerical domain, those line-solitons are
patched to the exact one-soliton solutions given by 
\[
u(x,\pm y_{\rm max},t)=A_0\sech^2\sqrt{\sf{1}{2}A_0}\,(x\mp c_0y_{\rm max}-\nu t)\,,
\]
with $\nu=\sf{3}{4}c_0^2+\sf{1}{2}A_0$. Notice that these solitons correspond to  the exact one-soliton solution
of the KdV equation. The numerical simulations are based on a spectral method with window-technique similar to the method used in \cite{TO:07}  (the details will be reported in \cite{KK:09}).
The V-shape initial wave was first considered by M. Oikawa and H. Tsuji (see for example \cite{PTLO:05,
TO:07}) in order to understand  the  generation of freak (or rogue) waves. They noticed  generations of
different types of asymptotic solutions, and found the resonant interactions which create localized high
amplitude waves.
%%%%%%%%%%%%%%%%%%%%%%%%%%%%%%%%%%%%%%%%%%
%%%%%%%%%%%%%%%%%%%%%%%%%%%%%%%%%%%%%%%%
In this paper, we present the results for the cases corresponding to $A_0=2$ and two different angles,
$\Psi_1$ and $\Psi_2$ with $\Psi_1<\Psi_c<\Psi_2$.
where the critical angle is given by $\Psi_c=\tan^{-1}\sqrt{2A_0}\approx 63.4^{\circ}$.
Then we explain these results in terms of certain $(2,2)$-soliton solutions discussed in the 
previous section, and in particular,
we describe the connection with the Mach reflection discussed in \cite{M:77}.

The main idea here is to consider the V-shape initial wave as the part of some $(2,2)$-soliton solution
listed in the previous section.  In order to identify those soliton solutions from the V-shape initial wave form, let us first denote
them as $[{i_1},{j_1}]$-soliton for $y\gg 0$ and $[{i_2},{j_2}]$-soliton for $y\ll 0$.
Then using the relations, $k_j-k_i=\sqrt{2A_0}$ and $k_j+k_i=\tan\Psi_0$, for $[i,j]$-soliton, we have
\begin{equation}\label{NumericalK}
\left\{\begin{array}{llll}
&\displaystyle{k_{j_2}=\frac{1}{2}(\tan\Psi_0+2)},\qquad  &k_{i_2}=\frac{1}{2}(\tan\Psi_0-2),\\[1.5ex]
&\displaystyle{k_{j_1}=-\frac{1}{2}(\tan\Psi_0-2)},\qquad  &k_{i_1}=-\frac{1}{2}(\tan\Psi_0+2).
\end{array}\right.
\end{equation}
Because of the symmetry in the initial shape, we also have $k_{j_2}=-k_{i_1}$ and $k_{i_2}=-k_{j_1}$. Moreover, 
at the critical angle, $\tan\Psi_c=\sqrt{2A_0}=2$,  we have $k_{i_2}=k_{j_1}=0$. 
We also identify $k_{i_1}$ as the smallest parameter and $k_{j_2}$ as the largest one, so that depending on the angle $\Psi_0$, we obtain the following ordering in the $k$-parameters:

For $0<\Psi_0<\Psi_c$ (i.e. $0<\tan\Psi_0<2$), we have
\[
k_{i_1}<k_{i_2}<0<k_{j_1}<k_{j_2},
\]
implying that the corresponding chords of the $[i_1,j_1]$- and the $[i_2,j_2]$-solitons overlap.
This means that the two solitons can be identified as part of either the $(3412)$-type (T-type) or the $(3142)$-type
solution (see Figure \ref{fig:chords}). That is, $[1,3]$ chord appears on the upper side of the diagram,
and $[2,4]$ chord on the lower side.  As will be shown below, the numerical simulation suggests
that the solution converges asymptotically to a $(3142)$-type soliton solution. This situation is related to
the Mach reflection discussed in \cite{M:77, F:80}.

For $\Psi_c<\Psi_0<\sf{\pi}{2}$ (i.e. $2<\tan\Psi_0$), we have
\[
k_{i_1}<k_{j_1}<0<k_{i_2}<k_{j_2}.
\]
In this case, the corresponding chords are separated implying that the two
solitons form part of either $(2413)$- or $(2143)$-type (O-type) solution. That is,  
 $[1,2]$- and $[3,4]$-chords appear on the upper and lower sides of the chord diagram, respectively.
 The numerical simulation seems to indicate that the solution converges asymptotically to an
 O-type solution, and this corresponds to the regular reflection \cite{M:77, F:80}.

We can thus predict the following types of the asymptotic solutions depending on the values
 $\Psi_0$:
\begin{itemize}
\item[(a)] If the angle is $\Psi_0=0$, then the initial wave is just one-soliton solution. This soliton
is parallel to the $y$-axis, so $k_1=-k_2$ showing no $y$-dependency.
\item[(b)] If the angle is in the range $0<\Psi_0<\Psi_c=\tan^{-1}\sqrt{2A_0}$, then the asymptotic
solution is the $(3142)$-type soliton solution (not T-type)
\item[(c)] If the angle satisfy $\Psi_c<\Psi_0<\sf{\pi}{2}$, then the asymptotic solution is the $(2143)$-type
soliton solution (i.e. O-type, and not $(2413)$-type).
\end{itemize}

In the next two subsections, we present the results for the cases (b) and (c) (the case (a) is obvious).

\subsection{Numerical simulation for $\Psi_0>\Psi_c$}
\begin{figure}[t!]
\centering
\includegraphics[scale=0.75]{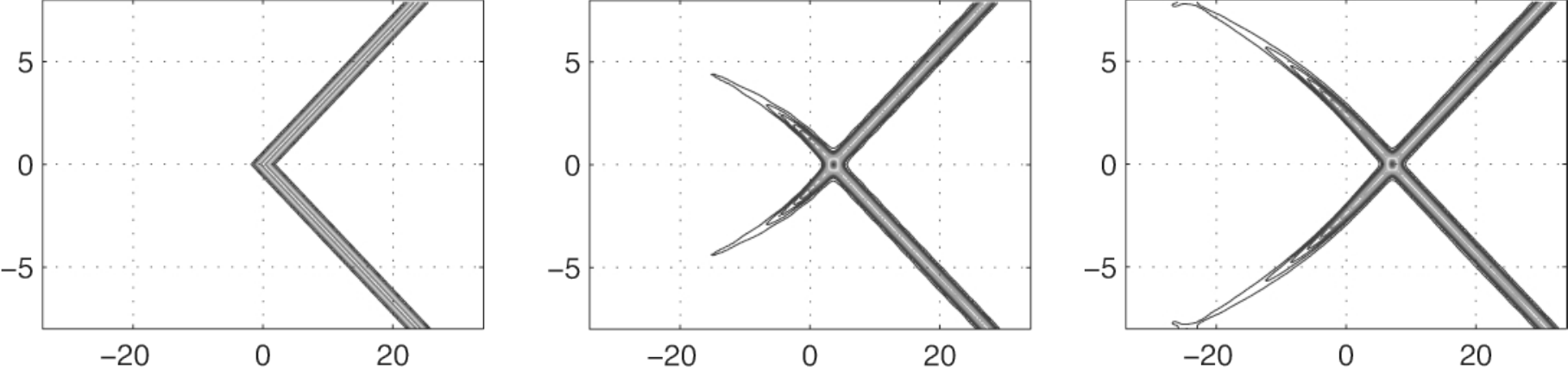} \\
\vskip 0.5cm
\includegraphics[scale=0.75]{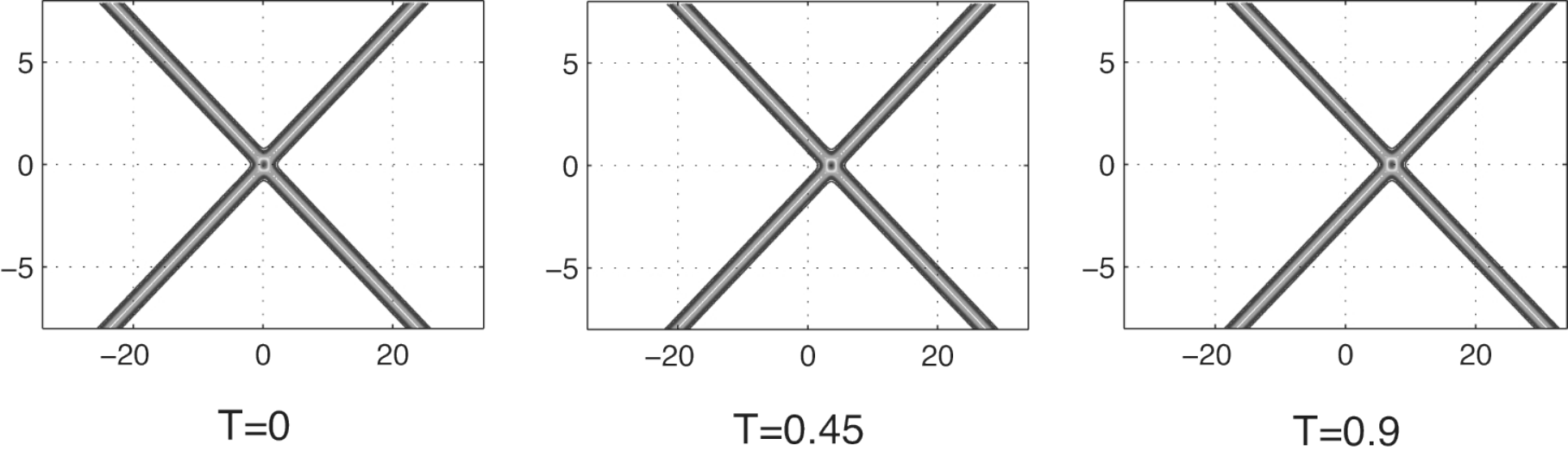} \\
\caption{Numerical simulation with the V-shape initial wave and an exact solution
corresponding to the asymptotic solution.
The parameters are given by $A_0=2$ and $\Psi_0=\tan^{-1}(3)=71.6^{\circ}>\Psi_c=\tan^{-1}(2)\approx 63.4^{\circ}$.
The solution converges asymptotically to the O-type solution shown in the three figures in the second line.
The parameters are given by $(k_1,k_2,k_3,k_4)=(-\sf{5}{2},-\sf{1}{2},\sf{1}{2},\sf{5}{2})$.
\label{fig:ON}}
\end{figure}

In Figure \ref{fig:ON}, the upper three figures illustrate the numerical simulations of the initial value
problem for the V-shape initial wave with $A_0=2$ and $\Psi_0=\tan^{-1}(3)\approx 71.57^{\circ}$.
Here the critical angle is $\Psi_c=\tan^{-1}(2)\approx 63.4^{\circ}$, and we observe 
 asymptotically an O-type soliton solution. The corresponding $k$-parameters are
obtained from (\ref{NumericalK}), i.e.
\[
(k_1,k_2,k_3,k_4)=(-\sf{5}{2},-\sf{1}{2},\sf{1}{2},\sf{5}{2}).
\]
Since two solitons $[1,2]$-and $[3,4]$-solitons for $x>0$ in the initial data are $[1,2]$- and $[3,4]$-solitons, which 
meet at the origin, i.e. $\theta_{12}^+=\theta_{34}^+=0$,
we take the $A$-matrix in the form,
\[
A=\begin{pmatrix}
1 & \sf{5}{3} &0 & 0 \\0 & 0 & 1 & \sf{1}{3}
\end{pmatrix}\,,
\]
where we have used (\ref{Oshifta}) and (\ref{Oshiftb}), i.e.
$a=\frac{k_4-k_1}{k_4-k_2}=\frac{5}{3}$ and $b=\frac{k_4-k_2}{k_4-k_2}=\frac{1}{3}.$
The corresponding O-type exact soliton solution is illustrated  in the lower figures
 in Figure \ref{fig:ON}.  We also confirm that the numerical solution is indeed converging to
this exact solution (details of these results will be reported in \cite{KK:09}).

%%%%%%%%%%%%%%%%%%%%%%%%%%%%%%%%%%%%%%%%%%
%%%%%%%%%%%%%%%%%%%%%%%%%%%%%%%%%%%%%%%%

\subsection{Numerical simulation for $\Psi_0<\Psi_c$}
\begin{figure}[t]
\centering
\includegraphics[scale=0.75]{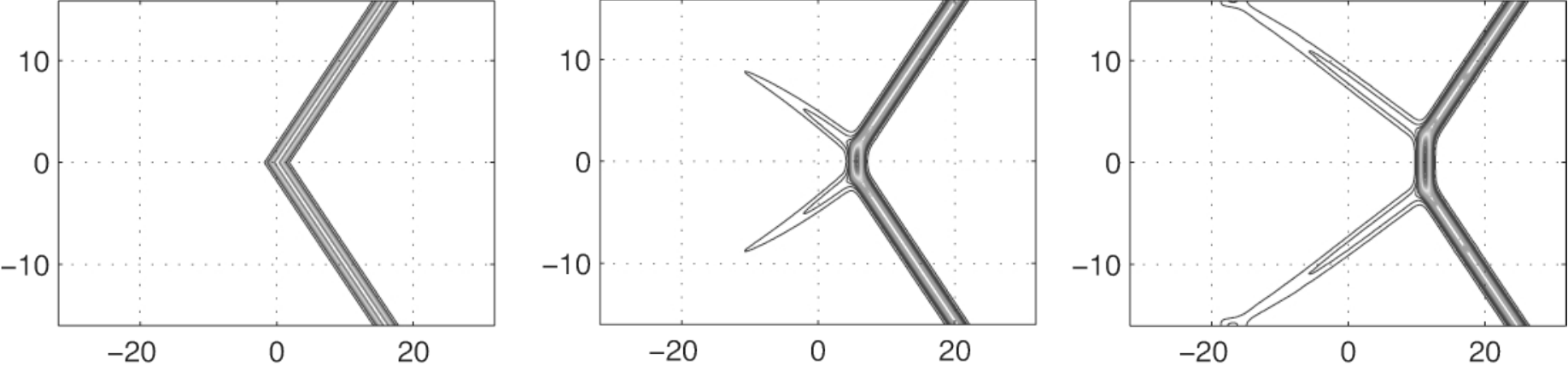} \\
\vskip 0.5cm
\includegraphics[scale=0.75]{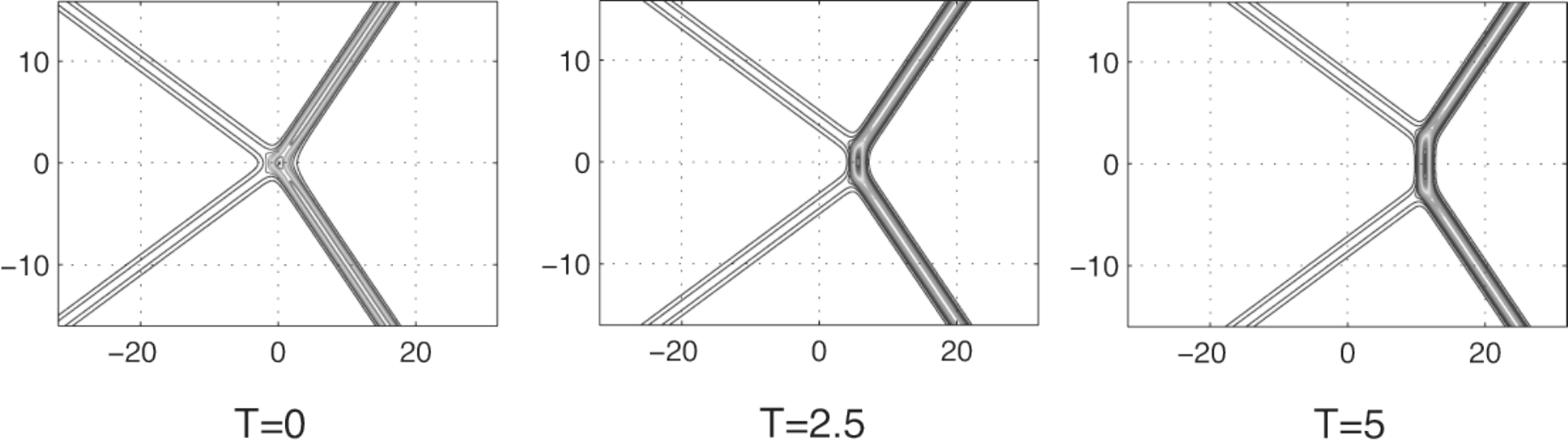} \\
\caption{Numerical simulation with the V-shape initial wave and an exact solution
corresponding to the asymptotic solution.
The parameters are given by $A_0=2$ and $\Psi_0=45^{\circ}$.
The intermediate wave has the amplitude about $4.5$.
This solution seems to converge asymptotically to the $(3142)$-type solution
shown in the second line which is generated by the $A$-matrix
with $a=2, b=\sf{2}{3}$ and $c=2$ (i.e. all four line-solitons meet at the origin at $t=0$).
The parameters are given by $(k_1,k_2,k_3,k_4)=(-\sf{3}{2},-\sf{1}{2},\sf{1}{2},\sf{3}{2})$.
\label{fig:Vinitial}}
\end{figure}

In Figure \ref{fig:Vinitial}, the upper figures illustrate the result of a simulation
with the initial V-shape wave (\ref{initialdata}) with $A_0=2$ and $\Psi_0=45^{\circ}$.
The angle $\Psi_0$ is now less than the critical angle $\Psi_c$.
The asymptotic solution is expected to be of $(3142)$-type whose parameters are determined as follows:
\begin{itemize}
\item[(i)] The $k$-parameters can be obtained from (\ref{NumericalK}) above,
\[
(k_1,k_2,k_3,k_4)=(-\sf{3}{2},-\sf{1}{2},\sf{1}{2},\sf{3}{2}).
\]
The angles of the line-solitons of $[1,2]$- and $[3,4]$-types in $x\ll 0$ are then given by
\[
\Psi_{[3,4]}=-\Psi_{[1,2]}=\Psi_c=\tan^{-1}(2)\approx 63.4^{\circ}.
\]
\item[(ii)]
The elements of the $A$-matrix may be obtained by the condition that
the intersection point at $t=0$ is located at the origin for the initial solitons
identified as $[1,3]$- and $[2,4]$-types.
Then from the phase shift formulas (\ref{shift1324}), we choose
$\frac{c}{b}=3$ and $\frac{c}{a}=1$,
so that $[1,3]$- and $[2,4]$-solitons meet at the origin at $t=0$.
We also expect that all four solitons (two original in $x>0$ and
two other solitons in $x<0$) may be considered to intersect at the origin at $t=0$.
This assumption leads to $s=1$ in (\ref{shiftab}), and
we obtain
\begin{equation*}\label{A3142}
A=\begin{pmatrix}
1&2&0&-2\\0&0&1&\sf{2}{3}
\end{pmatrix}.
\end{equation*}
However this choice of the $s$-parameter may not be correct, since the generation of the line-solitons
in $x<0$ affects the interaction timing at $t=0$ (it seems that a correct choice is $s<1$, and
this will be discussed further in \cite{KK:09}).
\end{itemize}

The lower figures in Figure \ref{fig:Vinitial} illustrate the corresponding exact solution with this $A$-matrix.
One can observe that the asymptotic solution with the V-shape initial wave with $\Psi_0=45^{\circ}<\Psi_c$ converges to the corresponding $(3142)$-type soliton solution. 
An explicit error analysis for the results will be also reported in \cite{KK:09}.

%%%%%%%%%%%%%%%%%%%%%%%%%%%%%%%%%%%%%%%%%%
%%%%%%%%%%%%%%%%%%%%%%%%%%%%%%%%%%%%%%%%

\subsection{The Mach reflection}
\begin{figure}[t]
\centering
\includegraphics[scale=0.48]{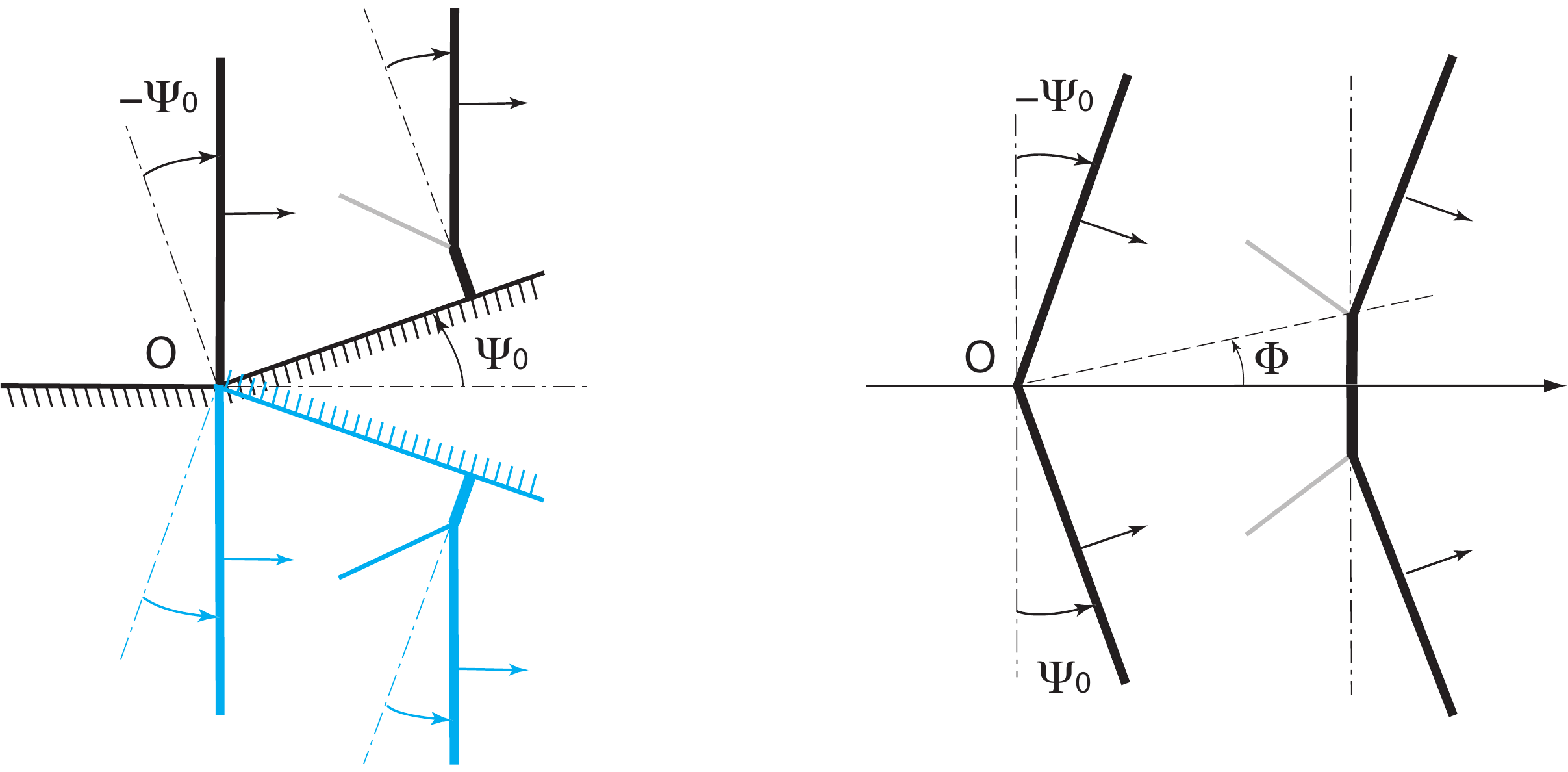} 
\caption{The Mach reflection. The left figure illustrates a semi-infinite line-soliton propagating
parallel to the wall. The right figure is an equivalent system to the left one when we ignore
the viscous effect on the wall. The resulting wave pattern is of $(3142)$-soliton solution.
\label{fig:MR}}
\end{figure}

In \cite{M:77}, J. Miles considered an oblique interaction of two line-solitons  using O-type solutions.
He observed that resonance occurs at the critical angle $\Psi_c$, and when the angle $\Psi_0$
is smaller than $\Psi_c$, the O-type solution becomes
singular (recall that at the critical angle $\Psi_c$, one of the exponential term in the $\tau$-function vanishes). 
He also noticed a similarity between this resonant interaction and the Mach reflection
found in a shock wave interaction. This may be illustrated by the left figure of Figure \ref{fig:MR},
where an incidence wave shown by the vertical line is propagating to the right, and it hits a rigid wall
with the angle $-\Psi_0$ measured counterclockwise from the axis perpendicular to the wall
(see also \cite{F:80}).
If the angle of the incidence wave (equivalently the inclination angle of  the wall) is large, 
the reflected wave behind the incidence
wave has the same angle $\Psi_0$, i.e. a regular reflection occurs. However, if the angle is small,
then an intermediate wave called the Mach stem appears as shown in Figure \ref{fig:MR}.
The critical angle for the Mach reflection is given by the angle $\Psi_c$.
The Mach stem, the incident wave and the reflected wave interact resonantly, and
those three waves form
a resonant triplet. The right one in Figure \ref{fig:MR} illustrates an equivalent system of
the wave propagation in the left figure (one should ignore the effect of viscosity on the wall, i.e. no boundary layer). At the point $O$, the initial wave has V-shape
with the angle $\Psi_0$, which forms the initial data for the simulation.
Then the numerical simulation describes the reflection 
of line-soliton with an inclined wall, and these results explain well the appearance of the Mach
reflection in terms of the exact soliton solution of $(3142)$-type.

%%%%%%%%%%%%%%%%%%%%%%%%%%%%%%%%%%%%%%%%%%
\begin{figure}[t!]
\centering
\includegraphics[scale=0.42]{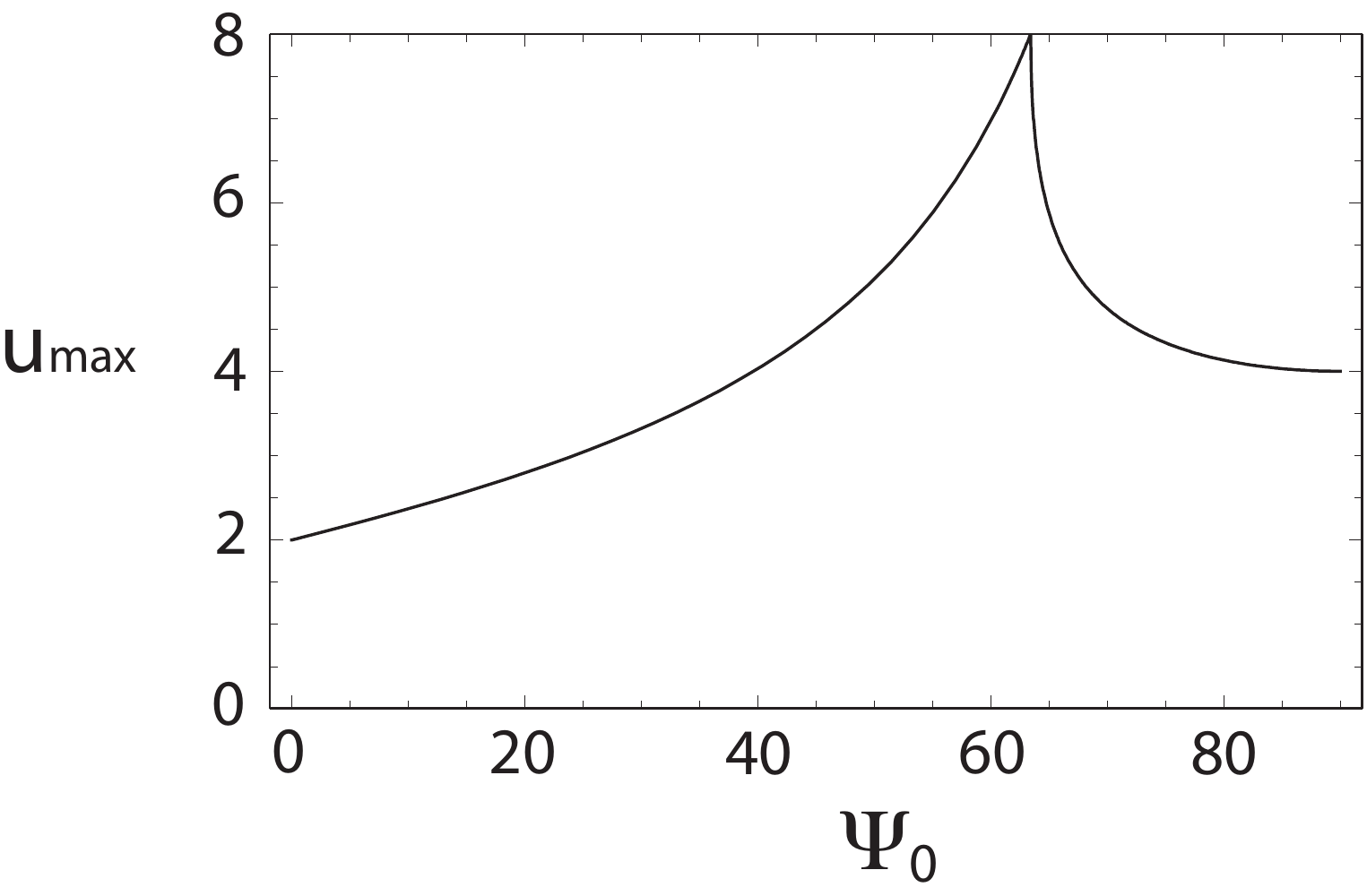} 
\caption{The maximum amplitude obtained from the O-type soliton for
$\Psi_0<\Psi_c$ and the $(3142)$-type soliton for $\Psi_0>\Psi_c$.
The critical angle is $\Psi_c=\tan^{-1}\sqrt{2A_0}\approx 63.4^{\circ}$.
The highest amplitude is obtained at the critical angle, and it is 4 time larger than the
initial soliton.
\label{fig:MaxA}}
\end{figure}

%%%%%%%%%%%%%%%%%%%%%%%%%%%%%%%%%%%%%%%%
The maximum amplitude for this problem occurs at the wall, called the maximum run-up,
and this can be obtained by the formula (\ref{Omax}) for the O-type solution, i.e. 
$\Psi_0>\Psi_c$, and the formula (\ref{stemA}) for $(3142)$-type solution.
Figure \ref{fig:MaxA} illustrates the maximum amplitude for various angles $\Psi_0$.
Notice that at the critical angle $\Psi_c\approx 63.4^{\circ}$, the amplitude takes four times
higher than that of the incident wave (this figure was first obtained in \cite{M:77}).
One can also find the length of the Mach stem (i.e. the intermediate soliton of $[1,4]$-type)
from $(3142)$-soliton solution, that is, for the point $(x,y)$ of the resonant interaction of
the triplet, we have
\[
x=\frac{1}{4}\left(\tan\Psi_c+\tan\Psi_0\right)^2 t,\qquad y=\frac{1}{2}\left(\tan\Psi_c-\tan\Psi_0\right)t,
\]
where $\tan\Psi_c=\sqrt{2A_0}$ and the point O is assigned as $(0,0)$. The angle $\Phi$ in Figure \ref{fig:MR} is then given by
\[
\tan\Phi=\frac{2(\tan\Psi_c-\tan\Psi_0)}{(\tan\Psi_c+\tan\Psi_0)^2}\,.
\]
This formula might be useful to determine the $s$-parameter for $(3142)$-type soliton solution.
%%%%%%%%%%%%%%%%%%%%%%%%%%%%%%%%%%%%%%%%%%
%%%%%%%%%%%%%%%%%%%%%%%%%%%%%%%%%%%%%%%%

\vskip 1cm
\leftline{\bf Acknowledgements.}
\smallskip

YK would like to thank the organizers, M. J. Ablowitz, D.-Y. Hsieh and J. Yang,
for the invitation. He enjoyed very much the conference and he thinks that the
conference is one of the best ones he has ever attended.

We would like to thank M. Oikawa, H. Tsuji at Kyushu University for many valuable discussions on their works of the initial value problem with V-shape initial waves.
We would also like to thank our colleague, C.-Y. Kao, for letting us to use her remarkable results of the numerical
simulations of the KP equation.

%%%%%%%%%%%%%%%%%%%%%%%%%%%%%%%%%%%%%%

\end{document}